\documentclass[twocolumn,showpacs,preprintnumbers,amsmath,amssymb,floatfix]{revtex4} 
\usepackage{amssymb,amsmath,eepic,epsfig,psfrag}
\usepackage{epsfig,graphics}
\usepackage{dcolumn}
\usepackage{bm}

\usepackage{citesort}
\begin{document}
\title{Conceptual aspects of line tensions}
\author{L. Schimmele$^{1,2}$}
\author{M. Napi\'orkowski$^{3}$}
\author{S. Dietrich$^{1,2}$} 
 \affiliation{$^{1}$Max-Planck-Institut f\"{u}r
 Metallforschung, D-70569 Stuttgart, Heisenbergstr. 3, Germany, \\$^{2}$ Institut f\"{u}r
 Theoretische und Angewandte Physik, Universit\"{a}t Stuttgart, Pfaffenwaldring 57,
 D-70569 Stuttgart, Germany,\\
 $^{3}$ Instytut Fizyki Teoretycznej, Uniwersytet Warszawski, 00-681 Warszawa, Ho\.za 69, Poland 
 }
%


%
\begin{abstract} 
We analyze two representative systems containing  a three-phase-contact line:
a liquid lens at a fluid--fluid interface and a liquid
drop in contact with a gas phase residing on a solid substrate.
In addition we study a system containing a planar liquid--gas interface in contact
with a solid substrate. We discuss to which extent the decomposition of the
grand canonical free energy of such systems into volume, surface, and line
contributions is unique in spite of the freedom one has in  positioning the
Gibbs dividing interfaces.
Before being able to define the line contribution to the grand canonical free energy per unit length, 
the so-called line tension, first 
certain properties have to be attributed    
to the bulk phases and to the interfaces; these are chosen in agreement with
what is known for spherical drops and for planar interfaces.
In the case of a lens it is found that the line tension is independent
of arbitrary choices of the Gibbs dividing interfaces.
In the case of a drop, however, one arrives at two
different possible definitions of the line tension. One of them corresponds seamlessly to that applicable
to the lens. The line tension defined this way turns out
to be independent of choices of the Gibbs dividing interfaces. In the case of the second definition,
however, the line tension does depend on the choice of the Gibbs dividing interfaces.
We also provide equations for the equilibrium contact angles which are form-invariant with respect to  
notional shifts of dividing interfaces which change the description of the system but leave
the density configurations unchanged. In order to reconcile the transformation laws for the line tensions
under notional changes of dividing interfaces, which follow from
the decomposition of the grand canonical free energy, with the principle of form invariance and
moreover with the invariance of observables under notional changes,
additional stiffness constants attributed to the line must be
introduced. The choice of the dividing interfaces influences the actual values of the stiffness
constants. We show how these constants transform as a function of the relative displacements
of the dividing interfaces.
Our formulation provides  a clearly defined scheme to determine line properties
from  measured dependences of the contact angles on lens or drop volumes.
This scheme implies relations different from the modified
Neumann or Young equations, which currently are the basis for extracting line tensions from
experimental data.
These relations show that the experiments do not render the
line tension alone but a combination of the line tension, the Tolman length, 
and the stiffness constants of the line.
In contrast to previous approaches our scheme works consistently for any
choice of the dividing interfaces. It further allows us to compare results obtained  by different
experimental or theoretical methods, based on different conventions of
choosing the dividing interfaces.
\end{abstract}
\maketitle{} 
\section{Introduction} 
\renewcommand{\theequation}{1.\arabic{equation}} 
The handling of  small liquid droplets in contact with a gas phase on top of a solid substrate, 
or of liquid lenses at the 
interface of two other fluid phases plays an important role in the context of microfluidics 
(see, e.g., Refs. \cite{Mifl1,Mifl2,Mifl3,Mifl4}).  For 
droplets or lenses with linear extensions in the nanometer regime a proper  
thermodynamic description requires to  account for not 
only the bulk and surface properties but also for the special properties of matter in the vicinity of three-phase contact. In 
order to capture the contribution of the three-phase-contact 
region  to the relevant thermodynamic potential,   a certain excess contribution to the appropriate free energy that scales  
with the linear extension of the system is associated with the three-phase-contact line,  defined by a common intersection of 
the interfaces meeting in the region of three-phase contact.  
\par 
The liquid lens is an example of an inhomogeneous system in which   
three thermodynamic phases, say $\alpha$, $\beta$, and $\gamma$, coexist in a (constrained) equilibrium.
The thermodynamic coexistence of bulk phases is provided by specific choices of the thermodynamic state
of the system whereas their spatial coexistence follows from appropriate boundary conditions.  
We consider the following set-up. A drop of the non-wetting $\beta$ phase is placed at the interface between 
two other fluids. A microscopically thin equilibrium wetting film of $\beta$ phase is formed at the 
interface and for suitably chosen substances and conditions a surplus of $\beta$ phase forms a lens 
at the $\alpha$--$\gamma$ interface.
(In principle the exploration of the configuration space allows
for shifting the lens laterally along the interface; however, for the purposes of the present paper one may
disregard this degree of freedom.) 
For such a system two basic scenarios are possible.
The lens ($\beta$ phase) can exchange matter with the surrounding phases, so that the chemical
potentials in all phases are equal. There are cases in which the lens is in a stable equilibrium
with the surrounding phases; in other cases the lens is unstable but could be stabilized by imposing
suitable constraints.     
Alternativly, one may consider a nonvolatile liquid ($\beta$ phase), i.e., one constrains the volume of the liquid 
while chemical equilibrium is not attained.    
\\  
In addition to the concept of 
interfaces separating the bulk phases and of interfacial or surface tensions, the systematic thermodynamic description of such 
systems leads to considering the contact line ${\cal{L}}_{\alpha\beta\gamma}$ along which the three interfaces meet. The 
line tension  $\tau$ is attributed to the 
contact line ${\cal{L}}_{\alpha\beta\gamma}$. It is defined as 
the 'line contribution' to the grand canonical free energy $\Omega$ of the inhomogeneous system per unit length of the 
contact line \cite{Gib1,Lang1,RowWid,Wid1,Wid5} which is the leading term left after subtracting volume
and surface or interface contributions from $\Omega$, i.e.,  
\begin{multline} 
 \frac{\Omega - 
(\sum\limits_{\kappa=\alpha,\beta,\gamma} V_{\kappa} \,\,\omega_{\kappa}) -  A_{\alpha\beta}\,\sigma_{\alpha\beta} - 
A_{\alpha\gamma} \,\sigma_{\alpha\gamma} -A_{\beta\gamma}\,\sigma_{\beta\gamma}}{L_{\alpha\beta\gamma}} 
  \\
 = \, \tau \, + \, {\mathrm{ s.l.t.}} \quad.     
\label{eta}       
\end{multline} 
With s.l.t. we denote subleading terms which vanish for $L_{\alpha\beta\gamma} \longrightarrow \infty$. 
The symbol $\omega_{\kappa}$ denotes the grand canonical free energy density of the homogeneous bulk phase 
$\kappa$ ($\kappa =\alpha, \beta, \gamma$), $V_{\kappa}$ is the volume assigned to this phase,   $A_{\kappa\kappa'}$ is 
the area of the $\kappa$--$\kappa'$ interface, $\sigma_{\kappa\kappa'}$ is the corresponding interfacial tension, and 
$L_{\alpha\beta\gamma}$ is the length of the three-phase-contact line. The definition of $\tau$ in Eq. (\ref{eta})  
refers to a reference state in which 
uniform bulk phases are extrapolated right up to the interfaces, and analogously the interface or surface properties of 
laterally homogeneous interfaces or surfaces are extrapolated right up to the contact line ${\cal{L}}_{\alpha\beta\gamma}$.  In 
Eq. (\ref{eta}) we do not take into account contributions to $\Omega$ related to the presence of walls enclosing the whole system;  
the only inhomogeneities of the system which 
are relevant in the present analysis are those related to the spatial coexistence of the three thermodynamic phases $\alpha$, $\beta$, and 
$\gamma$.  The above decomposition allows one to calculate the line tension for a given thermodynamic system  provided 
the quantities on the lhs of Eq. ({\ref{eta}}) are determined in separate preceding steps involving similar considerations of suitable  
thermodynamic limits.  
\par 
A different typ of an inhomogeneous fluid system is a sessile drop on a solid substrate. In this case only two 
thermodynamic phases 
coexist whereas the substrate acts as an inert spectator phase. In such cases the three-phase-contact line 
corresponds to the  region where the interface between two coexisting thermodynamic phases meets the substrate 
\cite{Pom1,JoGe1,PomVan1,Gen01,D01,Wid2t,Ind2t,Wid3,GD1,BD,Dob2,Dob3,White1,Wayn1,Dus1,AmirNeu}.
In the literature the term line energy or line elasticity is also frequently used in connection with an extra surface energy
associated with the deformation of the three-phase-contact line as the result, e.g., of interactions with 
surface defects (see, e.g., Refs. \cite{JoGe1,PomVan1,Gen01,Raph1t,Pom1}).
This quantity has to be clearly distinguished from the   
line tension as introduced via Eq. ({\ref{eta}}).   
\par 
In the literature the notion of line tension is also used to describe the one-dimensional interface of two coexisting, intrinsically 
two-dimensional phases such as liquid- and vapor-like phases in Langmuir--Blodget films. This, however, corresponds 
to the lower-dimensional version of an interfacial tension and not to the coexistence of three phases as considered here.
 \\
There is a growing body of literature describing both experimental and theoretical investigations of the line tension 
and already a number of reviews have been published on that subject \cite{Ind2t,Drelich,AmirNeu,Tosh1e,Rusanov-0,Rusanov-2}.
Experimental investigations were carried out for drops on solid substrates 
 \cite{Law2e,Wang1,Pom2e,Pom3,Bue1,GaNe1,Tosh1,Dun1e,Dus1e,Pom1e,Amir3e,Herm1e,Herm2e,See1,Daillant,Checco,Hoorfar,StaRaSchu,Amirfazli,GaNe2,Wang2}, for liquid lenses at the interface of two fluids
\cite{Tosh1,StaRaSchu,Dussaud,Chen4,Chen5,Aveyard,Li1,Takata}, and for spherical particles or bubbles at the interface of two
fluids \cite{Schel,Plat1,Dim1,Ave1,JensLi,Iva1,Broch10,Butt1,Butt2,Butt3}; also the role of line tension in epitaxial growth
has been discussed \cite{Goe}. On the theoretical side the theory of capillarity has been extended in a phenomenological
way by taking into account line contributions and the consequences of these extensions have been explored 
\cite{Wid1,Buk3,BorNeu,Pethica,Iva1,Chen3,Chen1,Marmur-n1,Marmur-n2,Babak,SoloWhite,Xia,KubNap1,GaNe2,Wid3,Li1,Brink,Brink-b,Brink-c,Gretz,Buff1,Guzz}.
Furthermore there are microscopic calculations of the line tension  
\cite{Wid5,GD1,BD,Dob2,Dob3,WidCl1,WidWid1,VarRob1,Perk1,Buk1,Ker1,Nav1,Tar1,JoGe2,Dob1,Gand} 
and many studies, in most cases based on microscopic theories as well, concentrate on the behavior of
the line tension in the vicinity of a wetting transition 
\cite{Wid2t,Ind2t,Szl1,Ind1,IndBacLan1t,IndRob1t,VarRob2,DobInd1t,IndDob1t,RobInd0,Blos1,Perk2,Widom-n4}. 
Theoretical studies are also devoted to the examination of the influence of the droplet size on the line tension 
\cite{Jakub} and of the electrostatic contributions to the line tension which arise in the presence of surface charges
\cite{Chou1}.   
The boundary tension between two coexisting wetting films of different thicknesses was studied as well 
(see, e.g., Refs. \cite{Perk1,Perk2,Perk3,Erring}).
A drop-size dependence of the contact angle was studied in molecular dynamics simulations without deducing
values for the line tension (see, e.g., Ref. \cite{Guo}). 
Finally, the line tension was studied also via molecular dynamics simulations (see, e.g., Refs. \cite{Bres4,Werder})
or via an analysis of probability distributions as obtained from Monte Carlo simulations at 
three-phase contact \cite{Djikaev-n1}.   
\par 
A closer inspection of these results reveals that certain aspects of the line tension either give rise to conflicting 
statements or 
remain unaddressed, leading us to set out to clarify the following basic questions: 
\begin{itemize}
\item { {In which sense is the concept of the line tension well defined? }}
\item { { Is it sufficient for the determination of equilibrium shapes to characterize the thermodynamics of the contact 
line by a line tension only?}}
\end{itemize}
These questions arise because it is not obvious that the definition of $\tau$ via 
Eq. (\ref{eta}) leads to a unique result. The reason lies in the arbitrariness in the definition of the position of the interfaces 
between 
adjacent phases. This arbitrariness is due to the smooth spatial transition between the adjacent phases.
Once the density distributions of the fluid components across the interfacial region are known, for example on the 
basis of scattering experiments or atomic force microscopy \cite{Pom2e,Pom3,Bue1}, or theoretically by simulations or 
density-functional calculations, a criterion has to be applied which fixes the interface 
position somewhere in the transition region. However, there exists a multitude of sensible choices. The 
arbitrariness in the definition of the interface positions leads to an arbitrariness in the definitions of the volumes, 
areas, lengths,  
and to some extent even interfacial tensions (see, e.g., Refs. \cite{RowWid,Kalikman,Tolman,Hill,Buff,Kondo,Hend}).  
Although it was shown by Widom \cite{Wid1,RowWid} that 
$\tau$ is independent of the particular choice of the dividing interfaces in the case of a straight three-phase-contact line in 
a system with three thermodynamically coexisting fluid phases separated by planar interfaces, it is not clear that the same will 
be true in other cases as well, e.g., if two fluid phases are in contact with an inert solid phase, or for systems in 
which the 
curvatures of the interfaces and of the contact line plays a role.
A first discussion, which naturally raises the issues, concerning the uniqueness of the line tension for the case 
in which two fluid phases are in contact with and meet at an inert solid phase, is given in Sect. 3. 
This discussion, however, turns out to be incomplete and it leads to a contradiction with one of 
the results obtained in Sect. 5, which states that the line tension {\em as defined there} is unique. 
A thorough discussion
and the resolution of the contradiction is given in Sect. 6. 
In Sect. 5 we mainly investigate the three-phase-contact line in systems with curved interfaces.  
On the other hand, just these systems 
are studied experimentally if one attempts to determine the value of the line 
tension, e.g., via measuring the dependence of the contact angle on droplet size (see, e.g., Refs. \cite{Pom1e,Pom2e,Herm1e,Herm2e,Drelich,Dussaud,Chen4,Chen5,Aveyard,Amirfazli,Law2e,Li1,Takata}).  
\par 
The curvature dependence of interfacial tensions leads to contributions to the free energy which scale with the linear  
extension of the system, i.e., very much like the line contribution. Since in many cases interface curvature and the length of the three-phase-contact line 
cannot be varied independently, these two contributions are not distinguishable per se. Thus, line tension and 
curvature effects on the interfacial tensions have to be discussed simultaneously. Moreover, the curvature expansion 
of the surface tension has to be known in advance before the line tension may be determined from Eq. 
(\ref{eta}). \\ 
Since in the present context it is unavoidable to consider curvature effects on the interfacial tensions and because
it is known that a consistent 
description of curved (spherical) interfaces must take into account --- for general dividing interfaces --- 
derivatives of the surface 
tensions with respect to their radii of curvature, i.e., their bending rigidities (see, e.g., Refs.
\cite{RowWid,Kondo,Hend,Kalikman,Tolman,Hill,Buff,RejNap,T1,T101,T2,T3,T4,T5,T6,T7,T8,T9,T10,T11,T12,T14}), 
we expect that a complete thermodynamic 
description of three-phase-contact lines requires the introduction of further material parameters 
in addition to the line tension. 
Indeed, 
a number of theoretical analyses have appeared in the literature in which additional material parameters were 
introduced in order to characterize the three-phase-contact line \cite{Marmur-n1,Marmur-n2,Babak,SoloWhite,BorNeu}. 
Recent studies have been carried out by Rusanov et al. \cite{Rusanov-1} for a 
drop on a solid substrate, and by Widom and coworkers in connection with the line analogue of the 
Gibbs adsorption equation \cite{Widom-n1,Widom-n2,Widom-n3,Widom-n5,Widom-n6,Widom-n7} 
for a straight contact line at a genuine 
three-phase contact.    
In Sect. 5 we systematically study these issues guided by the concept of form-invariance of 
the basic equations under so-called notional shifts of dividing interfaces (for a definition see, c.f., Sects. 2 and 5). 
\par 
As a prerequisite to a thorough study of these issues we also have to answer similar questions  
related to the concept of surface tensions.  However, before discussing them and in particular the main 
questions concerning the line tension, we point at related issues recurring in 
the literature which can be judged only after the basic questions raised above have found a satisfactory answer. These 
issues are: 
\begin{itemize}
\item What is the typical magnitude of the line tension ? \\
Of course, the line tension depends on the thermodynamic state of 
the system and the number of relevant thermodynamic degrees of freedom varies from system to system. For example, for 
certain systems the temperature dependence of the line tension  becomes especially pronounced close to wetting 
transitions \cite{Wid2t,WidCl1,WidWid1,Szl1,Ind1,Ind2t,Law2e,Wid3,BD,Dob2,Dob3,Sch1t,IndBacLan1t,IndRob1t,VarRob1,VarRob2,Abr1t,DobInd1t,Perk1,IndDob1t,RobInd0,Blos1,Buk1}.  
On the other hand, for many systems away from such special 
thermodynamic states there is now widespread agreement that the value of the line tension is of the order of $10^{-11}$ N 
\cite{Pom2e,Pom3,GD1,BD,Daillant}. With the experimental accuracy available at present the corresponding prefactor distinguishes 
between different systems. 
\item What is the sign of the line tension? 
\\Various experimental and theoretical investigations lead to values of the line 
tension  which include both signs \cite{Clar1,GD1,Tosh1,Herm1e,Rosso,Chen4,Chen5,BD,Wang1}. 
Since - contrary to the interfacial tension  - 
there exists no thermodynamic argument for a specific sign of  
the line tension \cite{Wid1,Guzzardi}, so that experimental findings for the line tension with a certain sign cannot be discarded 
from the outset.
(Even in this respect a contrary statement may be found in the literature \cite{Li2}.)  
On the other hand, if one is interested in the temperature dependence of the line tension for a 
specific system which undergoes a first-order wetting transition, both theoretical and experimental results show 
that upon increasing the temperature towards the wetting transition temperature the line tension changes sign from 
negative to positive values \cite{Ind2t,Law2e,BD,Wang1,Wang2}. 
\item Is a negative line tension compatible with the structural stability of drops? \\ 
The mesoscopic analysis of line tensions including  
small wavelength fluctuations confirms that negative values of the line tension  do not lead to instabilities 
\cite{Dob2,Rosso,Guzzardi}. 
\item What does the line tension depend on? \\
Similarly to the interfacial tension  it is a function of the 
thermodynamic state of the system. For straight contact lines, defined by intersecting planar interfaces,
the phases involved must be at stable thermodynamic coexistence. For example, 
one-component fluids in contact with an inert
substrate must be at liquid--vapor coexistence $\mu = \mu_0 (T)$, where $\mu$ denotes the chemical potential.
Thus in this case the line tension is a function of temperature only. If the fluid in contact with a 
substrate consists of a binary mixture
of A and B particles, the fluid must be at fluid--fluid coexistence, i.e., $\mu^{\mathrm{A}} = \mu^{\mathrm{A}} (\mu^{\mathrm{B}},T)$
which leaves two thermodynamic variables free. For three-phase contact among three fluid phases the system must be at the 
triple line ($\mu^{\mathrm{A}} =  \mu^{\mathrm{A}}_0(T), \mu^{\mathrm{B}} = \mu^{\mathrm{B}}_0(T)$) of A-rich liquid,
B-rich liquid, and vapor coexistence so that in this case the line tension is again a function of temperature only.
This dependence can be reparametrized in terms of the temperature dependent contact angle $\theta (T)$.  
The situation is somewhat different for curved contact lines defined by intersecting curved interfaces.
Droplets of finite size residing on a substrate or liquid lenses formed at planar fluid--fluid interfaces 
are examples in
which the contact lines are curved. 
Due to their curvature droplets or lenses remain in a (constrained)  
equilibrium with their surrounding phases,  
which takes place at chemical potentials slightly off their values $\mu_0^i$ 
at stable thermodynamic coexistence with planar interfaces. Under these conditions 
the pressure is different inside and outside the droplet or lens. This pressure difference and at the same time 
the size of the droplet or lens are prescribed by the chosen chemical potentials or alternatively by the 
temperature and the ambient pressure which deviates from that for stable coexistence at the same temperature.    
Thus in principle the line tension may now depend on a further thermodynamic variable in addition to $T$.
This dependence can be reparametrized in terms of the size of the droplet or lense.  
  \\
In addition, from the theoretical point of view the line tension $\tau$ is a functional of both the interaction  
potentials between the fluid particles and the substrate potential. If the microscopic forces are too
long-ranged $\tau$ becomes ill defined while the corresponding interfacial and surface tensions retain their validity 
\cite{Ind2t,DN1}, i.e., in this case the size dependence of $\Omega$ cannot be described in terms of a bulk,
surface, and line contribution with a line tension which is size- and shape-independent in the thermodynamic limit.  
 \\
It appears that in extracting line tensions from experimental data so far the curvature dependence of the 
interfacial tension, characterized by the  
Tolman length \cite{T0,ChenTrein,T1,T1b,T101,T2,T3,T4,T5,T6,T7,T8,T9,T10,T11,T12,T14,Anisimov},  
has been completely neglected. Therefore    
the question arises to which extent the experimental determination of the line tension is affected by the 
Tolman length.  
\item Can one measure the line tension? \\
In the last chapter we shall point at possible difficulties to determine line tensions uniquely. 
\end{itemize}
\vspace*{0.5cm}
\section{The substrate--fluid surface tension}
\renewcommand{\theequation}{2.\arabic{equation}}
\setcounter{equation}{0}
\vspace*{0.5cm} 
In the following we discuss a possible source of ambiguity in the definition of the line tension
(see Eq. \eqref{eta}) in the case that one of the 
phases considered in Sec. 1 is an inert substrate, i.e., we consider two-phase  coexistence in the 
presence of an inert wall rather than coexistence of three genuine thermodynamic phases.  We focus on 
the issue of non-uniqueness of the liquid--substrate or gas--substrate interfacial tensions.  
To this end we first briefly mention 
the related question of the thermodynamic determination of the interfacial tension $\sigma_{\kappa\kappa'}$ in a system in which 
two  coexisting {\it fluid phases}, say  $\kappa$ and $\kappa'$, meet along a planar interface \cite{RowWid,Wid1}.  
Using the formula 
\begin{equation}
\label{sigma}
\sigma_{\kappa\kappa'} = \lim_{V_{\kappa,\kappa'},A_{\kappa \kappa'} \rightarrow \infty} \,
                               \frac{\Omega -V_{\kappa} \,\omega_{\kappa} - 
V_{\kappa'} \,\omega_{\kappa'} }{A_{\kappa\kappa'} } \quad, 
\end{equation} 
the question appears whether the corresponding value of the 
interfacial tension depends on the arbitrary choice of the position of the fluid--fluid interface 
and the corresponding volumes $V_{\kappa}$ and $V_{\kappa'}$.   
In the case of a planar interface the value of the interfacial tension does 
not depend on the position of the interface, i.e.,  Eq. (\ref{sigma}) leads to a unique result. The reason 
is that the total volume $V_{\kappa} + V_{\kappa '}$ as well as the surface area $A_{\kappa \kappa '}$ are independent 
of the location of the dividing surface; in addition due to thermal equilibrium one has $\omega_{\kappa} = 
\omega_{\kappa '}$. (For a complete discussion of this issue see, e.g., Refs. \cite{RowWid,Wid1}.)   \\  
However, in situations in which one of the phases, say phase $\gamma$, is an inert substrate with a planar 
surface, a difference to the case of fluid--fluid interfaces arises because the inert substrate is not one of 
the thermodynamically 
coexisting phases. Although it is not a priori obvious whether one should treat the substrate as a part of the system or not, it 
is usually considered as an external object which just provides a steep external potential defining the boundaries of the 
system.  \\ 
It might appear that the freedom in positioning the dividing interface is less obvious for the solid--fluid interface than for the 
fluid--fluid interface because a solid surface is defined rather sharply, say by the positions of the nuclei
of the atoms forming the 
topmost layer. However, even then the position at which an actual experimental technique
(e.g., atomic force microscopy, optical methods, etc.)  
will locate the surface of the solid will certainly deviate  from the definition given above, because of the smooth  
decay of the substrate  potential and of the finite extension of the electron cloud of the substrate, and because 
the fluid 
phase in contact with the solid only gradually attains its bulk properties.  Depending on the kind of experiment and on the 
way the data are analyzed, relative shifts in the location of the dividing surface which are of the order of one or even several 
atomic radii are conceivable. This shift multiplied by the surface tension under consideration yields a force
which is comparable with the magnitude of the line tension. 
Moreover, a solid substrate in contact with a vapor phase might be covered with a  
thin liquid-like wetting film which is in thermal equilibrium with the bulk vapor phase. In a thermodynamic description in terms of interfacial and line tensions  
as considered here, this thin wetting film is not treated as a separate entity but as part of the solid--vapor
interface and as such it contributes to the actual  
solid--vapor ({\em g}as) surface tension $\sigma_{\mathrm{sg}}$. Accordingly the question arises, 
where to place the solid--vapor 
interface. It could be somewhere in the transition region from the liquid-like film to the vapor or at the transition 
region from the solid to the liquid-like film. It is very likely  
that different experimental techniques imply different conventions. 
(If nonequilibrium situations are considered $\sigma_{\mathrm{sg}}$ may be different from its equilibrium value
 if the aforementioned thin wetting film -- present at thermal equilibrium -- has not yet formed.)  
\par  
In order to explore the consequences of the freedom in choosing the dividing surface, we consider a system in which a planar 
inert substrate is exposed to the fluid $f$, i.e., the liquid or gas phase (see Fig. 1). 
\begin{figure}[h] 
\includegraphics*[scale=.30]{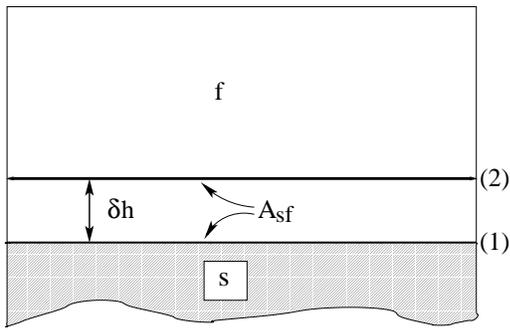}
\caption{\label{fig1} A fluid phase (f) in the presence of a planar substrate (s).
Two choices for the position of the planar substrate--fluid interface are marked as (1) and (2); they are shifted
with respect to one another by a distance $\delta h$. In each case the area $A_{\mathrm{sf}}$ of the
substrate--fluid interface is the same.}
\end{figure}
If we consider the solid as a part
 of our 
system the value of the substrate--fluid surface tension  $\sigma_{sf}$ follows from the equation 
\begin{equation}
\label{sf}
\sigma_{sf} = \lim_{V_{\mathrm{f},\mathrm{s}},A_{\mathrm{s} \mathrm{f}} \rightarrow \infty} \,
                      \frac{\Omega - V_{f} \,\omega_{f}\, - V_{s} \,\omega_{s}}{A_{sf}}  \quad, 
\end{equation}
where $\Omega$ denotes the grand canonical potential of the fluid plus that of the substrate, $V_{f}$ is the value of the fluid 
volume compatible with the chosen location of the planar substrate--fluid interface,  $V_{s}$ is the corresponding volume of
the solid, and $\omega_{f}$ and $\omega_{s}$ are the grand canonical free energy
densities of the fluid and solid, respectively.  $A_{sf}$ is the area of the planar solid--fluid 
interface. If the substrate is in a constrained equilibrium (e.g., no interdiffusion of solid and fluid particles),
one has  $\omega_{f} \neq \omega_{s}$.  \\
Since the position of the substrate--fluid interface is not unique we can consider another position of this 
interface which is parallel to the first one and located a distance $\delta h$ above it (see Fig.1). In this case $V_{f}^{(2)} = 
V_{f}^{(1)} - \delta h \,A_{sf}$,  $V_{s}^{(2)} = V_{s}^{(1)} + \delta h \,A_{sf}$, whereas the surface area 
$A_{sf}$ does not change upon the vertical shift of the interface. It follows from Eq. (\ref{sf}) that 
the  values of the substrate--fluid surface tensions  corresponding to these two choices differ by 
\begin{equation}
\label{deltasigmasf}
\sigma_{sf}^{(2)} - \sigma_{sf}^{(1)} = \left ( \omega_{f}\, - \omega_{s}\, \right )  \delta h \quad. 
\end{equation}
Thus a different choice of the location of the substrate--fluid interface corresponds to a redistribution of the free energy 
between the bulk and the surface terms, and to different values of the substrate--fluid surface tension. Note that in the case 
of {\it thermodynamically coexisting \rm} liquid and gas phases the grand canonical free energy densities of the two 
phases (i.e., the negative pressures) are equal ($\omega_{l} = \omega_{g} $) and thus 
the well known conclusion $\sigma_{lg}^{(2)} = \sigma_{lg}^{(1)}$ follows 
(see Refs. \cite{RowWid,Wid1}).  
At the same time we emphasize that the difference $\sigma_{sg} - \sigma_{sl}$ (which, e.g., enters into 
Young's law for the contact angle (see, c.f., Eq. (\ref{Young})) does not depend on this choice of  
the dividing surface provided the dividing interfaces between solid and gas 
on one hand and between solid and liquid on the 
other hand are chosen to be at the same height above the solid, and provided the solid is in the same state in 
both cases.   
 
\section{Three-phase-contact line at intersecting planar interfaces: uniqueness of the line tension}
\renewcommand{\theequation}{3.\arabic{equation}} 
\setcounter{equation}{0}\vspace*{0.5cm} 
In the case of three genuine thermodynamically coexisting phases the issue analogous to the one raised in Sect. 2 is 
whether the value of the line 
tension $\tau$ determined from Eq. (\ref{eta}) depends on the location of the dividing interfaces which in turn determine the 
location of the contact line. One can show \cite{RowWid} that in the case that these phases meet along a {\it straight line 
\rm} parallel shifts of this line do not influence the value of the line tension. \\
In order to study whether the location of the substrate--fluid interface affects the value of $\tau$, we investigate the following 
configuration of two-phase coexistence in the presence of a planar solid substrate.  
Due to thermal equilibrium of the gas and liquid phase one can impose lateral boundary conditions such that far
to the left the substrate is exposed to the gas phase whereas far to the right the substrate is exposed 
to the liquid. This enforces the formation of the liquid--gas interface which meets the substrate with
a contact angle $\theta$ (see Fig. 2). On a macroscopic scales the liquid--gas interface is also planar. 
\begin{figure}[t,b] 
\includegraphics*[scale=.40]{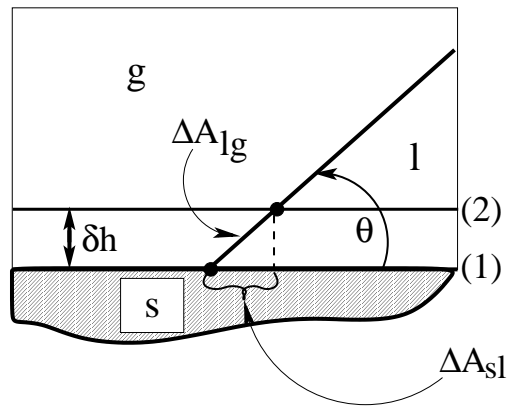}
\caption{\label{fig2} Coexisting gas (g) and liquid (l) phases in
the presence of a planar substrate (s). The planar liquid--gas interface meets the substrate with
a contact angle $\theta$. Far to the left (right) there is the substrate--gas (--liquid) interface.
Two choices of the position of the substrate--fluid interfaces are
marked by (1) and (2); their distance is denoted by $\delta h$.
This results in two parallel contact lines the cross sections of which are indicated by the dots.
$\Delta A_{\mathrm{lg}}$ and $\Delta A_{\mathrm{sl}}$
denote the corresponding changes in the liquid--gas and substrate--liquid interfacial areas, respectively.
For both choices the contact angle $\theta$ is the same.}
\end{figure}
\\ 
We again consider two parallel positions of the substrate--fluid interface at a distance $\delta h$ from each other. 
This results in two 
corresponding, parallel contact lines.  From simple geometrical considerations it follows that the values of the 
line tensions corresponding to these choices differ by 
\begin{equation}
\label{deltaeta1}
\tau^{(2)} - \tau^{(1)} \,=\, \frac{\sigma_{lg} + (\sigma_{sl}^{(1)} -\sigma_{sg}^{(1)})\cos\theta}
{\sin\theta} \,\delta h \quad.
\end{equation}
In the calculation leading to Eq. (\ref{deltaeta1}) we have used the aforementioned result 
$\sigma_{sg}^{(2)} - \sigma_{sl}^{(2)} = \sigma_{sg}^{(1)} -\sigma_{sl}^{(1)}$ and the fact that the freedom 
in the choice of the substrate--fluid 
interface position does not influence the value of the liquid--gas interfacial tension 
$\sigma_{lg}$, i.e.,  $\sigma_{lg}^{(2)}= \sigma_{lg}^{(1)}=\sigma_{lg}$.  
After using Young's equation \cite{RowWid,Gen01,D01} 
\begin{equation}
\label{Young}
\sigma_{sg} = \sigma_{sl} + \sigma_{lg} \cos\theta_{0}
\end{equation}
and identifying $\theta = \theta_{0}$ one obtains
\begin{equation}
\label{deltaeta2}
\tau^{(2)} - \tau^{(1)} = \sigma_{lg} \,\delta h \, \sin\theta_{0} \quad.
\end{equation}
This result reflects the fact that there is no explicit force balance perpendicular to the solid surface. Taken at face value the 
above result would show that the freedom in positioning the substrate--fluid interfaces with the consequential shift 
of the contact 
line is reflected in the change of the value of the line tension. This change is proportional to the distance $\delta h$ between 
the two arbitrarily selected positions of the substrate--fluid interfaces. Numerically, the rhs of Eq. (\ref{deltaeta2})
is comparable with $\tau$. However, Eq. (\ref{deltaeta2}) is in conflict with a result which will be obtained in 
Sect. V of the present work,
namely the invariance of the line tension with respect to notional changes of the system. This 
puzzle will be resolved in Sect. VI.
Here we only note that it turns out that the two conflicting statements are based on two different 
definitions of a line tension both of which seem to be absolutely compelling from the point of view of how
they are introduced. A relation between the two definitions of $\tau$ will be given. But we also point
out that the way how Eq. (\ref{deltaeta2}) has been obtained above has to be scrutinized carefully,   
because, while deriving Eq. (\ref{deltaeta2}), all contributions to $\Omega$ which are generated 
by separating a subsystem from its surrounding
and which are proportional to the linear extension of the system,  
have been disregarded. We shall show that in general this is not permissible.   \\
%
\section{ Line tension and contact angles: problems   
associated with the standard modified Neumann  and Young equations  }
\renewcommand{\theequation}{4.\arabic{equation}}
\setcounter{equation}{0}\vspace*{0.5cm} 
We start this section by considering genuine three-phase coexistence in which a lense consisting of a fluid phase $\beta$ is located 
at the interface between two fluid phases $\alpha$ and $\gamma$ (see Fig. 3(a)). 
\begin{figure}
\includegraphics*[scale=.30]{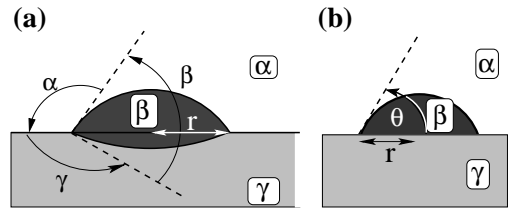}
\caption{\label{fig3} (a) A liquid lens (phase $\beta$) at the $\alpha$--$\gamma$
interface between two fluid phases $\alpha$ and $\gamma$; $r$ is the radius of the circular intersection
between the two spherical caps forming the lens. The various contact angles are denoted also as $\alpha$,
$\beta$, and $\gamma$.
(b) A sessile liquid drop (phase $\beta$) in contact with its vapor (phase $\alpha$) and a planar
substrate ($\gamma$) forming a contact angle $\theta$. The drop is a spherical cap. }
\end{figure}
The lense is taken to be formed by two 
spherical caps of different radii intersecting along a circle of radius $r$. The three-phase-contact line of circular shape is 
accompanied by the line tension $\tau$. To simplify the notation also the corresponding contact angles are denoted by the 
same symbols as the phases, i.e., $\alpha$, $\beta$, and $\gamma$, where $\alpha + \beta + \gamma = 2\pi$. In the absence of the 
line tension the contact angles  $\alpha_0$, $\beta_0$, and $\gamma_0$ fulfill the equation
(see, e.g., Ref. \cite{RowWid}) 
\begin{equation}
\sigma_{\alpha\gamma} + \sigma_{\alpha\beta} \, \cos\alpha_0 \, + \, \sigma_{\beta\gamma} \, \cos\gamma_0 \, = \, 0 \quad.
\end{equation}
If a  line-tension contribution to the constrained  
grand canonical free energy $\tilde{\Omega}$   
is included one obtains, from the minimization of 
$\tilde{\Omega}$ at a constant volume of the liquid phase $\beta$, the modified Neumann equation   
(see, e.g., Refs. \cite{Buff1,Dussaud}) 
\begin{multline}  
\label{mye1}
\sigma_{\alpha\beta} \,\left(\cos\alpha\,-\, \cos\alpha_0 \,\right)\, + \, \, \sigma_{\beta\gamma} \, \left(\cos\gamma\,-
\,\cos\gamma_0\right)\,=\, \frac{ \tau}{r}\, ,    
\\
\end{multline}  
provided neither bending rigidities of the interfaces nor further properties attributed to the contact line 
other than the line tension (such as rigidities against changes of contact angles) are taken into account.  
Equation (\ref{mye1}) can be equivalently rewritten as 
\begin{equation}
\label{mye10}
\cos\beta \,=\, \cos\beta_0 \,-\,\frac{\sin\beta_0}{\sin\alpha_0}\,\,\frac{\tau}{\sigma_{\alpha\beta}\,r} 
\quad.
\end{equation}
In particular, if one of the phases, say phase $\gamma$, is taken to represent an inert substrate with planar surface 
($\beta_{0}=\pi - \alpha_{0}$) (see Fig. 3(b)), the above equation turns into the modified Young equation
(see, e.g., Refs. \cite{Tosh1,Gretz,BorNeu}) 
\begin{equation}
\label{mye2}
\cos\theta = \cos\theta_0 - \frac{\tau}{\sigma_{\alpha\beta}\,r} \quad,
\end{equation}
where $r$ denotes the radius of the circular substrate--phase-$\alpha$--phase-$\beta$ contact line. Equations 
(\ref{mye10}) and (\ref{mye2}) represent asymptotic formulae valid in the limiting case of large lenses or drops,
i.e., $\tau$ governs the {\it leading} behavior for $r \rightarrow \infty$. 
  \\ 
We note that in many experiments involving sessile liquid drops a so-called line tension  $\tau$ is 
deduced from measurements 
of the contact angle $\theta$ as a function of the radius $r$ via fitting the modified Young equation (Eq. (\ref{mye2})) 
to the data  
\cite{RowWid,Pom2e,Pom3,Law2e,Wang1}. Similar experiments have been carried out with lens-like objects
 \cite{Dussaud,Chen4,Chen5,Aveyard,Li1,Takata}.  \\ 
However, a closer look at the procedure described above of determining $\tau$ reveals problems 
in using the modified Young 
(Eq. (\ref{mye2})) or Neumann equations (Eqs. (\ref{mye1}, \ref{mye10})) which are related to the freedom 
in positioning the 
dividing interfaces. A shift of the solid--liquid dividing interface (in the case of a sessile drop) by  
$\delta h$ or a change of the radius 
of a spherical interface (for the lens or the drop) by $\delta R$ leads to changes of the contact angles as well. From 
simple geometrical considerations one finds that the corresponding changes in $\cos \theta $ or  $\cos \beta $ 
are of the order 
$ \delta h / r$ or $ \delta R/ r$, i.e., they are of the same order as the corrections stemming from the 
presence of the line tension.  Therefore, upon  
applying the modified Neumann or Young equation to the same physical object but with different choices for the dividing 
interfaces one would have to introduce two different and suitably chosen values of $\tau$ for different dividing interfaces in order to obtain  
the correct relations between the two corresponding contact angles. (In the line-tension related correction terms 
in Eqs. (\ref{mye10}, \ref{mye2})  
the quantities other than $\tau$ are either independent of the choice of dividing interfaces or their 
changes with the dividing interfaces give rise to higher order corrections.)  On the other hand, by decomposing 
$\Omega$ in two different ways we had found that for a straight contact line at a genuine three-phase contact the line tension 
is independent of the choice of dividing interfaces. This contradicts the result of the previous argument. Finally, we can also 
look at what happens if we shift the substrate--fluid interface for the sessile drop. If we compute the 
difference $\tau^{(2)} - \tau^{(1)}$ enforced by the geometrical relations between $\theta^{(2)}$ and $\theta^{(1)}$ together with 
the requirement that both $\theta^{(2)}$ and $\theta^{(1)}$ fulfil the modified Young equation we 
obtain $\tau^{(2)} - \tau^{(1)} = - \sigma_{\mathrm{lg}}\delta \mathrm{h} \sin \theta _0 $ which has the same structure
as the 
value given in Eq. (\ref{deltaeta2}) (obtained from a comparison of two different ways of decomposing $\Omega$), but 
it has the opposite sign.
 \\
Summarizing these two findings it turns out that the relations between two line 
tension values obtained from two different decompositions of $\Omega$ for two different sets of dividing interfaces are at 
variance with the currently used modified Young and Neumann equations combined with elementary geometrical 
considerations.\\ 
In order to find equations free from the above inconsistencies we shall investigate in detail two representative 
systems: a liquid lens at the interface between two fluid phases and a sessile drop on a substrate. We shall include from the 
outset the effects of curvature on the interfacial tensions and we shall explicitly state all conventions used in defining different 
sets of dividing interfaces, and list all the properties which are assigned to the interfaces and to the reference bulk phases.    
\section{ A closer look at lenses and drops} 
\renewcommand{\theequation}{5.\arabic{equation}}
\setcounter{equation}{0}\vspace*{0.5cm} 

In this section we first study a lens-shaped fluid phase $\beta$ located at the planar interface between two other  
fluid phases $\alpha$ and 
$\gamma$, and secondly a sessile liquid drop of $\beta$ phase in contact with its vapor $\alpha$ on top of an undeformable 
inert solid substrate $\gamma$. 
\subsection{ General considerations} 
In this context two main questions are addressed. First, how does the line tension depend on `parallel` displacements 
of the positions of the dividing interfaces between two phases? Secondly, what can be learned from the fact that 
thermodynamic potentials of the total system must be independent of arbitrary conventions regarding the choice of dividing 
interfaces? In particular, we are interested in the resulting requirements with respect to the structure of the 
equations relating the contact angle(s) 
to the system size, e.g., to the radius $r$ of the three-phase-contact line. These 
questions are motivated by the inconsistencies encountered for the modified Young and Neumann equations   
in the previous chapter.  
\\
First, we specify the systems in more detail and give the rules that we have chosen to define the dividing 
interfaces. These rules still admit parallel shifts of the dividing interfaces. We also introduce the quantities that are  
used in order to describe the shifted dividing interfaces. It is further necessary to fix the properties attributed to the  (reference) bulk and the 
interfaces present in the lens or drop, and we do this in accordance with what is known about the bulk fluid and 
the interfacial properties of 
spherical drops. Only after this step is completed the properties attributed to the line may be extracted.  \\
The lens and the drop are completely characterized once the density distributions are known for all constituent species at 
given thermodynamic conditions. Only for small drops the line tension and curvature effects are expected to become 
relevant, and in this case gravity can be neglected. 
(The validity of Young's law in the presence of gravity is discussed in Ref. \cite{Blok3t}.) 
We also assume that no other external bulk forces act on the systems. 
Under these conditions, provided the volume of the $\beta$ phase is sufficiently large, one can expect that there exist regions  
of the lens (at some distance away from the three-phase contact lines) where the density distributions exhibit  
-- to a good approximation -- radial symmetry relative to one of the two centers of curvature 
$M_1$ and $M_2$; an additional 'center' at infinity characterizing the 
planar interface is present in the case of a lens (see Fig. 4). 
\begin{figure}[b,t]
\includegraphics*[scale=.30]{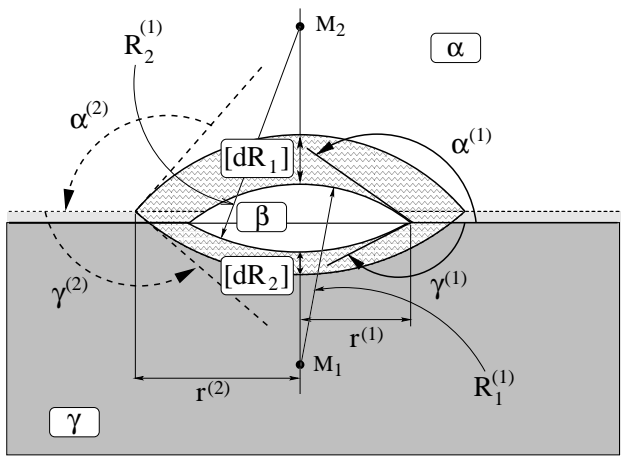}
\renewcommand{\figurename}{Fig.}
\caption{\label{fig4} A liquid lens at the planar $\alpha$--$\gamma$ interface (horizontal lines).
Two choices for the interfaces are marked by (1) and (2).
The relative concentric shifts of the spherical interfaces are characterized by $[\mathrm{d}R_1]$ and $[\mathrm{d}R_2]$.
The planar $\alpha$--$\gamma$ interface is shifted 
together with the spherical ones, as indicated by the dashed and solid horizontal lines and
the change in the amount of $\gamma$ phase shown in lighter gray.
The corresponding contact angles are denoted by ${\alpha}^{(1)}$, ${\gamma}^{(1)}$,
${\beta}^{(1)} = 2\pi - {\alpha}^{(1)} - {\gamma}^{(1)}$
and ${\alpha}^{(2)}$, ${\gamma}^{(2)}$, ${\beta}^{(2)} = 2\pi - {\alpha}^{(2)} - {\gamma}^{(2)}$;
the contact-line radii
are $r^{(1)}$ and $r^{(2)}$. M$_1$ and M$_2$ are the centers for the radii of curvature
$R_1^{(1)}$, $R_1^{(2)} = R_1^{(1)} + [\mathrm{d}R_1]$ and
$R_2^{(1)}$, $R_2^{(2)} = R_2^{(1)} + [\mathrm{d}R_2]$, respectively.}
\end{figure}
In the case of the drop we correspondingly expect 'radial' 
symmetries around 
one center $M$ plus an additional 'center' at infinity characterizing the  substrate--fluid interface (see Fig. 5).    
For the lens, the center $M_1$ characterizes the density distributions at the 
$\alpha$--$\beta$ interface, the center $M_2$ characterizes those at the $\beta$--$\gamma$ interface. 
\begin{figure}[t]
\includegraphics*[scale=.30]{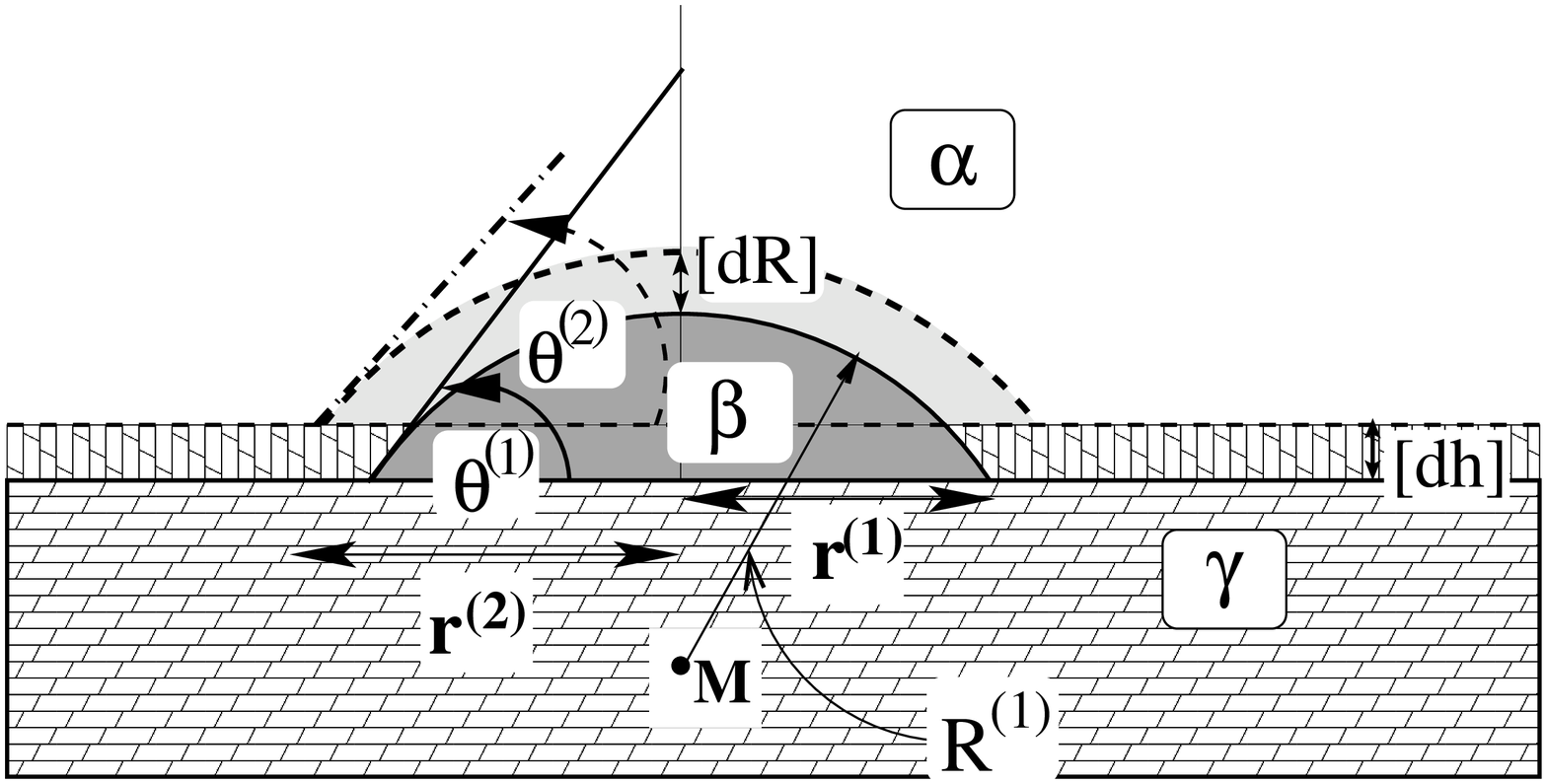}
\renewcommand{\figurename}{Fig.}
\caption{\label{fig5} A sessile liquid drop on a planar
substrate. Two choices for the liquid--vapor ($\beta$--$\alpha$), the substrate--liquid
($\gamma$--$\beta$), and substrate--vapor  ($\gamma$--$\alpha$) interfaces are denoted as (1) and (2);
the corresponding concentric shifts are characterized by $[\mathrm{d}R]$ and
$[\mathrm{d} h]$. The corresponding contact angles are denoted by ${\theta}^{(1)}$ and ${\theta}^{(2)}$,
and $r^{(1)}$ and $r^{(2)}$ are the corresponding contact-line radii. M is the center of the radii of
curvature.
}
\end{figure}
In view of these radial symmetries it makes only sense to consider concentric shifts of the interfaces
with respect to the fixed centers. The two phases 
$\alpha$ and $\gamma$ are assumed to be separated by a planar interface. If the $\alpha$--$\beta$ and $\beta$--$\gamma$ 
interfacial structures do not overlap except near the three-phase region, the interior of the lens is occupied by
an almost homogeneous $\beta$ phase.   
This homogeneous $\beta$ phase interpolates smoothly between the radially symmetric density 
distributions around the two centers associated with the two interfaces.   Similar considerations apply to the drop except 
that the $\beta$--$\gamma$ interface is planar which can be regarded as a limiting case of a spherical interface with 
infinite radius of curvature. 
\\
From the previous remarks it follows that to a large extent the isodensity contours
of the $\alpha$--$\beta$ and $\beta$--$\gamma$ interfaces  
are segments of spherical surfaces.  In order to define  Gibbs dividing interfaces separating 
two adjoining phases we use the spherical parts of the isodensity contours in a two-phase region and extrapolate 
them into the three-phase-contact region where surfaces of constant densities actually are no longer spherical. Which of the 
infinite number of surfaces of constant densities is chosen in order to construct a Gibbs dividing interface 
is of course a matter of convention.  \\ 
For the lens, once the centers $M_1$, $M_2$ are given and the radii $R_1$, $R_2$ are chosen according 
to a certain convention we can define a three-phase-contact line of circular shape by the intersection of the two spheres 
($M_1$,$R_1$) and ($M_2$,$R_2$) as indicated in Fig. 4. The third, planar Gibbs dividing interface between the phases $\alpha$ 
and 
$\gamma$ is placed then in such a way that it coincides with the plane determined by the previously defined circular 
three-phase-contact line.  
Again this choice is a mere convention but deviating from it would create three different lines of intersection 
between three pairs of interfaces ($\alpha$--$\beta$ intersecting with $\beta$--$\gamma$, $\alpha$--$\beta$ with $\alpha$--$\gamma$, and 
$\alpha$--$\gamma$ with $\beta$--$\gamma$). Furthermore, for such a deviating choice a ring-shaped volume, 
with triangular cross section defined by these three lines and the 
connecting interfaces, could not be assigned to any of the three phases and would have to be treated separately as 
a new ''line phase''. 
This would create an unrewarding complication because the central idea of introducing the mathematical 
Gibbs dividing surface is to allow for a natural extrapolation of  
the properties of the adjacent bulk-like phases from both sides right to the dividing surface were they are 
assumed to change discontinously. Within this approach there is no room for additional phases. A consistent description of the 
actual system follows by defining appropriate surface excess quantities and their densities, like the surface tension. We 
stick to Gibbs' idea and we use the convention described above within which all three lines of intersections 
coincide, and no volume filled with a phase of an unassigned character is left over. \\
The procedure for the drop is analogous. The sphere ($M,R$) defines the $\alpha$--$\beta$ Gibbs dividing surface.  Again,  
the spherical part of the isodensity contour is extrapolated into the three-phase-contact region. 
In the same spirit as above the $\alpha$--$\gamma$  
and the $\beta$--$\gamma$ Gibbs dividing interfaces are chosen to lie in the same plane. The three-phase-contact 
line is defined by the intersection of the $\alpha$--$\beta$ Gibbs dividing surface with the common $\alpha$--$\gamma$ and 
$\beta$--$\gamma$ plane (see Fig. 5). Using this convention has the advantage of being in agreement with the one chosen for the lens in 
the limit of one of the curvature radii becoming infinite. Furthermore, this construction creates 
only one line of intersection instead of two. 
\\ 
Once the dividing interfaces are defined the total volume $V$ is subdivided into domains assigned properly to the 
phases $\alpha$, $\beta$, and $\gamma$. The grand canonical potential of the system is decomposed into the 
corresponding bulk, interfacial, and line contributions. For the lens one has     
\begin{eqnarray} 
\Omega & = & - \, \sum_{\kappa=\left\{ \alpha,\beta,\gamma \right\} } p_{\kappa} V_{\kappa} +  A_{\alpha 
\beta}\sigma_{\alpha \beta} +  A_{\beta \gamma}\sigma_{\beta \gamma}  
 \nonumber\\  
      & & +  \left(A - \pi r^2\right)\sigma_{\alpha \gamma} \,+\, 2\pi r \tau \quad,   
\label{lensdrop1}
\end{eqnarray} 
where $p_{\kappa}$ is the pressure in the bulk-like phase $\kappa$ (= $\alpha$, $\beta$, or $\gamma$) and $V_{\kappa}$ 
is the corresponding volume. $A$ is the area of the planar $\alpha$--$\gamma$ interface in the absence 
of the lens.  
The values of $V_{\kappa}$ depend on the way the dividing interfaces are chosen. The total volume $V = \sum_{\kappa } V_{\kappa}$ of 
the system is independent of this choice, and independent of the physical size of the lens or the drop. \\
For the drop one has 
\begin{eqnarray} 
\Omega & = & \sum_{\kappa=\left\{ \alpha,\beta \right\} } - p_{\kappa} V_{\kappa} +   \omega_{\gamma} 
V_{\gamma} +                    A_{\alpha \beta}\sigma_{\alpha \beta} +  \pi r^2 \sigma_{\beta \gamma} 
 \nonumber\\ 
     & & +  \left(A - \pi 
r^2\right)\sigma_{\alpha \gamma} +                 2\pi r \tau \quad,
\label{lensdrop1b}
\end{eqnarray} 
where $A_{\alpha \beta}$, $A_{\beta \gamma}$ (= $\pi r^2$ for the drop) are the areas of the  $\alpha$--${\beta}$ 
and the $\beta$--$\gamma$ 
interfaces, respectively. $A$ is the area of the planar $\alpha$--$\gamma$ interface (in the absence of the 
drop) and it is independent of the choice of the dividing interfaces. The area $\pi r^2$ 
denotes the $\alpha$--$\gamma$ interfacial area replaced 
by the drop. The radius of the circular three-phase-contact 
line is denoted by $r$ and $\sigma_{\alpha \beta}$, $\sigma_{\beta \gamma}$, and $\sigma_{\alpha \gamma}$ 
denote the surface tensions of the $\alpha$--${\beta}$, $\beta$--$\gamma$, and  $\alpha$--$\gamma$ interfaces, 
respectively. Finally, $\tau$ 
is the line tension, i.e., the excess free energy per unit length of the three-phase-contact line   
we are interested in. 
Strictly speaking the line tension defined via Eqs. (\ref{lensdrop1}) and (\ref{lensdrop1b}) is not identical
to its definition via Eq. (\ref{eta}), because in Eqs. (\ref{lensdrop1}) and (\ref{lensdrop1b}) we first keep
the subleading terms and treat them as subleading contributions to $\tau$, whereas in Eq. (\ref{eta}) we isolate
and drop these terms from the outset. However, if in the following we speak about the line tension itself,
as deduced from the decomposition of $\Omega$ via Eqs. (\ref{lensdrop1}) and (\ref{lensdrop1b}), we shall always
drop the subleading terms. In this sense the definitions in Eqs. (\ref{lensdrop1}) and (\ref{lensdrop1b})
and that via Eq. (\ref{eta}) do agree and we therefore do not introduce a different notation to 
distinguish between these two expressions of the line tension.  
\par 
For the lens the pressures in the $\alpha$ and $\gamma$ phases are equal,   
\begin{equation}
\label{lensdrop1c} 
p_{\alpha} =  p_{\gamma} = p  \/ ,
\end{equation} 
because the coexisting phases $\alpha$ and $\gamma$ are separated by a planar interface. 
For the drop the role of $p_{\gamma}$ is played  
by $ - \omega_{\gamma}$, i.e., the grand canonical free energy density of the solid while for the phase $\alpha$  we set 
$p_{\alpha} = p $. 
The pressure of the $\beta$ phase deviates from $p$ and is written as 
(see, c.f., the discussion following Eq. (\ref{lensdrop4b}))   
\begin{equation}
\label{lensdrop1d}
p_{\beta} = p + \Delta p  . 
\end{equation}
\\
The expression for the grand potential $\Omega$ can be regrouped as  
\begin{equation} 
\label{lensdrop2}
\Omega = \Omega_0 + \Delta \Omega   \quad,
\end{equation} 
where 
\begin{equation}
\label{lensdrop3}
\Omega_0 = -p \,V + A \,\sigma_{\alpha \gamma} 
\end{equation}
for the lens, and  
\begin{equation}
\label{lensdrop3b}
\Omega_0 = -p \left ( V_{\mathrm{\alpha}} + V_{\mathrm{\beta}} \right ) + \omega_{\gamma} V_{\gamma} +   A \sigma_{\alpha 
\gamma}
\end{equation} 
for the drop. Consequently one obtains   
\begin{equation}
\label{lensdrop4}
\Delta \Omega = - \Delta p \,V_{\beta} + A_{\alpha \beta } \sigma_{\alpha \beta }   + 
A_{ \beta \gamma}  \sigma_{ \beta \gamma} -\pi r^2 \sigma_{\alpha \gamma}                + 2 \pi r \tau   
\end{equation}
for the lens, and  
\begin{equation}
\label{lensdrop4b}
\Delta \Omega = - \Delta p \,V_{\beta} + A_{\alpha \beta } \sigma_{\alpha \beta }                + \left ( \sigma_{ \beta \gamma} - 
\sigma_{\alpha \gamma} \right ) \pi r^2                 + 2 \pi r \tau  
\end{equation} 
for the drop.   \\
The equations given above hold for the two following scenarios. In the first scenario, the lens or drop
can exchange matter with the surrounding phase so that the chemical potentials in all phases are 
equal and the pressure in the $\beta$ phase is determined by the chemical potentials.  
In the second scenario, nonvolatile lenses or drops are investigated and the volume of the lens or
drop is prescribed. In this case  
the pressure in the $\beta$ phase is not an independent thermodynamic variable
but is determined by the amount of $\beta$ phase.  
The term $\Omega_0$ (Eq. (\ref{lensdrop3}) ) is independent of the lens size 
and of the choice of dividing interfaces. This holds also for the drop (Eq. \ref{lensdrop3b}))   
due to the transformation law given in Eq. (\ref{deltasigmasf}) which is valid if the $\alpha$--$\gamma$ and 
$\beta$--$\gamma$ interfaces form a common plane, as is the case for the convention we have chosen. Since 
$\Omega$ as a physical quantity and $\Omega_0$, as just argued, are independent of choices of dividing interfaces,   
also $\Delta \Omega$ is independent of such choices. Thus in the following we can focus on  $\Delta \Omega$. 
In order to extract the value of $\tau$ from 
a known, e.g., calculated $\Delta \Omega$, or to infer information about possible changes of $\tau$ upon shifts of the 
dividing interfaces, at first it is necessary to specify all other quantities appearing in 
Eqs. (\ref{lensdrop4}) and (\ref{lensdrop4b}). 
The geometrical quantities like volumes, areas, and lengths are defined once the dividing interfaces are chosen. 
The surface tension $\sigma_{\alpha \gamma}$ is that of a planar interface between phases $\alpha$ and 
$\gamma$. If both phases $\alpha$ and $\gamma$ are fluid it is known (see, e.g., Ref. \cite{RowWid}) that $\sigma_{\alpha \gamma}$ is independent of the 
position of the dividing surface between these phases.  If one of the phases, e.g., $\gamma$ is an inert solid phase, 
$\sigma_{\alpha \gamma}$ must be analyzed more closely; but we postpone this discussion to a later subsection
(see, c.f., Subsec. 5.3.1) in which sessile drops are  
discussed separately. The quantities $\sigma_{\alpha \beta }$ and $\sigma_{\beta \gamma}$ are 
surface tensions of curved surfaces -- again the case of a drop requires a separate discussion --  
and thus they depend on 
the corresponding radii of curvature.  
We indicate this dependence  $\sigma_{\alpha \beta }(R_1)$, $\sigma_{\beta \gamma}(R_2)$ explicitly whenever it is 
necessary in order to avoid confusion.  As already discussed above we postulate that the interfacial 
tensions appearing in Eqs. (\ref{lensdrop4}) and (\ref{lensdrop4b}) have the same properties as their spherically 
closed counterparts, such as  
\begin{equation}
\label{lensdrop5}
\sigma_{\alpha \beta }(R_1) = \sigma_{\alpha \beta }(\infty) \left( 1 - \frac{2\delta_{\alpha \beta }^{\mathrm T}}{R_1} + s.l.t. 
\right)  \quad,
\end{equation} 
were $\delta_{\alpha \beta }^{\mathrm T}$ is the Tolmann length for the $\alpha$--$ \beta$ interface; 
subleading terms in $1/R_1$ are not considered here. An analogous equation holds for 
$\sigma_{\beta \gamma}(R_2)$. Although in the following these expressions will not be be used explicitly it will be  always 
understood that surface tensions depend on the radius of curvature in such a way 
(concerning the physical radius, see the corresponding remarks below).  
A related property of the pressure difference $\Delta p$ is described by the generalized Laplace equation \cite{RowWid,Kondo,Hend,Kalikman}  
\begin{eqnarray} 
\Delta p & = &\frac{2\sigma_{\alpha \beta }(R_1)}{R_1} + 
\left[\frac{\mathrm{d}\sigma_{\alpha \beta }}{\mathrm{d} R_{1}} \right] \quad, \nonumber \\ 
         & = & \frac{2\sigma_{\beta \gamma}(R_2)}{R_2} +
\left[\frac{\mathrm{d}\sigma_{\beta \gamma}}{\mathrm{d} R_{2}} \right] \quad,
\label{lensdrop6}
\end{eqnarray}   
where the terms in square brackets are quantities termed by Rowlinson and Widom \cite{RowWid} notional derivatives.  
Such a term multiplied by a small [d$R_{i}$] approximates the change in 
surface tension upon increasing the radius of the dividing surface by that value [d$R_i$] 
without changing the physical system, i.e., the density distributions and all 
thermodynamic variables remain fixed whereas the 
description of the system has been changed by shifting the dividing surface. In the following square brackets will be 
always used in order to characterize notional changes, e.g., [d$R_i$] denotes the 
notional change of the radius whereas we would write d$R_i$ if we speak about a physical change of the 
radius at a fixed convention of choosing the dividing surface (interface). The quantity 
$\left[\frac{\mathrm{d}\sigma_{\alpha \beta }}{\mathrm{d} R_i} \right]$ 
depends on the choice of the dividing surface. The so-called surface of tension is defined as that dividing surface  
for which $\left[\frac{\mathrm{d}\sigma_{\alpha \beta }}{\mathrm{d} R_{i}} \right]$ is zero.  
For the equimolar dividing surface the notional derivative coincides with the derivative of the surface tension 
with respect to the physical drop size, i.e., the derivative of Eq. (\ref{lensdrop5}). The second term on the rhs of Eq. 
(\ref{lensdrop6}) renders the rhs invariant with respect 
to changes of the dividing 
surface. This must be the case because $\Delta p$ defined as the pressure difference  between two bulk phases at given 
thermodynamic conditions is a measurable and therefore invariant quantity.  
Below, notional derivatives will be characterized   
by putting them into square brackets. 
\par
A property that will also be used extensively in the following is the fact,  
that $\sigma_{\xi \nu}(R_i)$ ($i \in \{1,2\}, \, \xi, \nu \in \{\alpha, \beta, \gamma\}$) is independent 
of the particular choice of the dividing surface up to and including the order $1/R_i$. At some occasions we 
shall use the relation  
\begin{equation}
\label{lensdrop7} 
\left[\frac{\mathrm{d}\sigma_{\xi \nu }}{\mathrm{d} R_i} \right]^{(2)} -  \left[\frac{\mathrm{d}\sigma_{\xi \nu }}{\mathrm{d} R_i} 
\right]^{(1)} =  \frac{2\sigma_{\xi \nu }\left[\mathrm{d}R_i\right]}{R_i^2} + s.l.t.  
\end{equation} 
between notional derivatives of surface tensions for two differently chosen dividing interfaces.  
The two dividing interfaces are 
denoted by the superscripts (1) and (2) and their radii are related via $\left[R_i\right]^{(2)} = \left[R_i\right] ^{(1)} +  
\left[\mathrm{d}R_i\right]$; $\left[\mathrm{d}R_i\right]$ could be also understood as a differential. It 
could be also finite but then we assume that it is small compared with  
$R_i$. In Eq. (\ref{lensdrop7}) we write $R_i$ 
and do not introduce $\left[R_i\right]^{(1)}$ or $\left[R_i\right]^{(2)}$ because using one or the other convention
would lead to expressions which only differ in subleading terms which are neglected anyway.  Equation (\ref{lensdrop7})
is valid if the values of both $\left[R_i\right]^{(1)}$ and $\left[R_i\right]^{(2)}$ 
are close to that of the radius corresponding
to the surface of tension and it is a direct consequence of well-known results obtained for spherical interfaces 
(see, e.g., Refs. \cite{RowWid,Hend}). The neglected s.l.t. contain terms of the order $(1/R_i)^3$ or higher and 
in addition they contain $\left[\mathrm{d}R_i\right]$ or differences between $\left[R_i\right]^{(i)}$ and
the radius corresponding to the surface of tension raised to the second or higher power.  
Most of the following conclusions, however, do not make explicit use of Eq.   
(\ref{lensdrop7}), but only of the fact that $\left[\frac{\mathrm{d}\sigma_{\xi \nu }}{\mathrm{d} R_i} \right]^{(2)} - 
\left[\frac{\mathrm{d}\sigma_{\xi \nu }}{\mathrm{d} R_i} \right]^{(1)}$ is of higher order than $1/R_i$. We also use 
(in the second part of our reasoning) the following relation -- known for spherical drops \cite{RowWid,Hend} -- 
between the stiffness against changes of the radius of  
curvature at given thermodynamic conditions (i.e., fixed temperature and chemical potentials), 
denoted here as $\frac{\mathrm{d}\sigma_{\xi \nu }}{\mathrm{d} R_i} \vert$,  
and the notional derivative of the surface tension:  
\begin{equation}
\label{lensdrop8} 
\frac{\mathrm{d}\sigma_{\xi \nu }}{\mathrm{d} R_i} \Big\vert = 
\left[\frac{\mathrm{d}\sigma_{\xi \nu }}{\mathrm{d} R_i} \right]  \quad. 
\end{equation}
In the following we discuss lenses and drops. Although the arguments proceed similarly in both cases, 
we discuss them separately 
because we want to address certain special problems connected exclusively with the solid--fluid interfaces. 
%
\subsection{The lens} 
%
First, we investigate the consequences of the requirement that the grand canonical 
potential of the total system is independent of particularly chosen dividing interfaces. Secondly, we derive -- 
from a variational procedure at fixed lens volume -- 
the equations yielding the contact angles as a function of the radius $r$ characterizing the lens size. The variational 
problem is set up in a way which guarantees that the contact angles transform correctly upon notional shifts of the Gibbs dividing 
interfaces. The results following from these two procedures are then compared.  \\
\subsubsection{Notional variation of the grand canonical potential}  
Notional variations (indicated by square brackets) leave the grand canonical potential 
unchanged which in ''differential'' form is written as 
\begin{equation}
\label{lensdrop9}
\left[ \mathrm{d} \Delta \Omega \right] = 0
\end{equation} 
with $\Delta \Omega$ defined in Eq. (\ref{lensdrop4}).
The left-hand side of Eq. (\ref{lensdrop9}) gives the notional change of $\Delta \Omega$ upon notional 
shifts of the dividing interfaces away from some given -- but essentially arbitrary -- set of dividing
interfaces, in the limit of very small notional shifts, characterized by, e.g., $[\mathrm{d}R_i]$ ($i = 1,2$) 
(see below).   
In the above equation all notional 
variations of geometrical quantities are then expressed in terms of the notional variations $[\mathrm{d}R_i]$ ($i = 1,2$) of 
the two radii of curvature $R_1$ and $R_2$ such as, for example, 
\begin{equation}
\label{lensdrop10} 
\left[  \mathrm{d}V_{\beta} \right] =  \left[ \frac{ 
\mathrm{d}V_{\beta} }{ \mathrm{d} R_1 } \right] \left[ \mathrm{d} R_1 \right] + 
\left[ \frac{ \mathrm{d}V_{\beta} }{ \mathrm{d} R_2 } 
\right] \left[ \mathrm{d} R_2 \right] \quad.
\end{equation}  
Similar equations hold for the notional variations of interfacial 
areas $\left[\mathrm{d} A_{\alpha \beta} \right]$, 
$\left[\mathrm{d} A_{\beta \gamma} \right]$, $\left[\mathrm{d} ( \pi r^2 )  \right]$,  
and for the length of the three-phase-contact line $\left[\mathrm{d} ( 2\pi r )  \right]$ as well as for the surface 
and line tensions.  
The notional derivatives $\left[\frac{\mathrm{d}\sigma_{\alpha \beta }}{\mathrm{d} R_1} \right]$ and 
$\left[\frac{\mathrm{d}\sigma_{\beta \gamma }}{\mathrm{d} R_2} \right]$ of the surface tensions are -- as already stated -- taken  
to be identical to those defined for completely spherical drops (see Eqs. (\ref{lensdrop6} - \ref{lensdrop8})). 
In addition we demand that $\left[\frac{\mathrm{d}\sigma_{\alpha \beta }}{\mathrm{d} R_2} \right] = 0$ and 
$\left[\frac{\mathrm{d}\sigma_{\beta \gamma }}{\mathrm{d} R_1} \right] = 0$ because a notional change of the radius $R_1$ of the 
$\alpha$--$\beta$ interface should not lead to notional changes of 
the $\beta$--$\gamma$ interfacial tension, i.e., of the interface on the opposite side of the lens, and vice versa (see Fig. 4). We further 
introduce two notional derivatives of the line tension: $\left[\frac{\mathrm{d}\tau}{\mathrm{d} R_1} \right]$ and 
$\left[\frac{\mathrm{d}\tau}{\mathrm{d} R_2} \right]$. These are actually defined by Eq. (\ref{lensdrop9}) 
because all other quantities in this equation are either given by geometry or fixed via the set of definitions given 
above. Certain properties of these new quantities will follow from the analysis given below.   
\\
The pressure difference $\Delta p$ 
is related to the surface tensions and their notional derivatives via Eq. (\ref{lensdrop6}). 
Since $\Delta p$ is unique the right hand sides of these two equations in Eq. (\ref{lensdrop6})  
must be equal. 
(The validity of the Laplace equation for even small lenses has been checked by simulations, e.g., in Ref. \cite{Bres3}.) 
After using the geometrical relations $R_1 = r/\sin \alpha$ and $R_2 = r/\sin \gamma$ we obtain a first 
equation relating $\alpha$, $\gamma$ and $r$:   
\begin{equation}
\label{lens1}
\sigma_{\alpha \beta} \sin \alpha + \frac{r}{2}          
\left[ \frac{\mathrm{d} \sigma_{\alpha \beta }}{\mathrm{d} R_1}  \right] = \sigma_{\beta \gamma} \sin \gamma + \frac{r}{2} \left[ 
\frac{\mathrm{d} \sigma_{\beta \gamma }}{\mathrm{d} R_2}  \right]  \quad. 
\end{equation} 
Because the notional changes $[\mathrm{d} 
R_1]$ and $[\mathrm{d}  R_2]$ are independent, the prefactors of $[\mathrm{d} R_1]$ and $[\mathrm{d}  R_2]$, obtained after 
expressing Eq. (\ref{lensdrop9}) in terms of these variables, must both vanish. 
In this way two further equations are obtained leading to three 
equations in total. On the other hand only two equations are required in order to determine the contact 
angles for a given physical 
lens size, because the geometric parameters $\alpha$, $\beta$, and $r$ are already related through a prescribed volume of the $\beta$ phase. 
Therefore the two aforementioned equations following from Eq. (\ref{lensdrop9}) have to be identical. 
This leads to the following consistency condition:  
\begin{equation}
\label{lens2}
\begin{split} 
\cos \gamma  \left[ 
\frac{\mathrm{d} \tau }{\mathrm{d} R_2}  \right] - \cos \alpha  
\left[ \frac{\mathrm{d} \tau }{\mathrm{d} R_1}  \right] = \frac{r}{2} 
\left\{  \left[ \frac{\mathrm{d} \sigma_{\beta \gamma}}{\mathrm{d} R_2} \right] -  
\left[ \frac{\mathrm{d} \sigma_{\alpha \beta 
}}{\mathrm{d} R_1} \right] \right\}  \, .
\end{split}  
\end{equation}
The remaining third equation can be written in a symmetrized form,    
\begin{multline}
\label{lens3}
 \sigma_{\alpha \beta } \cos \alpha + \sigma_{\beta \gamma} \cos \gamma + \sigma_{\alpha \gamma } 
   =   \frac{\tau}{r}             \\ 
 + \frac{ r \left\{ \sin \gamma \cos \alpha - \sin \alpha \cos \gamma \right\} }{4 \cos 
\alpha \cos \gamma }  \left\{ \left[ \frac{\mathrm{d} \sigma_{\beta \gamma }}{\mathrm{d} R_2}  \right] -  \left[ \frac{\mathrm{d} 
\sigma_{\alpha \beta }}{\mathrm{d} R_1}  \right] \right\}              \\    
\hspace*{-0.10cm}  
  + \frac{ \left\{ \sin \gamma \cos \alpha + \sin 
\alpha \cos \gamma \right\} }                 { 2 \cos \alpha \cos \gamma}   
\left\{ \cos \gamma \left[ \frac{\mathrm{d} \tau }{\mathrm{d} 
R_2}  \right] +         \cos \alpha  \left[ \frac{\mathrm{d} \tau }{\mathrm{d} R_1}  \right] \right\} \hspace*{-0.10cm} ,   
 \\ 
\end{multline} 
or, using the consistency condition given in Eq. (\ref{lens2}), it can be put into the following form:  
\begin{multline}
\label{lens4}
\hspace*{-0.40cm}  
\sigma_{\alpha \beta } \cos 
\alpha + \sigma_{\beta \gamma} \cos \gamma + \sigma_{\alpha \gamma } = \frac{\tau}{r} + \sin \gamma 
          \hspace*{-0.10cm}  \left[ \frac{\mathrm{d} \tau }{\mathrm{d} R_2}  \right] +    
  \sin \alpha \hspace*{-0.10cm} \left[ \frac{\mathrm{d} \tau }{\mathrm{d} R_1}  \right] \hspace*{-0.10cm}.
\\
\end{multline}
Since the contact angles $\alpha$ and $ \gamma$ are natural variables characterizing the 
three-phase-contact region we express notional 
derivatives of $\tau$ with respect to the radii in terms of notional derivatives with respect to the contact angles: 
$\left[\frac{\mathrm{d} \tau }{\mathrm{d} R_i}  \right] =   
   \left[ \frac{\mathrm{d} \tau }{\mathrm{d} \alpha}  \right] 
\left[ \frac{\mathrm{d} 
\alpha }{\mathrm{d} R_i }  \right]  +  \left[ \frac{\mathrm{d} \tau }{\mathrm{d} \gamma}  \right]      
\left[ \frac{\mathrm{d} \gamma 
}{\mathrm{d} R_i }  \right]$, 
where  
$\left[ \frac{\mathrm{d} \alpha }{\mathrm{d} R_i }  \right]$ and $\left[ \frac{\mathrm{d} \gamma 
}{\mathrm{d} R_i }  \right]$ 
describe notional changes of $\alpha$ and $\gamma$, respectively, upon notional changes of $R_i$.  
For given centers $M_1$ and $M_2$, the descriptions in terms of the pairs ($\alpha$, $\gamma$) or ($R_1$,$R_2$) are 
equivalent. However, we do not transform the notional derivatives of the surface tensions, 
because the radii of curvature $R_i$ are the natural variables 
of the curved interfaces. In the new variables Eqs. (\ref{lens2}) and (\ref{lens4}) take the form 
\begin{equation}
\label{lens5} 
\begin{split} 
\left[ \frac{\mathrm{d} \tau 
}{\mathrm{d} \alpha}  \right] \sin ^2 \alpha -  \left[ \frac{\mathrm{d} \tau }{\mathrm{d} \gamma}  \right] \sin ^2 \gamma  
= \frac{r^2}{2} 
\left\{  \left[ \frac{\mathrm{d} \sigma_{\beta \gamma}}{\mathrm{d} R_2} \right]    -   \left[ \frac{\mathrm{d} \sigma_{\alpha \beta 
}}{\mathrm{d} R_1} \right] \right\}  
\end{split} 
\end{equation}
and 
\begin{eqnarray} 
\sigma_{\alpha \beta } \cos \alpha + \sigma_{\beta 
\gamma} \cos \gamma + \sigma_{\alpha \gamma } & = & \frac{\tau}{r} + \frac{\sin \alpha \cos \alpha}{r}     \left[ \frac{\mathrm{d} \tau 
}{\mathrm{d} \alpha }  \right] 
\nonumber \\ 
      +   \frac{\sin \gamma \cos \gamma}{r} \left[ \frac{\mathrm{d} \tau }{\mathrm{d} \gamma}  
\right] \quad.   &  & 
\label{lens6}
\end{eqnarray} 
\par 
The next question is whether the line tension $\tau$ depends on the choice of the dividing interfaces. In 
order to answer it we compare $\Delta \Omega$ (Eq. (\ref{lensdrop4})) evaluated for two different dividing interfaces 
and find {\it $\tau$ to be independent of the choices of the dividing interfaces within the leading order.} 
More precisely this statement says that notional shifts of the dividing interfaces change $\tau$ only by contributions which
decrease as $1/r$ with increasing $r$ or faster.  
(We recall our remarks after Eqs. (\ref{lensdrop1}) and (\ref{lensdrop1b})
saying that $\tau$ as defined by Eqs. (\ref{lensdrop1}) and (\ref{lensdrop1b}) may contain subleading terms.)
The {\em leading} contribution to $\tau$ which is independent of $r$, is independent of 
the choices of the dividing interfaces. Although the subleading contributions to $\tau$ itself
may be neglected in the term $\tau/r$ in Eq. (\ref{lens4}) if we want to keep only terms up to the order $1/r$, 
it is not permissible to neglect the terms containing the notional
derivatives  $\left[\frac{\mathrm{d} \tau }{\mathrm{d} R_i}  \right]$. If these derivatives are of the order $1/r$ 
 -- in fact we shall see below that notional shifts of the dividing interfaces lead to changes in
$\left[\frac{\mathrm{d} \tau }{\mathrm{d} R_i}  \right]$ which are of that order -- the pertinent
terms in Eq. (\ref{lens4}) are of the order of $1/r$, i.e., they are of the same order as  
$\tau/r$ with only the leading
term in $\tau$ kept.    
In proving the aforementioned results concerning the behavior of $\tau$ under notional shifts of the dividing interfaces,
we have used the fact that surface tensions are independent of the  
chosen dividing interfaces up to order $1/R$. Furthermore we have used the generalized Laplace equation
(\ref{lensdrop6}) and also the fact that the difference  
$\left[\frac{\mathrm{d}\sigma_{\xi \nu }}{\mathrm{d} R_i} \right]^{(2)} - \left[\frac{\mathrm{d}\sigma_{\xi \nu }}{\mathrm{d} R_i} 
\right]^{(1)}$ is a correction of higher order than we are interested in. Moreover, the contact angles and other 
geometrical quantities for a choice (2) for the dividing interfaces have to be expressed in terms of the 
corresponding quantities characterizing choice (1).   
This has been carried out up to the order giving rise to contributions of the order of the line-tension contribution.
 \\ 
Next we 
investigate the transformation behavior of the quantities 
$\left[ \frac{\mathrm{d} \tau }{\mathrm{d} R_i}  \right]$ or $\left[ 
\frac{\mathrm{d} \tau }{\mathrm{d} \alpha }  \right]$ and $ \left[ \frac{\mathrm{d} \tau }{\mathrm{d} \gamma}  \right]$ due to 
notional shifts of the dividing interfaces. For this purpose we consider the equations for the contact angles 
(Eqs. (\ref{lens1} - \ref{lens6})) 
for two different sets (1) and (2) of dividing interfaces. The geometrical quantities, 
say $\alpha ^{(2)}$, $\gamma ^{(2)}$, 
and $r ^{(2)}$, in equations valid for the set (2) are then  expressed in terms of $\alpha ^{(1)}$, $\gamma ^{(1)}$, and $r 
^{(1)}$ up to the required order. The same is done for the surface tensions and their notional derivatives. We also use 
the above result that $\tau$ is independent of the choice of dividing interfaces to leading order. From comparison of the two 
systems of equations (Eqs.  (\ref{lens1}),  (\ref{lens2}), and (\ref{lens4}))  
and the fact that both systems must yield the same relations between 
$\alpha ^{(1)}$, $\gamma ^{(1)}$, and $r ^{(1)}$ for a fixed physical system, by using Eq. (\ref{lensdrop7}) 
we obtain the following transformation rules:  
\begin{widetext} 
\begin{eqnarray} 
\left[ \frac{\mathrm{d} \tau 
}{\mathrm{d} R_1}  \right]^{(2)} -  \left[ \frac{\mathrm{d} \tau }{\mathrm{d} R_1}  \right]^{(1)} & 
= & - \frac{\sigma_{\alpha \beta} \sin 
\alpha \cos (\alpha + \gamma)}{r \sin (\alpha + \gamma)}       \left[\mathrm{d} R_1 \right] 
-   \frac{\sigma_{\beta \gamma} \sin \gamma 
}{r \sin (\alpha + \gamma)}       \left[\mathrm{d} R_2 \right] \/ ,  \nonumber  \\ 
             \left[ 
\frac{\mathrm{d} \tau }{\mathrm{d} R_2}  \right]^{(2)} -  \left[ \frac{\mathrm{d} \tau }{\mathrm{d} R_2}  \right]^{(1)} & = & - 
\frac{\sigma_{\alpha \beta} \sin \alpha }{r \sin (\alpha + \gamma)}       \left[\mathrm{d} R_1 \right] 
-   \frac{\sigma_{\beta \gamma} \sin 
\gamma \cos (\alpha + \gamma)}{r \sin (\alpha + \gamma)}       \left[\mathrm{d} R_2 \right]   \/ .  \nonumber  \\
    & &  
\label{lens7} 
\end{eqnarray}   
\end{widetext} 
Alternatively, this may be translated into the notional derivatives of $\tau$ with respect to the contact 
angles:  
\begin{eqnarray}
 \left[ \frac{\mathrm{d} \tau }{\mathrm{d} \alpha}  \right]^{(2)} -  \left[ \frac{\mathrm{d} \tau 
}{\mathrm{d} \alpha}  \right]^{(1)} & = & - \sigma_{\alpha \beta} \left[\mathrm{d} R_1 \right] \/ ,  \nonumber \\  
\left[ \frac{\mathrm{d} \tau 
}{\mathrm{d} \gamma}  \right]^{(2)} -  \left[ \frac{\mathrm{d} \tau }{\mathrm{d} \gamma}  \right]^{(1)} & = 
& - \sigma_{\alpha \beta} 
\left[\mathrm{d} R_2 \right] \quad.
\label{lens8} 
\end{eqnarray}
Equations (\ref{lens7}) and (\ref{lens8}) show that the notional derivatives of $\tau$ depend on where the dividing
interfaces are located. This is completely analogous to what is known for the notional derivative of the
surface tension for closed spherical interfaces.
\subsubsection{Variational treatment with constraint of fixed volume}  
A common procedure aimed at obtaining equations for the contact angles consists of minimization of the grand 
canonical potential $\Delta \tilde{\Omega}$ with the constraint of fixed volume $V_{\beta}$ (compare Eq. ((\ref{lensdrop4})): 
\begin{eqnarray} 
\Delta \tilde{\Omega} & = & A_{\alpha \beta } \sigma_{\alpha \beta }(R_1) + A_{ \beta \gamma}  \sigma_{ \beta 
\gamma}(R_2) -\pi r^2 \sigma_{\alpha \gamma} 
\nonumber\\
  & & + 2 \pi r \tau (\alpha,\gamma ,r)  + \lambda V_{\beta}  \quad.
\label{lens9}
\end{eqnarray} 
In Eq. (\ref{lens9}) no bulk contributions appear because the volume $V_{\beta}$, the bulk pressures $p_{\alpha}$  
and $p_{\beta}$, and thus the pressure difference $\Delta p = p_{\beta} - p_{\alpha} $ are kept fixed due to the fixed 
thermodynamic conditions specified by the temperature and the chemical potentials. The fixed volume condition is 
implemented via the last term in 
Eq. (\ref{lens9}) containing the Lagrange muliplier $\lambda$. The volume $V_{\beta}$ is the one enclosed
by the dividing interfaces. The independent variables of the variational problem are $\alpha$, 
$\gamma$, and $r$ ($\beta$ is determined via  $\alpha + \gamma + \beta = 2\pi$). In order to calculate 
the variation $\mathrm{d}\Delta \tilde{\Omega}$  
resulting from the variations of $\alpha$, $\gamma$, and $r$ we have to introduce -- in addition to 
the interfacial and the line tensions -- new material parameters. These are the stiffness constants 
$\frac{\mathrm{d}\sigma_{\alpha
\beta}}{\mathrm{d} R_1}  \vert \hspace{0.2cm} \mathrm{and}  \hspace{0.2cm}    \frac{\mathrm{d}\sigma_{\beta \gamma}}{\mathrm{d} R_2}
\vert $
of the interfaces 
describing the cost in interfacial free energy resulting from changes in the radii of curvature as well as  
the stiffnesses 
$ \frac{\mathrm{d}\tau}{\mathrm{d} \alpha}  \vert \hspace{0.1cm},
\hspace{0.1cm}   \frac{\mathrm{d}\tau}{\mathrm{d} \gamma}  \vert , \hspace{0.1cm} \mathrm{and}  \hspace{0.1cm}
\frac{\mathrm{d}\tau}{\mathrm{d} r}  \vert  $  
of the three-phase-contact line against changes of the contact angles  
$\alpha$ and $\gamma$, and the radius of curvature $r$, respectively.  
To be explicit: in the variation we introduce also contributions 
$A_{\alpha \beta } \frac{\mathrm{d}\sigma_{\alpha \beta}}{\mathrm{d} R_1}  \vert 
\frac{\partial R_1}{\partial r} \mathrm{d}r$ describing the cost in free energy  
associated with a variation of the               
curvature radius $R_1$, caused by a change of $r$ by $\mathrm{d}r$, without changing the interfacial area, or
 $2 \pi r \frac{\mathrm{d}\tau}{\mathrm{d} r}  \vert \mathrm{d}r$ etc..     
The symbol $\vert$ indicates that the stiffnesses are given by the respective  
derivatives evaluated {\em at fixed thermodynamic conditions.}  
\\ 
The stiffness constants of the interfaces, e.g., $\frac{\mathrm{d}\sigma_{\alpha \beta}}{\mathrm{d} 
R_1}  \vert$ should not be confused with the derivatives of surface 
tensions with respect to the physical radius. For example, $\frac{\mathrm{d}\sigma_{\alpha \beta}}{\mathrm{d} R_1}\vert$ 
does not coincide with the derivative of $\sigma_{\alpha\beta}$ given in Eq. (\ref{lensdrop5}) with 
respect to $R_1$ because drops of different physical sizes in an unstable or constrained equilibrium with their environment
do not correspond to the same thermodynamic conditions given by temperature and chemical potentials.  
The stiffness constants of the interfaces may be expressed via Eq. (\ref{lensdrop8}) in terms of the 
notional derivatives. This relation shows that the stiffness constant of an 
interface vanishes if the so-called surface of tension is chosen as the Gibbs dividing surface.
In the same way the stiffness constants attributed to the contact line should not be confused 
with similarly looking derivatives with respect to an implicit dependence of the line tension 
on contact angles. These implicit dependences reflect changes in the thermodynamic conditions.  
The meaning of the stiffness constants is completely different. 
In setting up the variational principle we explore  constrained equilibrium
configurations in the neighbourhood of {\it the} equilibrium configuration in order to find the 
equilibrium by minimizing the free energy at fixed thermodynamic conditions. Surface tensions,
the line tension, and the various stiffness constants describe the costs in free energy due
to virtual displacements of interfaces away from their equilibrium shape.    
(In the summary operational procedures will be discussed how to determine stiffness constants
 theoretically.) 
  \\
The equation 
\begin{equation}
\label{lens10}
\mathrm{d}
\Delta \tilde{\Omega} = \frac{\mathrm{d}\Delta \tilde{\Omega}}{\mathrm{d} r} \mathrm{d}r                   +  
\frac{\mathrm{d}\Delta \tilde{\Omega}}{\mathrm{d} \alpha} \mathrm{d}\alpha +  
\frac{\mathrm{d}\Delta \tilde{\Omega}}{\mathrm{d} 
\gamma} \mathrm{d} \gamma         = 0  
\end{equation} 
leads to three equations because $\alpha$, $\gamma$, and $r$ can be varied 
independently. They have the following form: 
\begin{widetext} 
\begin{eqnarray}
\hspace*{-0.60cm}
\lambda & = & - \frac{ 2 \sigma_{\alpha \beta} \sin \alpha }{r} -      \frac{\mathrm{d}\sigma_{\alpha 
\beta}}{\mathrm{d} R_1}  \Big\vert        +            \frac{ ( 1 - \cos \alpha )}{ ( 1 + \cos \alpha )}\left\{            
\frac{\mathrm{d}\sigma_{\alpha \beta}}{\mathrm{d} R_1}  \Big\vert         + \frac{ 2 \sin ^2 \alpha }{ r^2} 
\frac{\mathrm{d}\tau}{\mathrm{d} \alpha} \Big\vert     \right\}  \, , \, 
\label{lens11}
\end{eqnarray}
\begin{eqnarray}
\hspace*{-0.60cm} 
\lambda & = & - \frac{ 
2 \sigma_{\beta \gamma} \sin \gamma }{r} -      \frac{\mathrm{d}\sigma_{\beta \gamma}}{\mathrm{d} R_2}  \Big\vert +            
\frac{ ( 1 - \cos \gamma )}{ ( 1 + \cos \gamma )}\left\{ \frac{\mathrm{d}\sigma_{\beta \gamma}}{\mathrm{d} R_2}  \Big\vert         
+ \frac{ 2 \sin ^2 \gamma }{ r^2} \frac{\mathrm{d}\tau}{\mathrm{d} \gamma} \Big\vert     \right\} \, , \,   
\label{lens12}
\end{eqnarray}
and 
\begin{multline}
\label{lens13}
\sigma_{\alpha \beta} \cos \alpha  + \sigma_{\beta \gamma} \cos \gamma  + 
\sigma_{\alpha \gamma}   =   \frac{\tau}{r} + \frac{\mathrm{d} \tau}{\mathrm{d} r} \Big\vert +     \frac{\cos \alpha \sin \alpha}{r} 
\frac{\mathrm{d} \tau}{\mathrm{d} \alpha} \Big\vert  +  \frac{\cos \gamma \sin \gamma}{r} \frac{\mathrm{d} \tau}{\mathrm{d} 
\gamma} \Big\vert   \\
      + \frac{r(1 - \cos \alpha ) }{\sin \alpha }\left\{        \frac{\mathrm{d}\sigma_{\alpha 
\beta}}{\mathrm{d} R_1}  \Big\vert         + \frac{ 2 \sin ^2 \alpha }{ r^2} \frac{\mathrm{d}\tau}{\mathrm{d} \alpha} \Big\vert    
\right\} \\ 
       + \frac{r(1 - \cos \gamma ) }{\sin \gamma }\left\{ \frac{\mathrm{d}\sigma_{\beta \gamma}}{\mathrm{d} 
R_2}  \Big\vert + \frac{ 2 \sin ^2 \gamma }{ r^2} \frac{\mathrm{d}\tau}{\mathrm{d} \gamma} \Big\vert    \right\} \quad.
\end{multline} 
\end{widetext} 
The ensuing equality of the right hand sides of Eqs. (\ref{lens11}) and (\ref{lens12}) leads to one of two equations 
relating the variables  
$\alpha$, $\gamma$, and $r$. The second relation is provided by Eq. (\ref{lens13}). Finally, a third equation expresses
the fixed volume $V_{\beta}$ of the lens,     
\begin{eqnarray} 
V_{\beta} & = & \frac{1}{3}\pi r^3 \biggl \{ \frac{ (1 + \cos \alpha)(2- \cos \alpha) }{ (1 - \cos \alpha) \sin \alpha } 
 \nonumber\\ 
           & &   +  \frac{ (1 + \cos \gamma)(2- \cos \gamma) }{ (1 - \cos \gamma) \sin \gamma } \biggr \} 
\label{lens13-Z} 
\end{eqnarray} 
and thus finally determines $\alpha$, $\gamma$, $r$, and the Lagrange parameter $\lambda$ via 
Eqs. (\ref{lens11}) or (\ref{lens12}). 
\par 
We now compare two such sets of equations, each for a different choice of the dividing interfaces in order
to find relations between the stiffnesses of the line for two different choices for the dividing interfaces.  
Since the physical object is kept 
fixed, the relations between one set of variables ($\alpha^{(2)}$, $\gamma^{(2)}$, $r^{(2)}$) and another set 
($\alpha^{(1)}$, $\gamma^{(1)}$, $r^{(1)}$) follow from the definition of the lens by two intersecting
spheres $(M_1, R_1^{(i)})$ and  $(M_2, R_2^{(i)})$ as described in Sec. 5 and simple geometrical 
considerations. 
The volume $V_{\beta}$ attributed to the lens   
depends on the choice of the dividing interfaces, i.e., two different volumes $V_{\beta}^{(2)} \ne V_{\beta}^{(1)}$
are assigned to the same physical object. 
The relation between  $V_{\beta}^{(2)}$ and $V_{\beta}^{(1)}$ is known once the notional changes 
of the radii $R_1$ and $R_2$, fixing the relative positions of the dividing interfaces, are given.
Therefore Eq. (\ref{lens13-Z}) does not contain any new information that could be used in order
to relate the stiffnesses which have to be attributed to the line for two different choices for the
dividing interfaces. Thus in the following, we have to consider only two pairs of equations:
first, the equation resulting from the equality of the right hand sides of Eqs. (\ref{lens11}) and (\ref{lens12}), 
and secondly Eq. (\ref{lens13}), one pair for each choice of the dividing interfaces.        
The parameters $\alpha^{(2)}$, $\gamma^{(2)}$, and $r^{(2)}$ can be expressed in terms of 
$\alpha^{(1)}$, $\gamma^{(1)}$, and 
$r^{(1)}$, and of the notional changes of the radii $\left[ \mathrm{d} R_i \right] = R_i^{(2)} -  R_i^{(1)}$, 
$i = 1,2$.  
After expressing the pair of equations for convention 
(2) in terms of $\alpha^{(1)}$, $\gamma^{(1)}$, and $r^{(1)}$ up to the order of the line tension term, 
the resulting equations may be 
compared directly with the pair of equations obtained if convention (1) was chosen from the beginning. 
It is clear that the two pairs of 
equations must be equivalent since they describe the same physical system in terms of the same variables.   
In that comparison we further use  
the relation (\ref{lensdrop8}) between the stiffnesses of the interfaces 
against changes of the radii of curvature and 
the related notional derivatives. In addition we use that the mentioned notional 
derivatives written for two different dividing 
interfaces are related via Eq. (\ref{lensdrop7}) which in the chosen variables read 
(up to the relevant order it is not necessary to 
distinguish between $\alpha^{(1)}$ and $\alpha^{(2)}$ etc. and thus on the right hand side
we omit the superscripts distinguishing between these 
conventions): 
\begin{equation}
\label{lens14} 
\left[\frac{\mathrm{d}\sigma_{\alpha \beta }}{\mathrm{d} R_1} \right]^{(2)} -  
\left[\frac{\mathrm{d}\sigma_{\alpha \beta }}{\mathrm{d} R_1} \right]^{(1)} =  \frac{2\sigma_{\alpha \beta 
}\sin^2\alpha\left[\mathrm{d}R_1\right]}{r^2} + s.l.t.  
\end{equation} 
and 
\begin{equation}
\label{lens15} 
\left[\frac{\mathrm{d}\sigma_{\beta \gamma }}{\mathrm{d} R_2} \right]^{(2)} -  \left[\frac{\mathrm{d}\sigma_{\beta \gamma 
}}{\mathrm{d} R_2} \right]^{(1)} =  \frac{2\sigma_{\beta \gamma }\sin^2\gamma\left[\mathrm{d}R_2\right]}{r^2} + 
s.l.t.\hspace{0.3cm} \quad.   
\end{equation} 
As a result of the comparison and the requirement of equivalence of the two pairs of 
equations discussed above 
(rhs of Eq. (\ref{lens11}) equals rhs of Eq. (\ref{lens12}), and Eq. (\ref{lens13}) ) 
we obtain two coupled equations for the following three quantities: 
$\left( \frac{\mathrm{d}\tau}{\mathrm{d} 
\alpha} \vert^{(2)} -       \frac{\mathrm{d}\tau}{\mathrm{d} \alpha} \vert^{(1)} \right)$, $\left( \frac{\mathrm{d}\tau}{\mathrm{d} 
\gamma} \vert^{(2)} -       \frac{\mathrm{d}\tau}{\mathrm{d} \gamma} \vert^{(1)} \right)$, and $\left( 
\frac{\mathrm{d}\tau}{\mathrm{d} r} \vert^{(2)} -       \frac{\mathrm{d}\tau}{\mathrm{d} r} \vert^{(1)} \right)$ . 
These equations have a manifold of solutions. If we pick a particular solution with 
\begin{equation}
\label{lens16}
\frac{\mathrm{d}\tau}{\mathrm{d} r} \big\vert^{(2)} -       \frac{\mathrm{d}\tau}{\mathrm{d} r} \big\vert^{(1)}  = 
0
\end{equation}
the two remaining quantities are given by 
\begin{equation}
\label{lens17}
\begin{array}{l c l r }
\frac{\mathrm{d}\tau}{\mathrm{d} \alpha} \vert^{(2)} -       \frac{\mathrm{d}\tau}{\mathrm{d} \alpha} \vert^{(1)}  & = &                          
- \sigma _{\alpha \beta} \left[\mathrm{d}R_1 \right]  & \quad     \\
                                       & = &  \frac{ - r^2}{2\sin^2 \alpha}\left( \left[ \frac{\mathrm{d} 
\sigma _{\alpha \beta}}{ \mathrm{d}R_1} \right]^{(2)} -       \left[ \frac{\mathrm{d} \sigma _{\alpha \beta}} { \mathrm{d}R_1} 
\right]^{(1)}                                                                          \right)                                             &   \\ 
 \frac{\mathrm{d}\tau}{\mathrm{d} 
\gamma} \vert^{(2)} -       \frac{\mathrm{d}\tau}{\mathrm{d} \gamma} \vert^{(1)}  & = &   - \sigma _{\beta \gamma} 
\left[\mathrm{d}R_2 \right] &  
\\         
                                             & = &  \frac{ - r^2}{2\sin^2 \gamma}\left( \left[ \frac{\mathrm{d} 
\sigma _{\beta \gamma}}{ \mathrm{d}R_2 } \right]^{(2)} -       \left[ \frac{\mathrm{d} \sigma _{\beta \gamma}}{ \mathrm{d}R_2 
}\right]^{(1)}   \right)                                             & \quad.
\end{array} 
\end{equation}
For the special 
choice in Eq. (\ref{lens16}) leading to Eq. (\ref{lens17}) the Lagrange multiplier $\lambda$ 
(see Eqs. (\ref{lens11}) and (\ref{lens12})) becomes 
independent of the chosen dividing interfaces and therefore it allows for a physical interpretation. 
The conditions (\ref{lens17})  
are obviously fulfilled if the following relations between stiffness constants of the line and those of 
the interfaces hold:   
\begin{equation}
\label{lens18} 
\frac{\mathrm{d}\tau}{\mathrm{d} \alpha} \Big\vert              =           \frac{ - 
r^2}{2\sin^2 \alpha}        \frac{\mathrm{d} \sigma _{\alpha \beta}}{ \mathrm{d}R_1} \Big\vert =  
\frac{ - r^2}{2\sin^2 \alpha}         
\left[ \frac{\mathrm{d} \sigma _{\alpha \beta}} { \mathrm{d}R_1} \right]
\end{equation}  
and 
\begin{equation}
\label{lens19} 
\frac{\mathrm{d}\tau}{\mathrm{d} \gamma} \Big\vert  =   \frac{ - r^2}{2\sin^2 \gamma}        \frac{\mathrm{d} \sigma _{\beta 
\gamma}}{ \mathrm{d}R_2} \Big\vert   =  \frac{ - r^2}{2\sin^2 \gamma}        \left[ \frac{\mathrm{d} \sigma _{\beta \gamma}} { 
\mathrm{d}R_2} \right]   \quad .   
\end{equation} 
Additional terms which are independent of the choice of dividing interfaces could be added on the 
right hand sides of Eqs. (\ref{lens18}) and (\ref{lens19}) without violating the conditions  (\ref{lens17}).
However, keeping such terms would not lead to more general results, but to equations which are more complicated 
than those presented below. As we shall see, the special choices of the right hand sides of 
Eqs. (\ref{lens18}) and (\ref{lens19}) lead to equilibrium conditions for the contact angles which  
agree with Eqs. (\ref{lens1}) and (\ref{lens6}) which have been derived above from the principle 
of invariance of $\Omega$ under notional changes. This agreement is achieved for the particular 
choice  
$\frac{\mathrm{d}\tau}{\mathrm{d} r} \vert ^{(2)} - \frac{\mathrm{d}\tau}{\mathrm{d} r} \vert ^{(1)} = 0$, 
but $\frac{\mathrm{d}\tau}{\mathrm{d} r} \vert$ is still left undetermined by consistency requirements alone.      
Insertion of Eqs. (\ref{lens18}) and (\ref{lens19}) into Eqs. (\ref{lens11}) and (\ref{lens12}) leads to 
\begin{equation}
\label{lens20} 
\lambda 
= - \frac{ 2 \sigma _{\alpha \beta} \sin \alpha }{r} -  
\frac{\mathrm{d} \sigma _{\alpha \beta}}{ \mathrm{d}R_1} \Big\vert           =          
- \frac{  2 \sigma _{\alpha \beta}}{R_1}           - \left[ \frac{\mathrm{d} \sigma _{\alpha \beta}}{ \mathrm{d}R_1} \right ]  
\end{equation}
and  
\begin{equation}
\label{lens21} \lambda = - \frac{ 2 \sigma _{\beta \gamma}}{R_2} -            \left[ 
\frac{\mathrm{d} \sigma _{\beta \gamma}}{ \mathrm{d}R_2} \right ] \quad.
\end{equation} 
In other words, $\lambda$ equals minus 
the Laplace pressure expressed via the generalized Laplace equation (compare Eq. (\ref{lensdrop6}))
which supports the choice made in Eqs. (\ref{lens18}) and (\ref{lens19})   
above. Accordingly, the equations relating the contact angles to the radius $r$ read 
(recall that this holds only for choices which are in accordance with Eq. (\ref{lens16}) as well as 
Eqs. (\ref{lens18}) and (\ref{lens19})):  
\begin{equation}
\label{lens22}  
\sigma _{\alpha 
\beta} \sin \alpha  +     \frac{r}{2} \left[ \frac{\mathrm{d} \sigma _{\alpha \beta}}{ \mathrm{d}R_1} \right ]    
  =     \sigma _{\beta 
\gamma} \sin \gamma  
+     \frac{r}{2} \left[ \frac{\mathrm{d} \sigma _{\beta \gamma}}{ \mathrm{d}R_2} \right ]  
\end{equation} 
and
\begin{eqnarray} 
& & \sigma _{\alpha \beta} \cos \alpha  + \sigma _{\beta \gamma} \cos \gamma  + \sigma _{\alpha 
\gamma}    =   \frac{\tau}{r} + \frac{\mathrm{d} \tau}{\mathrm{d} r} \Big\vert 
\nonumber\\ 
    &  & + \frac{\sin \alpha \cos \alpha }{r} 
\frac{\mathrm{d} \tau}{\mathrm{d} \alpha} \Big\vert  
    +     \frac{\sin \gamma \cos \gamma }{r} \frac{\mathrm{d} \tau}{\mathrm{d} 
\gamma} \Big\vert      .
\label{lens23} 
\end{eqnarray} 
Comparison of  Eq.  (\ref{lens23}) with Eq. (\ref{lens6}) (Eqs. (\ref{lens22}) and (\ref{lens1}) 
are identical anyway) renders the relation
\begin{multline} 
\label{lens24}
\frac{\mathrm{d} \tau}{\mathrm{d} r} 
\Big\vert +    \frac{\sin \alpha \cos \alpha }{r} \frac{\mathrm{d} \tau}{\mathrm{d} \alpha} \Big\vert     +    \frac{\sin \gamma \cos 
\gamma }{r} \frac{\mathrm{d} \tau}{\mathrm{d} \gamma} \Big\vert   =  
  \\ 
\frac{\sin \alpha \cos \alpha }{r} \left [ \frac{\mathrm{d} 
\tau}{\mathrm{d} \alpha} \right ]      +    \frac{\sin \gamma \cos \gamma }{r} 
\left [ \frac{\mathrm{d} \tau}{\mathrm{d} \gamma} \right ] \quad.  
\end{multline} 
Adopting this relation for the surface of tension ${(\mathrm{s})}$ and using the relations 
in Eqs. (\ref{lens18}), (\ref{lens19}), and 
(\ref{lens16}) and noting further that 
$\left [ \frac{\mathrm{d} \sigma _{\alpha \beta }}{\mathrm{d} R_1} \right ]^{(\mathrm{s})} = 0 =    
 \left [ \frac{\mathrm{d} \sigma _{\beta \gamma }}{\mathrm{d} R_2} \right ]^{(\mathrm{s})} $ 
and thus $\frac{\mathrm{d} \tau}{\mathrm{d} \alpha} \vert ^{(\mathrm{s})} = 0 = 
\frac{\mathrm{d} \tau}{\mathrm{d} \gamma} \vert ^{(\mathrm{s})}$   we find 
(note that because of Eq. (\ref{lens16}) $\frac{\mathrm{d} \tau}{\mathrm{d} r} \vert$
 is independent of the choice of the dividing interfaces)   
\begin{equation}
\label{lens25}
\begin{split} 
\frac{\mathrm{d} \tau}{\mathrm{d} r} \Big\vert = \frac{\mathrm{d} \tau}{\mathrm{d} r} \Big\vert 
^{(\mathrm{s})}   =  \frac{\sin \alpha \cos \alpha }{r} \left [ \frac{\mathrm{d} \tau}{\mathrm{d} \alpha} \right ] ^{(\mathrm{s})}     +    
\frac{\sin \gamma \cos \gamma }{r} \left [ \frac{\mathrm{d} \tau}{\mathrm{d} \gamma} \right ] 
^{(\mathrm{s})} \hspace*{-0.20cm}.
\end{split}  
\end{equation}
On the rhs of Eq. (\ref{lens25}) superscripts $^{(\mathrm{s})}$ are omitted for $\alpha$, $\gamma$, and $r$ 
because differences 
between the values of $\alpha$, $\gamma$, and $r$ for different dividing interfaces 
give rise only to higher order corrections.  
Reinserting Eq. (\ref{lens25}) into Eq. (\ref{lens24}) we further find  
\begin{equation}
\label{lens26}
\frac{\mathrm{d} \tau}{\mathrm{d} \alpha } \Big\vert =    \left[ \frac{\mathrm{d} 
\tau}{\mathrm{d} \alpha } \right]   -    \left[ \frac{\mathrm{d} \tau}{\mathrm{d} \alpha } \right] ^{(\mathrm{s})}    
\end{equation}
and 
\begin{equation}
\label{lens27}
 \frac{\mathrm{d} \tau}{\mathrm{d} \gamma } \Big\vert =                \left[ 
\frac{\mathrm{d} \tau}{\mathrm{d} \gamma } \right] -    \left[ \frac{\mathrm{d} \tau}{\mathrm{d} \gamma } \right] 
^{(\mathrm{s})}      \quad. 
\end{equation}
Of course, due to these relations the transformation behavior for the stiffness constants of the line tension  
(Eqs. (\ref{lens16}) and (\ref{lens17})) is consistent with that for the notional derivatives (Eq. (\ref{lens8})). 
It should also be noted that 
$\frac{\mathrm{d} \tau}{\mathrm{d} \alpha } \vert ^{(\mathrm{s})} = 0$ and 
$\frac{\mathrm{d} \tau}{\mathrm{d} \gamma } \vert ^{(\mathrm{s})} = 0$ .  

\subsection{The drop}  
%
\subsubsection{Notional variation of the grand canonical potential} 
In the case of a drop our analysis also starts from the requirement that the grand potential must be 
invariant with respect to notional changes. They consist of a change 
of the radius of the 
$\alpha$--$\beta$ interface by $\left[ \mathrm{d} R \right]$ and a common shift of the  
$\alpha$--$\gamma$ and $\beta$--$\gamma$ interfaces  
by $\left[ \mathrm{d} h \right]$. From  the requirement $\left[ \mathrm{d} \Delta \Omega  \right] = 0$ 
we obtain two equations ($\left[ \mathrm{d} R \right]$ and $\left[ \mathrm{d} h \right]$ can be chosen 
independently) relating the contact angle $\theta$ to the radius $r$.  However, these equations have to be identical 
because already a single equation is sufficient to determine the relation between $\theta$ and $r$.  
This leads to the consistency relation 
\begin{equation}
\label{drop1} \left[ \frac{\mathrm{d} \tau }{\mathrm{d} h}  \right] +\cos \theta  \left[ \frac{\mathrm{d} \tau }{\mathrm{d} R}  \right] = - 
\frac{r}{2}  \left[ \frac{\mathrm{d} \sigma_{\alpha \beta }}{\mathrm{d} R} \right]   
\end{equation} 
and the following equation for the contact angle: 
\begin{equation}
\label{drop2}
\sigma_{\alpha \beta } \cos \theta + (\sigma_{\beta \gamma} - \sigma_{\alpha \gamma }) = - \frac{\tau}{r} - \sin \theta \left[ 
\frac{\mathrm{d} \tau }{\mathrm{d} R}  \right] \quad.
\end{equation}
Similarly as for the lens we have introduced notional derivatives 
$\left[ \frac{\mathrm{d} \tau }{\mathrm{d} h} \right]$ and $\left[ \frac{\mathrm{d} \tau }{\mathrm{d} R} \right]$. 
We also  make use of a notional derivative  
$\left[ \frac{\mathrm{d} \sigma_{\alpha \beta} }{\mathrm{d} R} \right]$ of the spherically shaped 
$\alpha$--$\beta$ interface but --- at the same time --- we do not take into account a notional derivative 
$\left[ \frac{\mathrm{d} (\sigma_{\beta \gamma} -  \sigma_{\alpha \gamma}) }{\mathrm{d} h} \right]$ of the difference in the 
interfacial tensions of the two planar substrate--liquid ($\gamma$--$\beta$) and substrate--vapor ($\gamma$--$\alpha$) 
interfaces. This calls for a comment. Here the quantity  $\sigma_{\beta \gamma} -  \sigma_{\alpha \gamma}$ denotes the 
difference of two interfacial tensions corresponding to a situation in which the same substrate phase $\gamma$ 
remains in contact with either the $\alpha$ or the $\beta$ phase. Both phases  $\alpha$ and $\beta$ are at the 
same pressure $p$. This quantity together with the $\alpha$--$\beta$  surface tension at infinite radius defines 
--- via the Young equation --- the contact angle $\theta_0$ of a macroscopicly large  
liquid drop (phase $\beta$) in contact with 
its vapor (phase $\alpha$) on top of the substrate $\gamma$. The angle $\theta_0$ is an observable quantity and 
does not depend on the choice of the dividing surface. Since $\sigma_{\alpha\beta}$ does not depend on the choice of 
the dividing interfaces either, it follows that $ \sigma_{\beta \gamma} -  \sigma_{\alpha \gamma} $ cannot depend on the 
chosen position $h$ of the common $\alpha$--$\gamma$ and  $\beta$--$\gamma$ dividing interfaces. 
\\
One arrives at the same conclusion by calculating the changes of $ \sigma_{\beta \gamma}$ and $\sigma_{\alpha \gamma}$ 
with respect to changes $\mathrm{d} h$ in the interface positions. In the difference 
$ \sigma_{\beta \gamma} -  \sigma_{\alpha \gamma} $ terms depending on $\mathrm{d} h$  
drop out provided the pressure in the phases $\alpha$ and $\beta$ is the same, and a common height 
'above' the substrate is chosen for the $\alpha$--$\gamma$ and for the $\beta$--$\gamma$ dividing interfaces (see, e.g., the 
discussion at the end of Sect. 2). (Both these assumptions are implicitly contained in the Young equation.) 
Actually, further complications would arise if, in an attempt to mimic the situation in a drop more closely, 
we would carry out the calculations at different pressures, e.g., for  $p_{\beta} =  p_{\alpha} + 
\Delta p$ with the Laplace pressure $\Delta p$. A quantity $ (\sigma_{\beta \gamma} -  \sigma_{\alpha \gamma})' $ 
defined in that way would depend on the choice of the dividing interfaces and additional terms of the same order as those 
introduced by the line tension would appear in the final equations. Furthermore  the difference 
$ (\sigma_{\beta \gamma} -  \sigma_{\alpha \gamma})' $ would not be related to  $\theta_0$ via the standard Young equation 
and thus could not be determined via an independent experiment of measuring the shape of a macroscopic drop. 
Therefore it does not appear to be useful
to introduce quantities like $ (\sigma_{\beta \gamma} -  \sigma_{\alpha \gamma})' $ (see, however, the
discussion in Subsect. VI C). 
\\ 
Equations (\ref{drop1}) and (\ref{drop2}) may be rewritten in terms of $\theta$ and $r$. 
In these variables the consistency equation 
and the modified Young equation take the following form:  
\begin{equation}
\label{drop3}  
\sin ^2 \theta \left[ \frac{\mathrm{d} \tau }{\mathrm{d} \theta}  \right]    = \frac{r^2}{2}  \left[ \frac{\mathrm{d} \sigma_{\alpha \beta} 
}{\mathrm{d} R} \right] \quad  
\end{equation}
and 
\begin{equation}
\label{drop4}
\begin{split} 
\sigma_{\alpha \beta } \cos \theta + (\sigma_{\beta \gamma} - \sigma_{\alpha \gamma })  =  
- \frac{\tau}{r} - \left[ \frac{\mathrm{d} \tau }{\mathrm{d} r }  \right]   
 - \frac{\sin \theta \cos \theta}{r} \hspace{-0.10cm} \left[ \frac{\mathrm{d} \tau }{\mathrm{d} \theta }  \right]
\hspace{-0.10cm}.
\end{split}                 
\end{equation}    
Since the $\gamma$--$\alpha$ and $\gamma$--$\beta$ interfaces are planar,  a notional derivative of $\tau$ with respect to the 
angle $\gamma$ as introduced for the lens cannot be defined in the present case; instead 
$\left[ \frac{\mathrm{d} \tau }{\mathrm{d} r }  \right]$ is used. As in the case of the lens the relations between 
($[\mathrm{d} R]$, $[\mathrm{d} h]$) 
and ($[\mathrm{d} \theta]$, $[\mathrm{d} r]$) are unique and the same is true for the 
corresponding notional derivatives.  
\\ 
Similarly as for the lens we find that  {\em $\tau$ is independent of the choice of dividing 
interfaces.} This statement is true in the same sense as the corresponding one discussed for the
lens below Eq. (\ref{lens6}). As in the case of the lens, higher order corrections to $\tau$ decreasing
with increasing $r$ like, e.g., $1/r$  -- such corrections are possible if $\tau$ is defined  
via Eq. (\ref{lensdrop1b}) instead of Eq. (\ref{eta}) -- depend on the choice of the  
dividing interfaces. Correspondingly, the notional derivatives of $\tau$ appearing in Eqs. 
(\ref{drop3}) and (\ref{drop4}) also depend on the choice of dividing interfaces and the corresponding 
terms  are furthermore of the same order as the term $\tau / r$ with only the leading term in $\tau$
being kept.     
\\ 
We now analyse the transformation of the notional derivatives upon notional shifts of the dividing interfaces. 
The same kind of arguments as for the lens leads to the following relations:  
\begin{eqnarray}
\left[ \frac{\mathrm{d} \tau }{\mathrm{d} R}  \right]^{(2)} - \left[ \frac{\mathrm{d} \tau }{\mathrm{d} R}  \right]^{(1)} & = &   
\frac{\sigma_{\alpha \beta} }{r } \left\{        \cos \theta \left[\mathrm{d} R \right] - 
\left[\mathrm{d} h \right] \right \}                                                             
\nonumber    \\ \left[ \frac{\mathrm{d} \tau }{\mathrm{d} h}  \right]^{(2)} - \left[ \frac{\mathrm{d} \tau }{\mathrm{d} h}  
\right]^{(1)} & = & - \frac{\sigma_{\alpha \beta}  }{r } \left \{ \left[\mathrm{d} R \right] - 
\cos \theta  \left[\mathrm{d} h \right] \right \}    \, .  
\nonumber    \\
   & &  
\label{drop5} 
\end{eqnarray} 
Changing the set of independent variables from ($h$, $R$) to ($r$, $\theta$) leads to the following form of the above 
transformation laws:   
\begin{eqnarray}
\left[ \frac{\mathrm{d} \tau }{\mathrm{d} \theta}  \right]^{(2)} - \left[ \frac{\mathrm{d} 
\tau }{\mathrm{d} \theta}  \right]^{(1)} & = &   
\sigma_{\alpha \beta} \left[\mathrm{d} R \right]     \nonumber                   \\ 
\left[ \frac{\mathrm{d} \tau }{\mathrm{d} r}  \right]^{(2)} - 
\left[ \frac{\mathrm{d} \tau }{\mathrm{d} r}  \right]^{(1)} & = & - \frac{\sigma_{\alpha \beta} 
\sin \theta }{r } \left[\mathrm{d} h \right]     
 \, . 
\label{drop6} 
\end{eqnarray} 
%
\subsubsection{Variational treatment with constraint of fixed volume }
%
Here the objective is to minimize the grand canonical potential $\Delta \tilde{\Omega}$ with the 
constraint of fixed volume $V_{\beta}$: 
\begin{equation}
\label{drop7}
\Delta \tilde{\Omega} =  A_{\alpha \beta } \sigma_{\alpha \beta }(R)    
-\pi r^2 (\sigma_{\alpha \gamma} - \sigma_{\beta  \gamma} )                + 
2 \pi r \tau (\theta ,r)  + \lambda V_{\beta}  
\end{equation}
in which no bulk terms are included since the Laplace pressure and the volume of the $\beta$ phase are considered to be 
fixed; the Lagrange multiplier is denoted by $\lambda$. We use $\theta$ and $r$ as the variables describing the system, 
and we introduce 
the stiffness constants $\frac{\mathrm{d}\sigma_{\alpha \beta}}{\mathrm{d} r}  \vert$ and $\frac{\mathrm{d}\sigma_{\alpha 
\beta}}{\mathrm{d} \theta}  \vert$ which are then expressed in terms of the 'natural' stiffness constant 
$\frac{\mathrm{d}\sigma_{\alpha \beta}}{\mathrm{d} R}  \vert$ with known properties.  In addition, similarly as for the 
lens, we introduce the stiffness constants $\frac{\mathrm{d}\tau}{\mathrm{d} r}  \vert$ and 
$\frac{\mathrm{d}\tau}{\mathrm{d} \theta}  \vert$ 
describing the cost in free energy attributed to the three-phase-contact line resulting from 
variational changes of $r$ and $\theta$ at fixed thermodynamic conditions. 
In the following we shall determine how these stiffness constants depend on the choice of dividing 
interfaces and how they are
related to notional derivatives.   
\\ 
If we impose the relation (for a justification see the discussion following, c.f., Eq. (\ref{drop11})) 
\begin{equation}
\label{drop8}  
\sin ^2 \theta  \frac{\mathrm{d} \tau }{\mathrm{d} \theta}  \Big\vert    = \frac{r^2}{2}   \frac{\mathrm{d} \sigma_{\alpha \beta} 
}{\mathrm{d} R} \Big\vert 
\end{equation} 
as in the case of the lens, the Lagrange multiplier $\lambda$  
acquires the meaning of the negative Laplace pressure: 
\begin{equation}
\label{drop9} 
\begin{split}  
\lambda = - \Delta p =  - \frac{ 2 \sigma _{\alpha \beta} \sin \theta }{r} -   
\frac{\mathrm{d} \sigma _{\alpha \beta}}{ \mathrm{d}R} 
\Big\vert    =   - \frac{  2 \sigma _{\alpha \beta}}{R} - 
\left[ \frac{\mathrm{d} \sigma _{\alpha \beta}}{ \mathrm{d}R} \right ]  . 
\end{split} 
\end{equation}  
Equation (\ref{drop8}) corresponds to the relations in Eqs. (\ref{lens18}) and (\ref{lens19}) which
were introduced in the case of the lens  
for similar reasons. We also remark that Eq. (\ref{drop8}) is formally identical to the consistency relation
in Eq. (\ref{drop3}). 
The second equation, obtained from minimizing $\Delta \tilde{\Omega} $ and due to the independence of the 
variables $\theta$ and $r$, has the following form (compare Eq. (\ref{lens13})):  
\begin{eqnarray} 
\sigma_{\alpha \beta } \cos \theta + (\sigma_{\beta \gamma} - \sigma_{\alpha \gamma }) = - \frac{\tau}{r} -  
\frac{\mathrm{d} \tau }{\mathrm{d} r }  \Big\vert    & &       
\nonumber \\ 
- 
\frac{r \sin \theta }{\left( 1 - \cos \theta \right )}     \frac{\mathrm{d} 
\sigma_{\alpha \beta } }{\mathrm{d} R }  \Big\vert         + 
\frac{ \left( 2 + \cos \theta \right ) \sin \theta}{r}     \frac{\mathrm{d} \tau 
}{\mathrm{d} \theta }  \Big\vert     \quad. & & 
\label{drop10}
\end{eqnarray} 
(An equation which expresses  
the fixed drop volume    
$V_{\beta} = (1/3)\pi r^3  [ (1 - \cos \theta)(2 + \cos \theta) ]/[ (1 + \cos \theta) \sin \theta ]$  
in terms of $r$ and $\theta$ and which together with the other equations fixes $r$, $\theta$, and the 
Lagrange parameter $\lambda$, does not contain information that could be used in the following.
The same has been found for the lens.) 
\\ 
Similarly as for the lens, we deduce from a comparison of the two equations, which follow from
Eq. (\ref{drop10}) for two differently chosen dividing interfaces,  
the transformation laws between the stiffness constants of the three-phase-contact line:  
\begin{eqnarray}
\frac{\mathrm{d} \tau }{\mathrm{d} \theta }  \Big\vert^{(2)} -  \frac{\mathrm{d} \tau }{\mathrm{d} \theta}  
\Big\vert^{(1)} & = &   
\sigma_{\alpha \beta} \left[\mathrm{d} R \right]  \nonumber                   \\  
\frac{\mathrm{d} \tau 
}{\mathrm{d} r}  \Big\vert^{(2)} -  \frac{\mathrm{d} \tau }{\mathrm{d} r}  
\Big\vert^{(1)} & = & - \frac{\sigma_{\alpha \beta} \sin 
\theta  }{r } \left[\mathrm{d} h \right]     \quad .  
\label{drop11}  
\end{eqnarray}  
Actually, the transformation behavior of $\frac{\mathrm{d} \tau }{\mathrm{d} \theta }  \vert$ 
is already fixed independently via Eq. (\ref{drop8}), 
$\frac{\mathrm{d} \sigma _{\alpha \beta}}{ \mathrm{d}R} \vert  =    
 \left[ \frac{\mathrm{d} \sigma _{\alpha \beta}}{ \mathrm{d}R} \right ] $,      
and Eq. (\ref{lensdrop7}). There is no contradiction between these two independent requirements 
because of the formal equivalence 
of the transformation laws in Eqs. (\ref{drop6}) for the notional derivatives and those for the stiffness 
constants in Eqs. (\ref{drop11}), and because Eqs. (\ref{drop3}) and (\ref{drop8}) are formally identical. 
In other words, one is indeed  
free to fix $\frac{\mathrm{d} \tau }{\mathrm{d} \theta}  \Big\vert$ by Eq. (\ref{drop8}) 
and thus to bestow  
the meaning of the negative Laplace pressure $- \Delta p$ on the Lagrange multiplier $\lambda$.  
\\ 
Equation (\ref{drop10}) can be simplified by using Eq. (\ref{drop8}):   
\begin{equation}
\label{drop12}
\begin{split}  
\sigma_{\alpha \beta } \cos \theta + (\sigma_{\beta \gamma} - \sigma_{\alpha \gamma }) = - \frac{\tau}{r} -  
\frac{\mathrm{d} \tau }{\mathrm{d} r }  \Big\vert       - \frac{ \sin \theta \cos \theta }{r}     
\frac{\mathrm{d} \tau }{\mathrm{d} \theta }  
\Big\vert  .
\end{split}  
\end{equation}  
From a comparison of Eqs. (\ref{drop12}) and  (\ref{drop4}) we obtain 
\begin{equation} 
\label{drop13}
\frac{\mathrm{d} \tau }{\mathrm{d} \theta }  \Big\vert = \left[ \frac{\mathrm{d} \tau }{\mathrm{d} \theta }  \right] 
\hspace{0.2cm} , 
\hspace{0.2cm}   \frac{\mathrm{d} \tau }{\mathrm{d} r}  \Big\vert =\left[ \frac{\mathrm{d} \tau }{\mathrm{d} r}  \right]  \quad.   
\end{equation} 
(Note that applying Eq. (\ref{drop8}) to Eq. (\ref{drop13}) in combination with Eq. (\ref{lensdrop8}) yields  
$ \frac{\mathrm{d} \tau }{\mathrm{d} \theta }  \vert ^{(\mathrm{s})} = 0 $.)  Again, these relations are compatible with the 
transformation laws for both the notional derivatives and the stiffness constants.   
\section{Possibility of choosing different definitions for line tensions}    
\renewcommand{\theequation}{6.\arabic{equation}}
\setcounter{equation}{0}\vspace*{0.5cm} 
At this point it is due to comment on the seeming contradiction between our statement 
that $\tau$ is independent of the 
choice of dividing interfaces and Eq. (\ref{deltaeta2}), following from a discussion of
a straight three-phase-contact line at a wedge shaped fluid volume,  
which states the opposite and tells how $\tau$ should change upon shifting the fluid--solid dividing interface. 
A question related to that issue is whether one can use the line tensions of a gas--liquid--solid contact  
calculated from a certain 
microscopic theory in combination with a decomposition scheme in a model system like that considered 
in Sect. III (i.e., for a fluid wedge geometry, see, e.g., Refs. \cite{GD1,BD}), 
and use them in the formulae derived above for the drop.
\\     
In order to answer these questions we reanalyze again the particular  
wedge geometry investigated already in Sect. III in order to find out whether implicit assumptions
in the analysis given there are at variance with prescriptions used in Sect. V  
for the drop. We also resume our discussion of the drop system
in order to see whether sensible alternative definitions of the line tension
in that system do exist and whether the contradiction mentioned above can be resolved by using an alternative
definition of $\tau$. If such an alternative exists the question about the relation between 
the two different line tensions has to be answered.    
\subsection{Decomposition and reassembly of a system:  \\ difficulties with edge terms}
In Sect. III we have calculated the changes of the areas of the liquid--gas
and of the solid--fluid interfaces due to a notional shift of the position of the solid--fluid
interface within the box with rectangular cross section, as 
shown in Fig. 2, implicitly assuming that these area changes are representative 
for an entire system from which Fig. 2 actually shows only
a small part. In this procedure a possible source of errors might be hidden. We therefore 
discuss now the underlying general procedure implicitly used in Sect. III. First, the total system
is decomposed into two subsystems I and II such that subsystem I contains  
the three-phase-contact line as the object of interest. On the other hand subsystem I is
surrounded by a subsystem II    
which itself does not contain any actual line contribution. (In the case of straight contact lines
this requires to use periodic boundary conditions in one direction; 
this is of no relevance for the following discussion.)
The contribution of subsystem I to the grand potential is then analyzed in terms of volume, 
interfacial, and line
contributions {\em ignoring unphysical edge terms}. The same is done for subsystem II 
except that no actual line contribution appears in the decomposition. 
At the end all contributions from both subsystems 
are put together in order to calculate the grand potential of the total system. 
It is expected that the grand potential (and the decomposition into volume, interfacial,
and line contributions) calculated in that way is equal to what one obtains  
if one decomposes the grand potential of the whole system into volume, interfacial, and line 
contributions and then adds up these contributions without first partitioning the system. 
In what follows we discuss why it may be sometimes misleading to ignore the unphysical edge 
terms. 
\par
\begin{figure}
\includegraphics*[scale=.32]{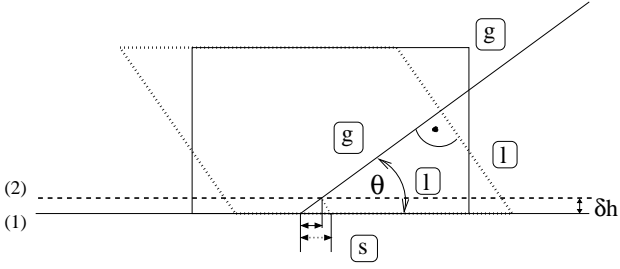}
\caption{\label{fig9} Two possibilities for cutting a subsystem I out of a total system
 containing a l--g interface and a three-phase-contact line with the solid s.
 These two possibilities are indicated by two boxes (full and dotted lines)
 enclosing the l--g--s contact line. The structure is assumed to be translationally invariant
 perpendicular to the plane of the figure. Two possible choices for the solid--fluid dividing interface
 are indicated by solid (1) and dashed (2) lines. The two dividing interfaces are separated by a
 distance $\delta h$. The areas of the solid--liquid (solid--gas) interfaces within subsystem I
 change when the solid--fluid dividing interface is shifted. The values of these area changes
 depend on the shape of the box defining subsystem I. This dependence is indicated by the two horizontal
 double arrows. The short double arrow indicates the changes, due to the shift $\delta h$,
 of the areas of the solid--gas
 and the solid--liquid interfaces within the box of rectangular cross section, the long
 double arrow the changes of these areas within the box with parallelogram-shaped cross section.
}
\end{figure}

 
The decomposition introduced above is not
unique. Two examples of such decompositions  
are indicated in Fig. \ref{fig9} by two boxes
of different shapes enclosing and defining the respective subsystem I. The structure is assumed to
be translationally invariant in the direction perpendicular to the plane shown in Fig. \ref{fig9}.
The cross section of the box has the shape of a rectangle for one of the chosen decompositions and 
in the other case it has the shape of a parallelogram such that two of the faces of the box  
are perpendicular to the liquid--gas interface. The embedding system II is not specified in 
any detail except that it is understood that the structures found within the box, e.g., the liquid--gas
interface, extend beyond the box boundaries into the embedding system II.   
For that reason interfacial energies at the boundaries of the boxes do not arise. (This statement
also comprises the solid--liquid and the solid--gas interfaces since the respective boundary of the
box can be placed anywhere inside the solid.) There are also no actual line contributions to the 
free energy from the edges of the box or from the line at which the liquid--gas interface intersects a 
face of the box. The environments of these lines are not different from bulk 
or from a corresponding region in an infinitely extended interface.  
Inspite of the absence of actual line contributions,  
artificial non-physical line contributions at the lines just mentioned have to be
included in the expressions for the grand canonical potentials characterizing the subsystems.
That there is a need to introduce these artificial line terms becomes very plausible if
one looks at the two possible decompositions of a system into two subsystems shown in Fig. \ref{fig9}.
Evidently interfaces are cut in different ways by the faces defining subsystem I for the
two considered options. Therefore, contributions to the grand canonical potential stemming from
certain parts of the interfaces are either attributed to subsystem I or subsystem II depending on how the 
decomposition is done. Only if one includes the artificial line terms in the analysis one can 
keep track of such traces of the decomposition; otherwise one runs into inconsistencies.
\par 
Before we proceed with this analysis, which ultimately aims at finding out whether and how
the true line tension may be determined from microscopic calculations for the most simple geometries,    
we continue with the additive decomposition of 
the grand canonical potential $\Omega$ of a system  
\begin{equation}
\label{decomp-gen-1}
\Omega = \Omega_{\mathrm{I}} + \Omega_{\mathrm{II}} \quad ,  
\end{equation}  
where $\Omega_{\mathrm{I}}$ and $\Omega_{\mathrm{II}}$ are the contributions 
to the grand canonical potential attributed to 
subsystem I and II, respectively, which, e.g., in the spirit of density functional
theory can be thought of to be calculated from the density distributions characterizing the 
total system.        
Already this decomposition is 
problematic and not unique although obvious choices do exist. 
For instance, if we base our theoretical description on a type of model in which 
the grand canonical potential is expressed in terms of a local functional of the number
density (of, e.g., the    
Ginzburg--Landau type), $\Omega_{\mathrm{I}}$
is obtained by integrating the local density of the grand canonical potential over the 
volume of the box defining the subsystem I.  
The local density of the grand canonical potential is obtained from particle-density distributions
which may be obtained from calculations restricted to the interior of the box. The boundary 
conditions have to correspond to a seamless continuation of the structures in the interior 
of subsystem I 
into the surrounding subsystem II. The decomposition procedure becomes less obvious if a non-local
density-functional description is used. In that case interactions across the boundaries of the
subsystems occur. In order to calculate the contribution to $\Omega_{\mathrm{I}}$   
coming from two-center integrals one might, e.g., restrict one of the integrations
to the interior of the box, whereas the second integration runs over the total volume of 
the whole system. The value obtained for $\Omega_{\mathrm{I}}$ is independent of size and 
shape of the surrounding subsystem II provided it is sufficiently big and the interactions 
between particles decay sufficiently rapidly as a function of their distance. Artificial 
interfacial contributions from the surfaces of the box are eliminated automatically by 
chosing the described procedure. By carrying out the integrations in the way outlined above 
we basically mimic a local density of  $\Omega$ allowing, however, for the embedding 
of the subsystem into a global system.        
\par
In the sense of the previous paragraph $\Omega$ can be 
decomposed into a sum of two contributions originating from two subsystems I and II. 
Since for this decomposition no use has been made of the concept of dividing interfaces between phases, in particular 
between solid and liquid or solid and gas,  
it is also clear that the two contributions $\Omega_{\mathrm{I}}$ and 
$\Omega_{\mathrm{II}}$ must be independent of choices for the dividing interfaces between 
solid and fluid.  
\\
In the next step $\Omega_{\mathrm{I}}$ and $\Omega_{\mathrm{II}}$ are further decomposed
into volume, interface, and line contributions, i.e., 
\begin{equation}
\label{decomp-gen-2}
\Omega_{\mathrm{I}} = V_{\mathrm{I}} \omega + A^{\mathrm{I}}_{\mathrm{lg}} \sigma_{\mathrm{lg}}
                      + A^{\mathrm{I}}_{\mathrm{sl}} \sigma_{\mathrm{sl}} + 
                        A^{\mathrm{I}}_{\mathrm{sg}} \sigma_{\mathrm{sg}} 
                      + L \tau + \sum _i L  _i \tau _i^{\mathrm{art,I}}  \quad ,  
\end{equation}  
and 
\begin{equation}
\label{decomp-gen-3}
\Omega_{\mathrm{II}} = V_{\mathrm{II}} \omega + A^{\mathrm{II}}_{\mathrm{lg}} \sigma_{\mathrm{lg}}
                      + A^{\mathrm{II}}_{\mathrm{sl}} \sigma_{\mathrm{sl}} + 
                        A^{\mathrm{II}}_{\mathrm{sg}} \sigma_{\mathrm{sg}} 
                      + \sum _i L  _i \tau _i^{\mathrm{art,II}}  \quad ,  
\end{equation}  
where $A^{\mathrm{I}}_{\mathrm{lg}}$ and $A^{\mathrm{II}}_{\mathrm{lg}}$ are the areas of 
the l--g interfaces within subsystem I and II, respectively, 
$A^{\mathrm{I}}_{\mathrm{sl}}$ and $A^{\mathrm{II}}_{\mathrm{sl}}$ are those of the s--l interfaces, and
$A^{\mathrm{I}}_{\mathrm{sg}}$ and $A^{\mathrm{II}}_{\mathrm{sg}}$ are the areas of the s--g interfaces 
(l: liquid, g: gas, s: solid or wall). 
$\tau$ is the line tension of the actual three-phase-contact line and thus it appears only in subsystem I.
The terms $\tau _i^{\mathrm{art,I}}$ or $\tau _i^{\mathrm{art,II}}$ characterize the artificial 
line contributions resulting from the decomposition into subsystems. The quantities $L  _i$ are the lengths of 
the artificial lines, which are equal in both subsystems.  Since 
$\Omega = \Omega_{\mathrm{I}} + \Omega_{\mathrm{II}}$ and because the only actual line inhomogeneity,  
which is present in the 
total system, is accounted for by the line tension $\tau$, one has  
\begin{equation}
\label{decomp-gen-4}
\sum _i L  _i \tau _i^{\mathrm{art,I}} = - \sum _i L  _i \tau _i^{\mathrm{art,II}} \quad ,
\end{equation} 
i.e., the artificial line contributions of the two subsystems cancel each other. 
\\
Since $\Omega_{\mathrm{I}}$ is independent of the choice of the dividing interfaces between phases,
in particular of the solid--fluid dividing interface,  
the same arguments as used in Sec. 3 can now be applied to the subsystem I and to 
$\Omega_{\mathrm{I}}$ 
in order to study the influence of a shift by an amount $\delta h$ of the s--g and s--l 
dividing interfaces. (We restrict the discussion to a structure that is translationally invariant
in the direction parallel to the line, i.e., $L  _i = L$.) 
If the rectangular box indicated in Fig. \ref{fig9} is chosen one obtains 
\begin{equation}
\label{decomp-gen-5}
 \left [ \tau + \sum _i L  _i \tau _i^{\mathrm{art,I}} \right ] ^{(2)} - 
 \left [ \tau + \sum _i L  _i \tau _i^{\mathrm{art,I}} \right ] ^{(1)} 
 = \sigma_{\mathrm{lg}} \sin \theta \delta h
\end{equation}  
for the difference of the expression 
in square brackets for the two choices (2) and (1) of the dividing interfaces (compare Fig. 2). 
The only but essential difference to Eq. (\ref{deltaeta2}) is that Eq. (\ref{decomp-gen-5}) gives the difference 
not only for $\tau$ but for the sum of the actual and the artificial line energies. There seems to be a chance   
that the seeming contradiction between the result that $\tau$
of the actual three-phase-contact line is independent of the choice of dividing 
interfaces and Eq. (\ref{deltaeta2}) can be resolved by including the artificial line contributions
and replacing Eq. (\ref{deltaeta2}) by Eq. (\ref{decomp-gen-5}).   
Of course Eq. (\ref{decomp-gen-5}) is only valid for the decomposition employing the rectangular box.   
If instead of a rectangular box a box with a cross section of the  
shape of a parallelogram (as shown in Fig. \ref{fig9}) is chosen Eq. (\ref{decomp-gen-5}) is replaced by  
\begin{equation}
\label{decomp-gen-6}
 \left [ \tau + \sum _i L  _i \tau _i^{\mathrm{art,I}} \right ] ^{(2)} - 
 \left [ \tau + \sum _i L  _i \tau _i^{\mathrm{art,I}} \right ] ^{(1)} 
 = 0   \quad ,                                       
\end{equation}  
because the changes of the solid--liquid (solid--gas) interfaces within subsystem I, resulting
from the shift of the solid--fluid dividing interface, are computed in the way indicated
by the long dotted double arrow in Fig. \ref{fig9} and not
in the way indicated by the short solid double arrow in Fig. \ref{fig9} 
which is the correct way to compute those changes within the rectangular box.   
From the comparison of Eq. (\ref{decomp-gen-5}) and Eq. (\ref{decomp-gen-6}) it becomes obvious that  
in order to deduce $\tau$ from a calulated expression for $\Omega_{\mathrm{I}}$ one has to subtract 
not only volume and interface contributions but also the artificial line contributions. 
Such artificial line contributions are generated if a box boundary cuts  
through the inhomogeneous structure of an interface in a non-adapted way. 
Within an interfacial region, which has a macroscopic extension in the lateral directions, the density
profiles of the constituents of the fluid only depend on the coordinate perpendicular to
the interface. If the box boundary defining subsystem I cuts perpendicularly through 
the interface, the cut is adapted to the density profiles and no artificial line contribution
is generated. If by contrast, a box boundary cuts through the interface with an angle 
deviating from $90^{\mathrm{o}}$, a spatial region of columnar shape, filled 
with an inhomogeneous fluid, is generated above or below the interface, which 
no longer belongs to subsystem I although according to the interfacial 
area attributed to subsystem I it should contribute to the interfacial energy within
that subsystem. Vice versa a volume of columnar shape is attributed to subsystem I 
although the fluid within that volume should contribute to the interfacial energy 
attributed to subsystem II.     
Since in general the free energy contributions from these two columnar spatial 
regions do not compensate one is left with an artificial line contribution.  
\par
\begin{figure}[h]
\includegraphics*[scale=.32]{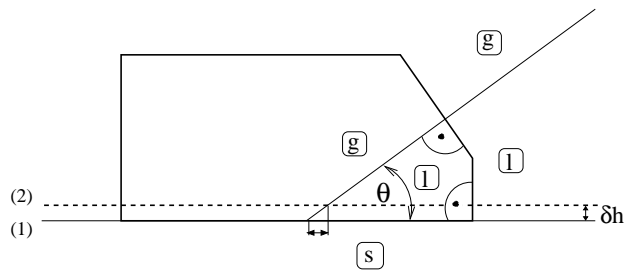}
\caption{\label{Fig-Wedge-b} Another possibility for cutting a subsystem I out of a total system
 containing a l--g interface and a three-phase-contact line with the solid s.
 The structure is assumed to be translationally invariant
 perpendicular to the plane of the figure. Two possible choices for the solid--fluid dividing interface
 are indicated by solid (1) and dashed (2) lines. The two dividing interfaces are separated by a
 distance $\delta h$. The areas of the solid--liquid (solid--gas) interfaces within subsystem I
 change (horizontal double arrow) if the solid--fluid dividing interface is shifted. These changes in areas are
 identical to those obtained if the box of rectangular cross section is chosen in Fig. \ref{fig9}.   
}
\end{figure}

 
For that reason we discuss still another possibility (Fig. \ref{Fig-Wedge-b})
for cutting a subsystem I out of a total system
containing a l--g interface and a three-phase-contact line with the solid s.  
For this choice the box boundaries cut perpendicularly, i.e., in an adapted manner  
through both interfaces. This choice avoids the appearance of artificial line terms
at the cuts of the box boundaries through the liquid--gas or the solid--fluid interfaces.  
An additional line generated by this particular choice of the box boundaries is placed
such, that the system is completely homogeneous in a big volume around that line. 
Therefore, no net artificial line contribution is associated with that line, too. 
Carrying out the same analysis as for the two other box shapes we again 
obtain Eq. (\ref{decomp-gen-5}) but the 
artificial line energies should now vanish leading back to Eq. (\ref{deltaeta2}).   
\par
After our previous analyses for different box shapes defining subsystem I one might
come to the following suggestion to cope with artificial line contributions.
For the rectangular box (which leads to Eq. (\ref{decomp-gen-5})) the only artificial
line contribution stems from the intersection of the liquid--gas interface 
with the box boundary. No artificial line contributions arise at the cuts of the 
box boundaries with the solid--fluid interface, because the box boundaries cut perpendicularly
through that interface. There are also no artificial line contributions from
the edges of the box located in homogeneous regions of the fluid.  
If we now argue that the artificial line contribution does not change upon a notional
shift of the solid--fluid interface because the corresponding line is far away from
that interface, the terms related to the artificial line contribution would cancel in 
Eq. (\ref{decomp-gen-5}) and we would again be led back to Eq. (\ref{deltaeta2}). 
In the case of the parallelogram shaped box and Eq. (\ref{decomp-gen-6}) we might argue 
using the same type of arguments, that only the two line contributions due to the 
intersection of the box boundaries with the solid--fluid interface do arise. But now 
one would be inclined to admit that these artificial line contributions might change
upon a a notional shift of the solid--fluid interface because the lines are spatially
associated with the shifted interface. Therefore, one would not draw conclusions about 
notional changes of $\tau$ from Eq. (\ref{decomp-gen-6}). However, one should not
rely too heavily on such type of arguments. After all, notional shifts of dividing interfaces
lead just to changes in interfacial areas and thus to changes in interfacial contributions
to the free energy which are then subtracted from the total free energy in order to
define the line tension(s). From the viewpoint of a macroscopic theory it 
is not clear how these changes of interfacial free energies should be split up  
among the real and the artificial line contributions.    
\\
In the present subsection we have discussed how a line tension $\tau$ and its
dependence on the choice of dividing interfaces can be determined unambigously 
from calculations for a simple model system, i.e., a liquid wedge in contact with a gas phase
on a solid substrate. 
In order to be able to proceed, one has to carry out the calculations  
within a finite subsystem (box) cut out from the unbounded model system. The 
seamless continuation into the embedding system has to be incorporated into the calculation
by choosing proper boundary conditions and by taking into account certain interactions 
of the finite subsystem with the embedding system. In addition, the box boundaries 
have to be choosen such, that they cut through interfaces perpendicularly and 
artificial box edges 
have to be placed into homogeneous regions of the total system (a possible choice of the 
box is shown in Fig. \ref{Fig-Wedge-b}).  
With all these precautions we find that the dependence of $\tau$ on the choice of the solid--fluid
dividing interface is definitely given by Eq. (\ref{deltaeta2}).  
\subsection{Finite containers filled with fluids }
\begin{figure}[h]
\includegraphics*[scale=.37]{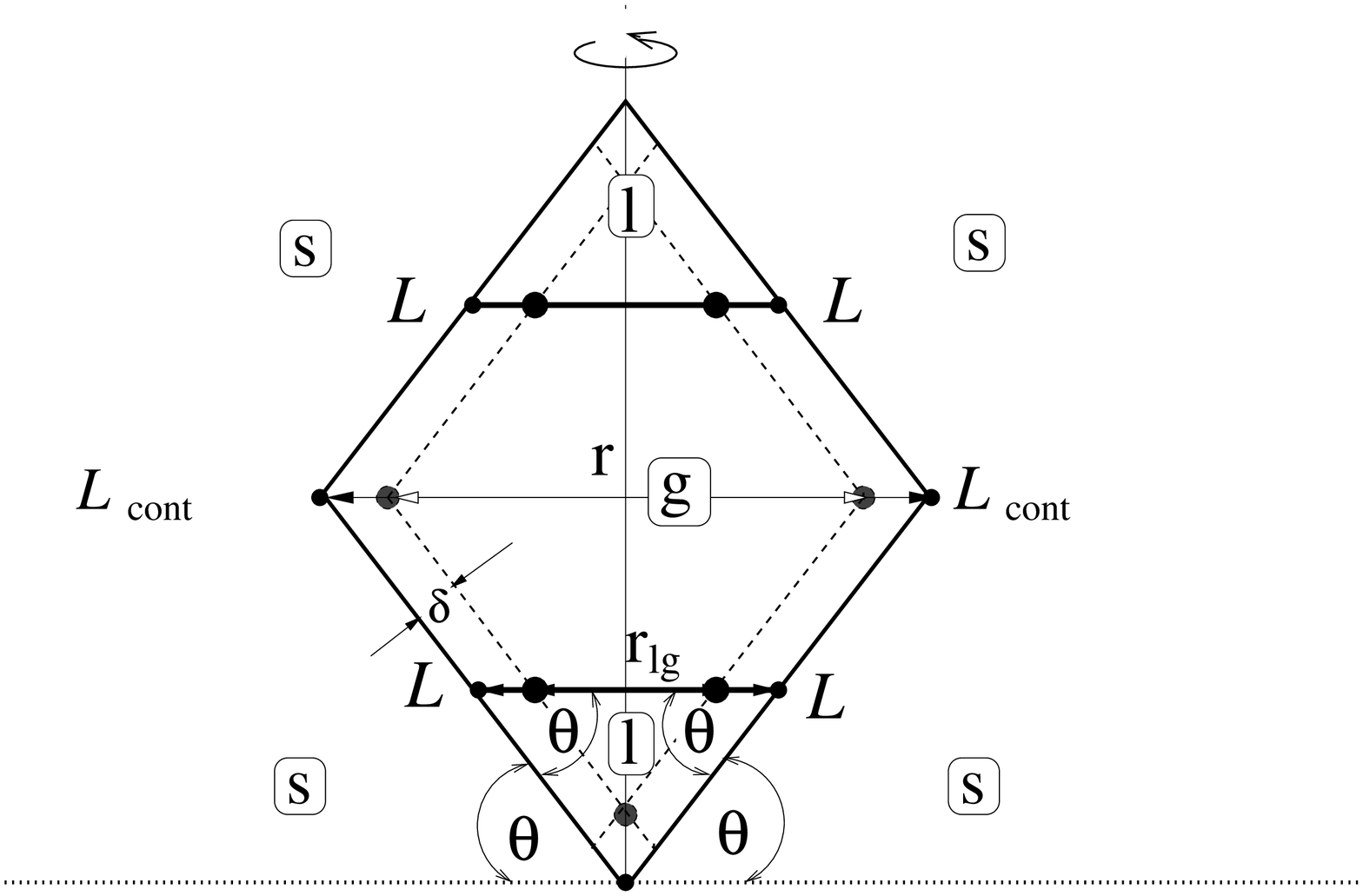}
\caption{\label{fig7} Cut through a biconical container (circular base) with an opening angle
 180$^{\mathrm{o}} - 2\theta$.  
 The apexes of the container are filled with the liquid phase (l) of a fluid up to 
 the planar liquid--gas interfaces. 
 The central part of the container is filled with the coexisting gas phase.
 The filling height is determined by a prescribed amount of liquid.
 Two possible choices for the
 dividing interface between solid (wall) and fluid are indicated by solid and dashed lines.
 The two possible dividing interfaces are separated by a distance $\delta$. The two
 corresponding positions of the liquid--gas--solid contact lines denoted as L are indicated
 by small and large dots. At the joint between the two cones there is another circular line 
 L$_{\mathrm{cont}}$ formed.
}
\end{figure}

After the discussions in the previous subsection we are still left with the contradiction
between Eq. (\ref{deltaeta2}), which tells us how $\tau$ should depend on the choice of 
dividing interfaces, and our statement in Sect. V  which says that $\tau$ should be
independent of that choice. We would like to stress that this latter statement
is based on a well defined decomposition scheme of the grand potential of a drop
on a substrate (or a liquid lens at a fluid interface).    
\\  
In order to gain further insight into this problem
we now discuss a finite fluid system enclosed in a container 
with solid walls. This way we avoid from the outset the necessity to cut out a finite
subsystem and we stay away from the danger to pick up unphysical edge or line contributions.
Our treatment of closed finite containers gives us also the opportunity to discuss
a further aspect of solid--fluid systems, i.e., the curvature of the solid--fluid
interface.  
We further chose the boundary conditions such that the liquid--vapor interface
is planar and thus no complications occur related to the Laplace pressure.
In addition there is no curvature correction to the liquid--vapor surface tension. 
On the other hand if a container with only planar walls is partially filled with 
a liquid in contact with its vapor phase, in addition to the liquid--vapor--solid
contact line further edge (line) contributions do appear which
scale with the same linear dimension of the container as the three-phase-contact line
we are interested in. Therefore, the different line (edge) contributions cannot be
separated unless these edge contributions have been determined independently from investigations
of reference configurations not containing a liquid--vapor--solid contact line. As a reference  
configuration the same container but completely filled with either liquid
or gas (vapor) is chosen. But then one encounters the problem that there is no unique description
how to extract all additional edge contributions individually and in particular it is 
impossible to find the transformations upon notional shifts of dividing
interfaces for each of these edge contributions individually.  
\\ 
In order to facilitate to vary the length of the liquid--vapor--solid contact line
indepently from the lengths of container edges we consider a biconical container 
composed of two identical but oppositely oriented right cones with circular base
which are glued together base to base along the circumference of the base 
as shown in Fig. \ref{fig7}.
All container walls are composed of the same solid material s. The cones are 
filled with liquid l up to a certain height, which is determined by the amount of liquid provided,
and the liquid is in contact with its
vapor phase g. The opening angle of the cone is chosen to be $180^{\mathrm{o}} - 2 \theta$
such that the liquid--vapor interface is planar. The system
contains two structural units characterized by lines. The first line is at the joint of the two   
cones and is termed L$_{\mathrm{cont}}$  (Fig. \ref{fig7}). It has the shape of a circle with a radius
denoted as $r$. In the following the related line (edge) tension will be called $\tau_{\mathrm{cont}}$.
The second line, the one we are
interested in, is the three-phase-contact line between solid, vapor, and liquid. It also has a circular
shape with a radius denoted as $r_{\mathrm{lg}}$. The related line tension will be called $\tau$. 
Again a parallel displacement of the dividing interfaces between solid and fluid
by an amount $\delta$ is considered. This parallel displacement is indicated in Fig. \ref{fig7}. 
This set-up allows one to vary the length of the three-phase-contact line
corresponding to $\tau$ independently
from the container dimensions.  It is tempting to conclude that thus the line tension $\tau$ 
and its dependence on the choice of dividing interfaces can be
separated unambigously from the line contributions related to the geometry of the container.
However, since the container walls are curved, we also have to take into account curvature corrections to
the solid--fluid interface tensions. As we shall see, these curvature corrections bring about contributions
to the grand potential which scale with the length of the liquid--vapor--solid contact line.   
Therefore, for this geometry one also has to study at first reference configurations not containing a 
liquid--vapor--solid contact line, i.e., the bicone being homogeneously filled with either liquid or vapor,
in order to determine how the curvature contributions depend 
on the choice of dividing interfaces. This is actually possible if an additional albeit  
plausible assumption is made (see below). In the next step the bicone filled partially with liquid
and partially with vapor 
is analyzed. What is found for the curvature terms in the previous step for the container
filled with an homogeneous fluid is then   
transferred to the new situation of a container filled with an inhomogeneous fluid without 
any modification.   
\subsubsection{Homogeneously filled bicone} 
We decompose the grand canonical potential into volume, interfacial, line contributions, 
and we include curvature contributions due to the curvature of the solid--fluid wall
(see, e.g., \cite{Roth1,Roth2,Roth3,Roth4}. The curvature
contributions could be also expressed as a curvature correction to the solid--fluid 
interfacial tension. Terms which do not scale at least with a linear container dimension
are disregarded:   
\begin{multline}
\label{bicone-hom-1}
\frac{1}{2}\Omega = -p V^{(i)}_{\mathrm{cone}} + A^{(i)}_{\mathrm{cone-ssf}} \, 
                                                \sigma^{(i)}_{\mathrm{sg(sl)}}
                      + \kappa^{(i)}_{\mathrm{sg(sl)}} C^{(i)}_{\mathrm{cone-ssf}}   \\
                      +  \frac{1}{2} L^{(i)}_{\mathrm{cont}} \tau^{(i)}_{\mathrm{cont}}
                        \quad ,
\end{multline}
which is the grand canonical potential for one half of the system and $p$ is the pressure
of the fluid, $V^{(i)}_{\mathrm{cone}}$ the volume of one cone, $A^{(i)}_{\mathrm{cone-ssf}}$
the area of the side surface of one cone, $\sigma^{(i)}_{\mathrm{sg(sl)}}$ is the interface
tension of a planar solid--gas or solid--liquid interface, respectively, $C^{(i)}_{\mathrm{cone-ssf}}$
is the mean curvature of the side surface integrated over the whole area of the side surface 
of one cone with $\kappa^{(i)}_{\mathrm{sg(sl)}}$ as the corresponding thermodynamic coefficient, 
and $\tau^{(i)}_{\mathrm{cont}}$
is the line (edge) tension associated with the joint between the two cones 
(see Fig. \ref{fig7}) of length $L^{(i)}_{\mathrm{cont}}$. The superscript
$^{(i)}$ indicates that geometric quantities as well as the interfacial and line tensions and
the coefficient of the curvature term depend on the choice ($i$) of the dividing interface
between solid and fluid. The integrated mean curvature of the side surface of a cone with
base radius $r^{(i)}$ is 
\begin{equation}
\label{curvature-1}
C^{(i)}_{\mathrm{cone-ssf}} = \frac{\pi r^{(i)} \sin \theta}{\cos \theta}  .  
\end{equation}   
If we again compare the decompositions for the two choices (1) and (2) of the dividing interface
shown in Fig. \ref{fig7} and if we use $\sigma^{(2)}_{\mathrm{sg(sl)}} - \sigma^{(1)}_{\mathrm{sg(sl)}} 
= -p \delta $ ( a special variant of Eq. (\ref{deltasigmasf}) not restricting the generality of
the following arguments) and if we neglect terms which do not scale at least with $r^{(i)}$,
we obtain the relation  
\begin{multline}
\label{bicone-hom-2}
\kappa^{(1)}_{\mathrm{sg(sl)}} C^{(1)}_{\mathrm{cone-ssf}} +  
                   \frac{1}{2} \tau^{(1)}_{\mathrm{cont}} L^{(1)}_{\mathrm{cont}}
\\ 
= 
\kappa^{(2)}_{\mathrm{sg(sl)}} C^{(1)}_{\mathrm{cone-ssf}} + 
                   \frac{1}{2} \tau^{(2)}_{\mathrm{cont}} L^{(1)}_{\mathrm{cont}}
\\ 
+ p \frac{\pi r^{(1)} }{\sin\theta \cos \theta } \delta ^2 - 
  \sigma^{(1)}_{\mathrm{sg(sl)}} \frac{\delta}{\sin\theta \cos \theta } 2 \pi r^{(1)}  .  
\end{multline}
Without a further assumption it is of course not possible to obtain the transformation behavior
upon shifting the dividing interface for $\kappa_{\mathrm{sg(sl)}}$ and 
$\tau_{\mathrm{cont}}$ independently. In order to proceed we consider a rounding of the edge 
at L$_{\mathrm{cont}}$ around the joint of the two cones and attribute to this
edge a total (integrated) mean curvature based on the following arguments.  
For a container with smooth walls without any edge, differential geometry provides the 
general relations (see, e.g., Refs. \cite{Smirnow, Hadwig})   
\begin{equation}  
\label{curvature-2} 
V^{(2)} = V^{(1)} - A \delta + C \delta ^2 - \frac{1}{3} X \delta ^3
\end{equation}
and 
\begin{equation} 
\label{curvature-3} 
A^{(2)} = A^{(1)} - 2 C \delta + X \delta ^2  \quad ,
\end{equation}  
expressing the changes in volume and surface area of a container, associated with            
an infinitesimal parallel shift $\delta$ of its surface,   
in terms of the area $A$, the total mean curvature $C$, and the total Gaussian
curvature $X$. From the calculated volume and area changes and from Eqs. (\ref{curvature-2}) and (\ref{curvature-3}),    
the total mean curvature for the bicone (with rounded edge) turns out to be 
\begin{equation}
\label{curvature-4}
C = \frac{2\pi r }{\sin \theta \cos \theta} \quad , 
\end{equation}
whereas adding up the integrated mean curvatures of the side surfaces of the two cones (Eq. (\ref{curvature-1}))
gives  
\begin{equation}
\label{curvature-5}
C_{\mathrm{bicone}} = \frac{2\pi r \sin ^2 \theta}{\sin \theta \cos \theta} \quad .
\end{equation}
This means that a contribution 
\begin{equation}
\label{curvature-6}
C_{\mathrm{seam}} = \frac{2\pi r \cos ^2 \theta}{\sin \theta \cos \theta} 
\end{equation}
to the total mean curvature of the container is missing. Obviously this contribution
can be attributed to 
the line L$_{\mathrm{cont}}$ at which the two cones are glued together
and the curvature attributed to that line can be realized by deforming the surface in 
its vicinity into one that is differentiable.    
We use now this observation in order to replace the line (edge) term by an equivalent 
curvature term:   
\begin{multline}
\label{bicone-hom-3}
 \frac{1}{2} \tau_{\mathrm{cont}} L_{\mathrm{cont}}  \longrightarrow 
 \frac{1}{2} \kappa_{\mathrm{sg(sl)}} C_{\mathrm{seam}}
 \\  
         =    \kappa_{\mathrm{sg(sl)}} \frac{\pi r \cos ^2 \theta}{\sin \theta \cos \theta}   
                 \quad .
\end{multline}
Combining Eq. (\ref{bicone-hom-2}) and Eq. (\ref{bicone-hom-3}) one finds  
\begin{equation}
\label{curvature-7}
\kappa^{(1)}_{\mathrm{sg(sl)}} = \kappa^{(2)}_{\mathrm{sg(sl)}} + p \delta ^2 - 
                                 2 \sigma ^{(1)}_{\mathrm{sg(sl)}} \delta  .   
\end{equation} 
\subsubsection{Bicone filled partially with liquid and vapor} 
The grand canonical potential of the bicone filled partially with liquid and partially
with vapor (see Fig. \ref{fig7}) decomposes as 
\begin{multline}
\label{bicone-inhom-1}
\frac{1}{2}\Omega = -p V^{(i)}_{\mathrm{cone}} + A^{(i)}_{\mathrm{cone-sg}} \,
                                                \sigma^{(i)}_{\mathrm{sg}}
                     + A^{(i)}_{\mathrm{cone-sl}} \, \sigma^{(i)}_{\mathrm{sl}}  
                                                                                  \\  
                      + \kappa^{(i)}_{\mathrm{sg}} C^{(i)}_{\mathrm{cone-sg}}   
                      + \kappa^{(i)}_{\mathrm{sl}} C^{(i)}_{\mathrm{cone-sl}}   
                      + A^{(i)}_{\mathrm{lg}} \, \sigma_{\mathrm{lg}}  
                                                                                  \\
                      +  \frac{1}{2} L^{(i)}_{\mathrm{cont}} \tau^{(i)}_{\mathrm{cont}}
                      + L^{(i)} \tau^{(i)} 
                        \quad ,
\end{multline}
where $A^{(i)}_{\mathrm{cone-sg}}$ and $A^{(i)}_{\mathrm{cone-sl}}$ are those areas of 
the cone side surface which are in contact with the gas and the liquid phase, 
respectively. $\sigma^{(i)}_{\mathrm{sg}}$ and $\sigma^{(i)}_{\mathrm{sl}}$ are the
interface tensions of the planar solid--gas and solid--liquid interfaces. 
$C^{(i)}_{\mathrm{cone-sg}}$ and $C^{(i)}_{\mathrm{cone-sl}}$ are the mean curvatures
of the cone side surface integrated over that part of the surface which is in contact
with the gas phase and the liquid phase, respectively. $\kappa^{(i)}_{\mathrm{sg}}$ and 
$\kappa^{(i)}_{\mathrm{sl}}$ are the corresponding curvature coefficients. 
$A^{(i)}_{\mathrm{lg}}$ is the area of the planar liqid--gas interface and 
$\sigma_{\mathrm{lg}}$ the surface tension of a planar liquid--gas interface.  
The last but one term in Eq. (\ref{bicone-inhom-1}), already familiar from the
homogeneously filled bicone,  gives the line contribution from
the edge along which the two cones are glued together. The last term is the contribution
from the solid--liquid--gas three-phase-contact line. The superscript $^{(i)}$ indicates
which of the quantities depends on the choice of the solid--fluid dividing interface   
($\sigma^{(i)}_{\mathrm{sg}} - \sigma^{(i)}_{\mathrm{sl}}$ is independent of such 
choices).  
As in the previous subsection we compare the decompositions for the two different choices of dividing interfaces
indicated in Fig. \ref{fig7}, we neglect all higher order terms which do not scale
with a characteristic system size, we further use Eq. (\ref{bicone-hom-2}) which 
has been obtained for the homogeneously filled bicone, and finally end up with the relation
\begin{multline}
\label{bicone-inhom-2}
\left( \kappa^{(1)}_{\mathrm{sl}} - \kappa^{(1)}_{\mathrm{sg}} \right) 
  \pi r^{(1)}_{\mathrm{lg}} \cot \theta   
         + 2 \pi r^{(1)}_{\mathrm{lg}} \tau ^{(1)} 
                                           \\ 
=
\left( \kappa^{(2)}_{\mathrm{sl}} - \kappa^{(2)}_{\mathrm{sg}} \right)
  \pi r^{(1)}_{\mathrm{lg}} \cot \theta 
         + 2 \pi r^{(1)}_{\mathrm{lg}} \tau ^{(2)}   . 
\end{multline}
Equation (\ref{bicone-inhom-2}) states that a certain combination of curvature and line 
contributions is invariant under a notional shift of the solid--fluid dividing interface. 
In order to fix the transformation behavior of the individual contributions under a   
notional shift, an additional convention is required. Insisting  
on using Eq. (\ref{curvature-7}), which appears to be a most plausible choice for the 
kind of systems considered, fixes the transformation behavior
of the difference of curvature coefficients which appear in Eq. (\ref{bicone-inhom-2}).  
This leads to Eq. (\ref{deltaeta2}) which tells that $\tau$ does depend on the 
choice of the solid--fluid dividing interface.  
\subsection{The drop revisited} 

The surface tensions at planar solid--fluid interfaces depend on the position of the dividing
interfaces as given in Eq. (\ref{deltasigmasf}). Therefore, the difference 
$\sigma_{\alpha \gamma} - \sigma_{\beta \gamma}$ of the gas--solid and the liquid--solid 
interface tensions at coexistence of the gas and the liquid phase does not depend on the 
choice of the solid--fluid dividing interface. In our analysis of the drop in Sect. V 
we could make use of that property because in defining $\tau$ we have used 
solid--fluid interface tensions at the very same pressure irrespective of whether 
the fluid phase is the gaseous phase outside the drop or the liquid phase in the interior of the drop.   
We have also discussed the possibility of introducing instead  
a quantity $(\sigma_{\alpha \gamma} - \sigma_{\beta \gamma})'$  in which the 
two interface tensions are meassured at different pressures. We also have given  
reasons why we have not pursued this possibility further. 
\begin{figure}[h]
\includegraphics*[scale=.30]{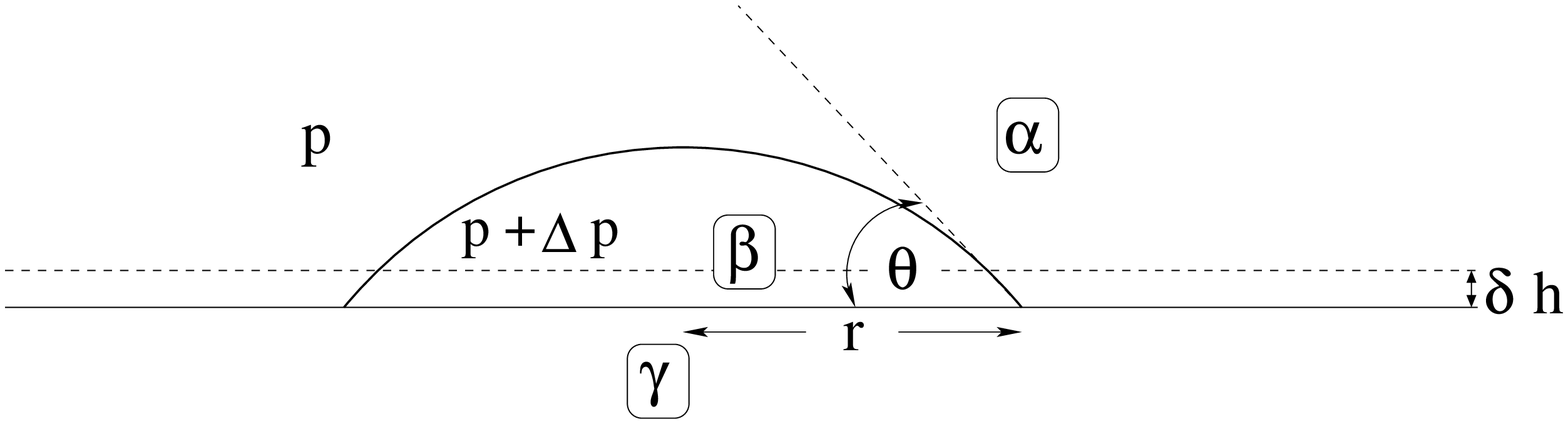}
\caption{\label{Fig-drop.eps} A sessile liquid drop on a planar substrate ($\gamma$). Two choices for the 
 substrate--fluid dividing interface separated by a distance $\delta h$ are shown.
 The contact angle  $\theta$ is only shown for that choice of the dividing interface, which is indicated
 by the solid horizontal line. The pressure in the liquid phase ($\beta$) differs by $\Delta p$ from that
 in the  gas phase ($\alpha$). The drop is a spherical cap with radius $R = r/\sin \theta$,
 area $A = 2\pi R^2 ( 1-\cos \theta)$, and volume $V = (\pi/3)R^3( 2 - 3\cos \theta + \cos ^3 \theta)$.  
}
\end{figure}
However, in view of the
conflicting results for the transformation behavior under notional shifts of dividing
interfaces between different definitions of line tensions we now seriously consider
this option. 
\\ 
In order to make our reasoning as transparent as possible we discuss now  
only a notional shift of the substrate--fluid dividing interface  as shown in Fig. \ref{Fig-drop.eps},
but leave the liquid--gas dividing interface fixed.
We again use the decomposition in Eq. (\ref{lensdrop4b}) for two positions of the 
substrate--fluid dividing interface and use at first the relation  
\begin{equation}
\label{interface-tension-contrast}  
  (\sigma^{(2)}_{\alpha \gamma} - \sigma^{(2)}_{\beta \gamma}) - 
  (\sigma^{(1)}_{\alpha \gamma} - \sigma^{(1)}_{\beta \gamma}) = 0 
\end{equation} 
which is valid if both interface tensions are measured at the same pressure (and if for both 
the $\alpha$--$\gamma$ and the $\beta$--$\gamma$ interface 
the dividing interface is positioned at the same height). After neglecting terms which do not scale with
at least a linear extension of the system one arrives at 
\begin{multline}
\label{drop-again-1}
2 \pi r \left( \tau^{(1)} - \tau^{(2)} \right)  = - \Delta p \left( V^{(2)} - V^{(1)} \right)     
                                                                              \\  
                 + \sigma_{\alpha \beta }(R) \hspace{-0.10cm} \left( A^{(2)} - A^{(1)} \right)   
         +  \pi \left( \sigma_{\beta \gamma} - \sigma_{\alpha \gamma} \right)
                   \hspace{-0.10cm}  \left( (r^{(2)})^2 - (r^{(1)})^2 \right) \hspace{-0.10cm}.
\\            
\end{multline}
Since in Eq. (\ref{drop-again-1}) neither the liquid--gas surface tension $\sigma_{\alpha \beta }(R)$
( in the relevant order) nor the difference $\left( \sigma_{\beta \gamma} - \sigma_{\alpha \gamma} \right)$ depend on the 
choice of the substrate--fluid dividing interface, the line tension $\tau$ is the only physical
quantity that can pick up all the notional changes of volume and interfacial contributions.
In particular, the notional change of the volume contribution  
(in leading order $  V^{(2)} - V^{(1)}  = -\pi r^2 \delta h  $ and $ \Delta p \propto 1/r$) 
does also contribute to the notional change of $\tau$. Evaluating Eq. (\ref{drop-again-1}) 
up to leading order, i.e., neglecting corrections to $\tau$ of the order of $1/r$,
one inevitably comes to the conclusion that $\tau$ is independent of the choice of the
substrate--fluid dividing interface. 
\\  
The only sensible way how an alternative definition of
the line tension $\tau$ could be introduced is to use in the decomposition scheme of  
Eq. (\ref{lensdrop4b}) a solid--liquid interfacial tension $\sigma_{\beta \gamma}(p + \Delta p)$
taken at the pressure $p + \Delta p$ of a bulk liquid in contact with the solid,
whereas the solid--gas interfacial tension $\sigma_{\alpha \gamma}(p )$ is taken at the 
pressure $p$. One may argue that this way one mimics the actual conditions at the interfaces 
of the liquid-drop--vapor--solid system. On the other hand the interfacial tensions 
$\sigma_{\alpha \gamma}$ and $\sigma_{\beta \gamma}$ are not
measurable individually, whereas the difference $\sigma_{\beta \gamma}(p) - \sigma_{\alpha \gamma}(p)$            
is a measurable quantity. Nevertheless, we now seriously consider that definition of $\tau$. 
In order to study the transformation of this newly defined $\tau$ under notional changes we   
proceed as at the beginning of this subsection with the only difference that instead of 
Eq. (\ref{interface-tension-contrast}) we have to use now 
\begin{equation}
\label{interface-tension-contrast-b}
  (\sigma^{(2)}_{\alpha \gamma}(p) - \sigma^{(2)}_{\beta \gamma}(p + \Delta p)) -
  (\sigma^{(1)}_{\alpha \gamma}(p) - \sigma^{(1)}_{\beta \gamma}(p + \Delta p)) = \Delta p \delta h 
                   .  
\end{equation}
If we carry out the analysis neglecting all terms which obviously can only give rise to
corrections to $\tau$ of the order $1/r$, i.e., terms which cannot contribute to
a line tension in its strict sense, we arrive at 
\begin{multline} 
\label{drop-again-2}
  \tau^{(2)} - \tau^{(1)} = 
                                     \\ 
                 \frac{\delta h}{\sin \theta} \left[ \sigma_{\alpha \beta}(R)  
          + ( \sigma^{(1)}_{\beta \gamma}(p + \Delta p) - \sigma^{(1)}_{\alpha \gamma}(p)) \cos \theta
                                                                           \right]  
                  \quad .
\end{multline}   
We now replace $\sigma^{(1)}_{\beta \gamma}(p + \Delta p)$ in Eq. (\ref{drop-again-2}))  by  
$\sigma^{(1)}_{\beta \gamma}(p)$ because the difference of the two quantities is of the order of $1/r$
(we then can skip the superscript $^{(1)}$ in the difference between the solid--liquid and the 
solid--gas surface tensions) and with the same argument we replace $\sigma_{\alpha \beta}(R)$
by its planar limit $\sigma_{\alpha \beta}^\infty$. We then can use Young's law and finally
arrive at      
\begin{equation}
\label{drop-again-3}  
\tau^{(2)} - \tau^{(1)} = \sigma_{\alpha \beta}^\infty \delta h \sin \theta 
                   \quad ,  
\end{equation}
i.e., we again find the transformation in Eq. (\ref{deltaeta2}), as we did  
repeatedly in discussing systems with planar liquid--gas interfaces.    
The reason why this second definition of the line tension leads us again to the transformation in  
Eq. (\ref{deltaeta2}) can be traced back to the fact that in the second definition a contribution
to the notional change of the volume contribution
$\Delta p \left( V^{(2)} - V^{(1)} \right)$ is absorbed by a corresponding notional change
of $\sigma_{\alpha \gamma}(p) - \sigma_{\beta \gamma}(p + \Delta p)$, whereas in the first 
definition this possibility does not exist.     
\subsubsection{Relation between the two line tensions}   
We now derive a relation between the two values of $\tau$ determined for the same system
under identical conditions, but using the two different definitions introduced above.    
In order to do so 
we compare the two decompositions of the grand canonical potential, on which  
the two definitions are based on, for a given, but otherwise arbitrary, choice of dividing
interfaces. In order to distinguish the two definitions we denote with $\tau$ the value
of the line tension determined according to the decomposition scheme introduced in Sect. V
(see Eq. (\ref{lensdrop4}) with $\sigma_{\alpha \gamma}(p)$  and $\sigma_{\beta \gamma}(p)$),  
which we sketched again at the beginning of this subsection. With $\tau_{\mathrm{w}}$
we denote the line tension determined via the second decomposition scheme 
(see Eq. (\ref{lensdrop4}) but with $\sigma_{\alpha \gamma}(p)$  and $\sigma_{\beta \gamma}(p+\Delta p)$). 
On the basis of its transformation behavior and because effects due to the Laplace pressure,
which do not occur for planar liquid--gas interfaces, are extracted from its definition, we identify 
$\tau_{\mathrm{w}}$ with
the line tension as defined for the previously discussed systems with planar liquid--gas 
interfaces (it shall be understood that the identification applies to the line tension in its
strict sense, i.e., disregarding subleading terms). The comparison
leads to the relation      
\begin{equation}
\label{drop-again-4}
\tau = \tau_{\mathrm{w}} + \frac{r}{2} \left[ \sigma_{\beta \gamma}(p + \Delta p) - 
                                                 \sigma_{\beta \gamma}(p) \right] 
                   \quad .  
\end{equation}
For a one-component fluid this relation can be re-expressed  
by using 
\begin{equation}
\label{drop-again-5}
\sigma(p + \Delta p) = \sigma(p) + \frac{\partial \sigma }{\partial \mu} 
                    \frac{\partial \mu}{\partial p} \Delta p  
    \quad ,
\end{equation}
where $\mu$ is the chemical potential, and  
\begin{equation}
\label{drop-again-6}
\frac{\partial \mu}{\partial p} =  \frac{1}{\rho_\mathrm{b}}
\quad ,
\end{equation}
where $\rho_\mathrm{b}$ is the bulk density of the fluid and  
\begin{equation}
\label{drop-again-7}
\frac{\partial \sigma }{\partial \mu} = -\Gamma =: - \frac{1}{A}\left[ \int _{V} \rho({\bf r})\mathrm{d}^3r -
                                                             \rho_\mathrm{b}V \right] 
\quad  ,
\end{equation} 
whith $\Gamma$ as the excess adsorption. 
Equation (\ref{drop-again-7}) can be easily derived from the definition of the interface tension
\begin{equation}
\label{drop-again-8}
\sigma = \frac{1}{A} \left( \Omega \left [ \rho _{\mathrm{eq}} ({\bf r}) \right ] + pV \right)  , 
\end{equation} 
where $\rho _{\mathrm{eq}} ({\bf r})$ is the equilibrium number density minimizing $\Omega$, 
and the general functional form of the grand canonical potential: 
\begin{equation}
\label{drop-again-9}
\Omega \left[ \rho ({\bf r}) \right] = {\cal{F}} \left( \rho ({\bf r}) \right) + 
                                 \int \rho ({\bf r}) \left( V_{\mathrm{ext}} - \mu \right)\mathrm{d}^3r  
\quad .
\end{equation} 
Combining Eqs. (\ref{drop-again-4}) -- (\ref{drop-again-9}) and expressing the Laplace pressure
$\Delta p$ in terms of the liquid--gas surface tension one obtains ($\rho_{\mathrm{b, \beta}}$
is the bulk density in the $\beta$ phase):
\begin{equation}
\label{drop-again-10}
\tau = \tau_{\mathrm{w}} - \frac{\Gamma_{\beta \gamma}}{\rho_{\mathrm{b, \beta}}} 
                                                 \sigma_{\alpha \beta} \sin \theta
\quad  .
\end{equation}
$\Gamma_{\beta \gamma}$ is the excess adsorption at the planar $\beta$--$\gamma$ interface. 
In Eq. (\ref{drop-again-10}) both terms on the right hand side depend on the choice of 
the liquid--solid dividing ($\beta$--$\gamma$) interface but these dependences cancel and thus $\tau$  
is independent of that choice as it should.
\\   
The discussion in the present section has shown 
that the contradiction between our statement in Sect. V,  
that $\tau$ is independent of the choice of dividing interfaces, and Eq. (\ref{deltaeta2}) is resolved 
as follows. We have found that the definitions of the line tension at a 
substrate--fluid--fluid interface, which have been chosen on one hand in Sect. V  
and on the other hand in Eq. (\ref{deltaeta2}), are different. The line tension,    
which appears in Eq. (\ref{deltaeta2}), is $\tau_{\mathrm{w}}$ and differs from $\tau$ as defined in Sect. V.
The possibility to define two different line tensions arises because it is not obvious how
to subtract the contribution to the grand canonical potential stemming from the planar
$\beta$--$\gamma$ (liquid--substrate) interface of a drop of $\beta$-phase in contact with
a substrate.
The liquid--substrate interface tension $\sigma_{\beta \gamma}$ is influenced by the Laplace pressure, but  
$\sigma_{\beta \gamma}$ is not accessible. However, the difference 
$\sigma_{\beta \gamma}(p)-\sigma_{\alpha \gamma}(p)$ of the fluid--substrate interface tensions
is measurable  
via the contact angle for large drops. If in the decomposition scheme defining the 
line tension we use an interface tension at a pressure deviating from the true pressure in 
the interior of the drop, i.e., $\sigma_{\beta \gamma}(p)$,  
in order to avoid the somewhat artificial combination 
$\sigma_{\beta \gamma}(p+\Delta p)-\sigma_{\alpha \gamma}(p)$,  one is led to the definition
$\tau$ for the line tension. If instead one uses $\sigma_{\beta \gamma}(p+\Delta p)$ 
in the decomposition one arrives at the definition $\tau_{\mathrm{w}}$.         
Corresponding alternative choices do not exist in the case of the contact of three fluid phases,
since all fluid--fluid interface tensions are measurable individually and 
since the dependence of interface tensions on curvature radii parametrize their dependence on
the Laplace pressure in a natural way. By contrast, the geometry of the solid--liquid interface
is not influenced by the Laplace pressure and it stays planar for all values of the Laplace pressure.     
\par
The definition $\tau_{\mathrm{w}}$ for the line tension has the merit that interface contributions
due to the modification of the $\beta$--$\gamma$ interface tension caused by the Laplace pressure, 
giving rise to a contribution proportional to the linear extension of the system, 
are not implicitly included in $\tau_{\mathrm{w}}$. However, in order to use $\tau_{\mathrm{w}}$ 
in equations determining the contact angle, it is necessary to re-express a number of equations 
given for the drop in Sect. V. Equation (\ref{drop1}) expressed in terms of $\tau_{\mathrm{w}}$ 
and its notional derivatives turns into    
\begin{equation}
\label{drop1-neu} 
\left[ \frac{\mathrm{d} \tau_{\mathrm{w}}}{\mathrm{d} h}  \right] +\cos \theta 
  \left[ \frac{\mathrm{d} \tau_{\mathrm{w}} }{\mathrm{d} R}  \right] =  
   \sin \theta \sigma_{\alpha \beta }
\end{equation}
and instead of Eq. (\ref{drop2}) one obtains  
\begin{equation}
\label{drop2-neu}
\sigma_{\alpha \beta } \cos \theta + (\sigma_{\beta \gamma}(p+\Delta p) - \sigma_{\alpha \gamma }(p)) 
         = - \frac{\tau_{\mathrm{w}}}{r} - \sin \theta \left[
\frac{\mathrm{d} \tau_{\mathrm{w}} }{\mathrm{d} R}  \right] ,
\end{equation}
or in an alternative notation 
\begin{multline}
\label{drop2-neu-b}
\sigma_{\alpha \beta } \cos \theta + (\sigma_{\beta \gamma}(p) - \sigma_{\alpha \gamma }(p))
         = -(\sigma_{\beta \gamma}(p+\Delta p) - \sigma_{\beta \gamma }(p)) \\ 
           - \frac{\tau_{\mathrm{w}}}{r} - \sin \theta \left[
\frac{\mathrm{d} \tau_{\mathrm{w}} }{\mathrm{d} R}  \right] \quad.
\end{multline}
The relations between the notional derivatives of $\tau$ and $\tau_{\mathrm{w}}$ are 
\begin{equation}
\label{drop2-neu-c}
\left[\frac{\mathrm{d} \tau }{\mathrm{d} R}  \right] = \left[\frac{\mathrm{d} \tau_{\mathrm{w}} }{\mathrm{d} R}  \right]
  + \frac{1}{2 \sin \theta} \left ( \sigma_{\beta \gamma}(p+\Delta p) - \sigma_{\beta \gamma }(p) \right )
\end{equation}
and 
\begin{equation}
\label{drop2-neu-d} 
\left[ \frac{\mathrm{d} \tau}{\mathrm{d} h}  \right] = \left[ \frac{\mathrm{d} \tau_{\mathrm{w}}}{\mathrm{d} h}  \right]
  - \frac{\cos \theta}{2 \sin \theta} \left ( \sigma_{\beta \gamma}(p+\Delta p) - \sigma_{\beta \gamma }(p) \right ) 
  - \frac{r \Delta p}{2}  . 
\end{equation}  
In terms of the variables $\theta$ and $r$ one obtains 
\begin{equation}
\label{drop3-neu}
 \left[ \frac{\mathrm{d} \tau_{\mathrm{w}} }{\mathrm{d} \theta}  \right]   = - \frac{r}{\sin \theta} \sigma_{\alpha \beta}
\end{equation}
and
\begin{equation}
\label{drop4-neu}
\begin{split}
\sigma_{\alpha \beta } \cos \theta + (\sigma_{\beta \gamma}(p+\Delta p) - \sigma_{\alpha \gamma }(p))  =  \\ 
- \frac{\tau_{\mathrm{w}}}{r} - \left[ \frac{\mathrm{d} \tau_{\mathrm{w}} }{\mathrm{d} r }  \right]        
 - \frac{\sin \theta \cos \theta}{r}    \left[ \frac{\mathrm{d} \tau_{\mathrm{w}} }{\mathrm{d} \theta }  \right] , 
\end{split}
\end{equation}
or 
\begin{equation}
\label{drop4-neu-b}
(\sigma_{\beta \gamma}(p+\Delta p) - \sigma_{\alpha \gamma }(p))  =  
  - \frac{\tau_{\mathrm{w}}}{r} - \left[ \frac{\mathrm{d} \tau_{\mathrm{w}} }{\mathrm{d} r }  \right] ,  
\end{equation}
which replace Eqs. (\ref{drop3}) and (\ref{drop4}).  
It is interesting to see that in Eq. (\ref{drop4-neu-b}) the term $\sigma_{\alpha \beta } \cos \theta$
drops out due to the identity Eq. (\ref{drop3-neu}), i.e., this term must now be included in the term 
$\left[ \frac{\mathrm{d} \tau_{\mathrm{w}} }{\mathrm{d} r }  \right]$. This is displayed in the 
relation 
\begin{equation}
\label{drop4-neu-c}
\begin{split}
\left[ \frac{\mathrm{d} \tau_{\mathrm{w}} }{\mathrm{d} r }  \right] = \sigma_{\alpha \beta } \cos \theta  
    - \frac{1}{2} (\sigma_{\beta \gamma}(p+\Delta p) - \sigma_{\beta \gamma }(p))  \\  + 
    \left[ \frac{\mathrm{d} \tau }{\mathrm{d} r }  \right] + \frac{\sin \theta \cos \theta}{r} 
    \left[ \frac{\mathrm{d} \tau }{\mathrm{d} \theta }  \right]  . 
\end{split}    
\end{equation}   
Equations (\ref{drop3-neu}) - (\ref{drop4-neu-c}) show that using $\tau_{\mathrm{w}}$ instead of $\tau$ 
spoils the clear hierarchy in the various terms describing notional changes. 
\\
For completeness we provide also the transformation laws for the notional derivatives of $\tau_{\mathrm{w}}$:  
\begin{equation}
\label{drop4-neu-d}
\left[ \frac{\mathrm{d} \tau_{\mathrm{w}} }{\mathrm{d} R }  \right]^{(2)} - 
\left[ \frac{\mathrm{d} \tau_{\mathrm{w}} }{\mathrm{d} R }  \right]^{(1)} = \frac{\sigma_{\alpha \beta }}{r} \cos \theta
                                                                            [\mathrm{d} R] ,     
\end{equation}  
\begin{equation}
\label{drop4-neu-e}
\left[ \frac{\mathrm{d} \tau_{\mathrm{w}} }{\mathrm{d} h }  \right]^{(2)} -
\left[ \frac{\mathrm{d} \tau_{\mathrm{w}} }{\mathrm{d} h }  \right]^{(1)} = - \frac{\sigma_{\alpha \beta }}{r} \cos \theta
                                                                            [\mathrm{d} h] ,
\end{equation}
\begin{equation}
\label{drop4-neu-f}
\left[ \frac{\mathrm{d} \tau_{\mathrm{w}} }{\mathrm{d} \theta }  \right]^{(2)} -
\left[ \frac{\mathrm{d} \tau_{\mathrm{w}} }{\mathrm{d} \theta }  \right]^{(1)} = - \sigma_{\alpha \beta }
                                                                            [\mathrm{d} R] ,
\end{equation}
and 
\begin{equation}
\label{drop4-neu-g}
\left[ \frac{\mathrm{d} \tau_{\mathrm{w}} }{\mathrm{d} r }  \right]^{(2)} -
\left[ \frac{\mathrm{d} \tau_{\mathrm{w}} }{\mathrm{d} r }  \right]^{(1)} = \frac{\sigma_{\alpha \beta } \sin \theta}{r} 
                                                                            [\mathrm{d} h] .
\end{equation}
\par 
We make now again contact with a variational approach with the constraint of fixed volume.
Of course one could stick to the variational approach as discussed in Sect. V without any modifications.
In that case one just might want to rewrite Eqs. (\ref{drop12}) and (\ref{drop13}) in terms of $\tau_{\mathrm{w}}$
and its notional derivatives. Alternatively, one may slightly modify the variational approach by
introducing a $\beta$--$\gamma$ interface tension measured at a pressure $p + \Delta p$ in the same 
way as in the definition of $\tau_{\mathrm{w}}$. As in Sect. V we further introduce stiffnesses 
$\frac{\mathrm{d} \tau_{\mathrm{w}} }{\mathrm{d} \theta}  \Big\vert $ and 
$\frac{\mathrm{d} \tau_{\mathrm{w}} }{\mathrm{d} r}  \Big\vert $ but we do not introduce any
further stiffness. (One might contemplate to endow the $\beta$--$\gamma$ interface with a stiffness  
but there is no geometric measure characterizing that interface which could be related to such a stiffness;
the area is already used and the radius of cirumference is better attributed to the contact line.) 
In that way one arrives at 
\begin{equation}
\label{drop8-neu}
\sin ^2 \theta  \frac{\mathrm{d} \tau_{\mathrm{w}}  }{\mathrm{d} \theta}  \Big\vert  
    = \frac{r^2}{2}   \frac{\mathrm{d} \sigma_{\alpha \beta}
        }{\mathrm{d} R} \Big\vert
\end{equation}
and 
\begin{equation}
\label{drop12-neu}
\begin{split}
\sigma_{\alpha \beta } \cos \theta + (\sigma_{\beta \gamma}(p + \Delta p) - \sigma_{\alpha \gamma }(p)) 
   = - \frac{\tau_{\mathrm{w}}}{r} -
\frac{\mathrm{d} \tau_{\mathrm{w}} }{\mathrm{d} r }  \Big\vert   \\
      - \frac{ \sin \theta \cos \theta }{r}
\frac{\mathrm{d} \tau_{\mathrm{w}} }{\mathrm{d} \theta }
\Big\vert  .
\end{split}
\end{equation}
One also obtains the relations 
\begin{equation}
\label{drop12-neu-b}
\frac{\mathrm{d} \tau_{\mathrm{w}} }{\mathrm{d} r }  \Big\vert = 
        \left[ \frac{\mathrm{d} \tau_{\mathrm{w}} }{\mathrm{d} r }  \right] 
       - \sigma_{\alpha \beta } \cos \theta - \frac{r \cos \theta}{2 \sin \theta}
       \left[ \frac{\mathrm{d} \sigma_{\alpha \beta }}{\mathrm{d} R} \right]   ,  
\end{equation} 
\begin{equation}
\label{drop12-neu-c}
\frac{\mathrm{d} \tau_{\mathrm{w}} }{\mathrm{d} r }  \Big\vert = \frac{\mathrm{d} \tau}{\mathrm{d} r }  \Big\vert
            - \frac{1}{2}\left( \sigma_{\beta \gamma}(p + \Delta p) -\sigma_{\beta \gamma}(p )\right)   , 
\end{equation}  
\begin{equation}
\label{drop12-neu-d}
\frac{\mathrm{d} \tau_{\mathrm{w}} }{\mathrm{d} \theta } \Big\vert = \frac{\mathrm{d} \tau}{\mathrm{d} \theta } \Big\vert 
   ,
\end{equation}
and the transformation law 
\begin{equation}
\label{drop12-neu-e}
\frac{\mathrm{d} \tau_{\mathrm{w}} }{\mathrm{d} \theta } \Big\vert ^{(2)} - 
\frac{\mathrm{d} \tau_{\mathrm{w}} }{\mathrm{d} \theta } \Big\vert ^{(1)} = 0 .
\end{equation} 
%
%
%
\section{Summary}
\renewcommand{\theequation}{7.\arabic{equation}}
\setcounter{equation}{0}\vspace*{0.5cm} 
We have discussed some conceptual issues that arise in a macroscopic thermodynamic
description of the three-phase contact of either three fluid phases or
of two fluid phases meeting an inert solid substrate. We have pointed
out that the concept of a line tension accompanying the
contact line has to be used with great care. The conceptual difficulties
arise because the interfaces between two phases are always diffuse and never
sharp. Therefore a certain degree of freedom exists in positioning, somewhere
in the transition region, an idealized mathematical interface (the so called
Gibbs dividing interface) separating the two adjacent phases.
As a consequence, a similar degree of freedom exists in the position of the contact line
defined as the common intersection of three dividing interfaces.
\par
We have analyzed implications for a consistent description of the contact line
following from the existence of these degrees of freedom.
For that purpose we have investigated two representative systems
of three-phase contact: a liquid lens at a fluid--fluid interface and a
liquid drop on top of a smooth substrate.  Both systems are  
used in experimental attempts to determine a line tension $\tau$ from the dependence
of contact angles on the system size.
We have defined a prescription for decomposing the grand canonical free energy
of a liquid lens or drop into volume, interface, and line contributions which
renders a line tension $\tau$ independent of a particular choice of
the dividing interfaces. The prescription rests on geometrical definitions and on the notion that those
interfaces of a lens or drop system, which are spherical segments, should be described in
the same way as the interfaces of completely spherical drops.
In particular this means that the pressure drop across the curved interface is
related to the surface tension (and surface stiffness against changes of the radius of curvature $R$)
via the generalized Laplace equation (\ref{lensdrop6}) and that the surface tension
of a spherical interface is independent of a notional shift $[\mathrm{d}R]$
of the dividing interface up to and including contibutions proportional to $ ~ 1/R$ for large $R$.
We have also used the fact that interface tensions of planar interfaces between two fluids
in thermal equilibrium are independent of the position of the dividing interfaces.  
The same is implied 
for the difference between the
substrate--gas (substrate--vapor) and the substrate--liquid interface tensions
although the substrate on one hand and the fluid phases on the other are not
in equilibrium. In fact this is true if the substrate--liquid and the substrate--gas interface 
tensions used in the decomposition scheme correspond to the same pressure, a prescription
which has the advantage of relating the difference of these two quantities directly to the 
contact angle of macroscopic drops which is a measurable quantity.   
Our result, that $\tau$ is independent of the choice of
dividing surfaces, is first of all a generalization to {\em curved} interfaces and {\em curved}
contact lines of what is known for the line tension of a {\em straight}
contact line at the intersection of three {\em planar} interfaces in a genuine three-phase contact \cite{RowWid,Wid1}.
A further generalization is the one from a genuine three-phase contact to a contact between
two fluid phases and an inert solid phase.
It should be noted, however, that in the latter case an alternative definition of the 
line tension, denoted as $\tau_{\mathrm{w}}$, is possible which seems to be more natural 
for systems containing planar liquid--gas interfaces and planar or curved solid walls and more useful 
for the purpose of computing line tensions by exploiting the simplest possible geometries. 
It should be emphasized that $\tau_{\mathrm{w}}$ does depend on the choice of dividing interfaces. 
The difference in the definition of this alternative line tension relative to the previous one  
rests on choosing in the decomposition scheme of the grand canonical potential a different   
substrate--liquid interface tension which is not taken at the pressure of the gas phase, 
which is implied for the substrate--gas interface tension,    
but at a pressure which is enhanced by 
the Laplace pressure.  
We have provided a simple relation between the values of the line tensions corresponding 
to the two alternative definitions (Eqs. (\ref{drop-again-4}) and (\ref{drop-again-10})).      
\par
We have further pointed out that the generalized Neumann or Young equations obtained from a minimization
principle by simply adding a line-tension contribution to the free energy
suffer from internal inconsistencies.
The purely geometric relations between two sets of contact angles obtained for two different
choices of the dividing interfaces are at variance with those equations and with our result from the
decomposition of the free energy which states that the line tension $\tau$ is
independent of such descriptive ambiguities. In the case of a drop on a solid substrate 
using the alternative definition $\tau_{\mathrm{w}}$ of a line tension cannot 
resolve those inconsistencies.   
\par 
In order to find equations for the contact angles which are internally consistent we have followed
two different routes. The first one is determined by the observation that the
grand canonical free energy must be independent of notional (descriptive) changes of the system, i.e.,
of 'parallel' shifts of the dividing interfaces at fixed physical configurations. In the mathematical formulation
we introduced  notional changes of the line tension and corresponding
notional derivatives of $\tau$ with respect to contact angle(s) and to the contact-line radius
in addition to the well established notional derivatives of surface tensions with respect to radii of curvature.
The second route follows a minimization of the grand canonical free energy under the constraint of fixed
drop or lens volume. In contrast to the previous formulations we have included into the free energy 
contributions from the stiffness of the interfaces with respect to a change of the radius of curvature 
and from stiffnesses of the contact line with respect to changes in contact angle(s) and to the contact-line radius,
all at fixed thermodynamic variables.
The equations following from these two routes have been compared and combined into one
(set) of equations. For the lens these are Eqs. (\ref{lens22}) and (\ref{lens23}) together with
the relations in Eqs. (\ref{lens25}) -- (\ref{lens27}) and also Eqs. (\ref{lens18}) and (\ref{lens19}) relating the
stiffness constants and notional derivatives. For the drop the corresponding equations are given
by Eq. (\ref{drop12}) and the relations in Eqs. (\ref{drop13}) and (\ref{drop8}).
Furthermore we have found that the actual values of the stiffness constants depend on the choice
of the dividing interfaces. The relations between two sets of stiffness constants for two different
sets of dividing interfaces are given by Eqs. (\ref{lens16}) and (\ref{lens17}) for the lens and by
Eq. (\ref{drop11}) for the drop. (In case of the drop one could also use the alternative Eqs. 
(\ref{drop8-neu}) to (\ref{drop12-neu-e}).) 
\par
At this point it is interesting to see that for reasons similar to those which compelled us to introduce
notional derivatives of  $\tau$ with respect to contact angles or stiffnesses of $\tau$ with respect
to changes of contact angles, Djikaev and Widom \cite{Widom-n2} (see also Ref. \cite{Widom-n3}) 
introduced a kind of 'derivative'
of $\tau$ with respect to contact angle in order to restore invariance against notional shifts of
dividing interfaces of their linear adsorption equation for a straight contact
line at a genuine three-phase contact.
Even closer to our discussion is the work by Rusanov et al. \cite{Rusanov-1}, which differs,
however, from ours in a number of important points. First, Rusanov et al. discuss only
drops on a substrate, whereas we discuss both lenses and drops. Second, in contrast to
Rusanov et al. we have
included notional shifts of the substrate--fluid interfaces. Third, Rusanov et al. have not
used a standard variational principle at constant volume 
(typical examples for the use of this principle are given, e.g., in Refs. \cite{Wid3,Nav1,Blok3t})
and they did not give transformation laws between values of stiffness constants for different
choices of dividing interfaces.
Finally we have tried to make clear at which points in our line of arguments there
is still room for chosing different conventions and what follows as a necessity for any sensible convention.
\par
Next we discuss the consequences of our investigations for interpreting experimental data.  
Before doing so, we provide explicit expressions for the change in
contact angles $\beta= 2\pi - (\alpha+\gamma)$ or $\theta$ with respect to
their limiting values $\alpha _0$, $\beta _0$, $\gamma_0$ or $\theta_0$ for macroscopicly large lenses or drops,
respectively (see Fig. 3).
From Eqs. (\ref{lens23}) and (\ref{lens22}) together with Eq. (\ref{lensdrop5}) and
the corresponding equation for $\sigma_{\beta \gamma}$ we obtain
\begin{equation}
\begin{split} 
\label{lens28}
        \cos \beta - \cos \beta _0           &  =       
      - \frac{\sin \beta _0}{r}  \biggl \{ 
        2 \delta _{\alpha \beta}^{\mathrm{T}} \cos \alpha _0 
          \\ 
     & +  2 \delta _{\beta \gamma}^{\mathrm{T}} \cos \gamma _0
       + \frac{ 2 \left ( \tau + r \frac{\mathrm{d} \tau}{\mathrm{d} r} \vert \right ) }
              {\sigma_{\alpha \beta}^\infty \sin \alpha _0 + \sigma_{\beta \gamma}^\infty \sin \gamma _0 }
           \\ 
      & + \frac{ \cos \alpha _0}{\sigma_{\alpha \beta}^\infty } \frac{\mathrm{d} \tau}{\mathrm{d} \alpha} \Big\vert
       + \frac{ \cos \gamma _0}{\sigma_{\beta \gamma}^\infty } \frac{\mathrm{d} \tau}{\mathrm{d} \gamma} \Big\vert
       \biggr \}   
 + \mathrm{O}\left ( \frac{\ln r}{r^2} \right ).
\end{split}
\end{equation}
Similarly, Eq. (\ref{drop12}) together with Eq. (\ref{lensdrop5}) yields  
\begin{equation}
\begin{split}
\label{drop14}
        \cos \theta - \cos \theta _0           &   =         
       \frac{1}{r \sigma_{\alpha \beta}^\infty } \biggl \{ 
                                \left ( 2 \delta _{\alpha \beta}^{\mathrm{T}} \sigma_{\alpha \beta}^\infty 
                                        - \frac{\mathrm{d} \tau}{\mathrm{d} \theta} \Big\vert \right )
                              \sin \theta _0 \cos \theta _0
            \\  
              &   -\tau -r \frac{\mathrm{d} \tau}{\mathrm{d} r} \Big\vert
                                                 \biggr \} 
           + \mathrm{O}\left ( \frac{\ln r}{r^2} \right ).
\end{split}
\end{equation}
The leading type, $ ~ \frac{\ln r}{r^2} $, of correction terms arises due to algebraically decaying dispersion
forces among the particles (see Ref. \cite{T0}). 
In Eqs. (\ref{lens28}) and (\ref{drop14}) $r$ is the radius of the circular contact line, 
$\sigma_{\xi \nu}^\infty$ the interface tension of the planar $\xi$--$\nu$ interface, 
$\delta _{\xi \nu}^{\mathrm{T}}$ the Tolman length of the $\xi$--$\nu$ interface,  
$\frac{\mathrm{d} \tau}{\mathrm{d} \alpha} \vert$ etc. are stiffnesses against changes of contact
angles or of the radius of the line, which are attributed to the line, and $\tau$ is the line tension.   
In the traditional analyses of size dependent contact angles only the term proportional to $\tau$ is 
included. However, the omitted terms give contributions to the right hand sides of Eqs. (\ref{lens28}) and
(\ref{drop14}) which are comparable in magnitude to that term. Although $\tau$ is independent of the
chosen dividing interfaces, the stiffnesses $\frac{\mathrm{d} \tau}{\mathrm{d} \alpha} \vert$ and
$\frac{\mathrm{d} \tau}{\mathrm{d} \gamma} \vert$ in the case of the lens and 
$\frac{\mathrm{d} \tau}{\mathrm{d} r} \vert$ and $\frac{\mathrm{d} \tau}{\mathrm{d} \theta} \vert$
in the case of the drop are not. Therefore, the additional terms in Eqs. (\ref{lens28}) and (\ref{drop14})
containing these stiffnesses even depend on 
the positions of dividing interfaces which may be chosen arbitrarily within a certain range.
This is in accordance with the fact that this dependence is also present for the contact angles on the lhs 
of Eqs. (\ref{lens28}) and (\ref{drop14}) (see step 1 in the protocol given below). 
Moreover, the changes in the values of these particular terms with the position of dividing interfaces
are as big as the term proportional to $\tau$ itself. 
In order to demonstrate this we compare $\tau$ with 
$ \frac{\mathrm{d} \tau}{\mathrm{d} \alpha} \vert$, $ \frac{\mathrm{d} \tau}{\mathrm{d} \gamma} \vert$,
$\frac{\mathrm{d} \tau}{\mathrm{d} \theta} \vert$, or $\frac{\mathrm{d} \tau}{\mathrm{d} r} \vert$
as well as with the terms of the form $ 2 \delta ^{\mathrm{T}} \sigma$ involving the Tolman length 
$\delta ^{\mathrm{T}}$.   
The change, e.g., of $ \frac{\mathrm{d} \tau}{\mathrm{d} \alpha} \vert$ with a shift of the 
$\alpha$--$\beta$ dividing surface by $[ \mathrm{d} R_1 ]$ is equal to $ - \sigma_{\alpha \beta} [ \mathrm{d} R_1 ]$
For typical values of
$\sigma_{\alpha \beta}$ of the order of $10^{-2}\, \mathrm{J/m^{2}}$, and for 
$[ \mathrm{d} R_1 ]$ of the order of $1\, \mathrm{nm}$ one obtains 
a change in the value of $ \frac{\mathrm{d} \tau}{\mathrm{d} \alpha} \vert$  
of the order of $10^{-11}\, \mathrm{J/m}$, which is just the typical value of calculated or 'measured' line tensions.
Similar estimates hold for the other terms in particular also for those of the form 
$ 2 \delta ^{\mathrm{T}} \sigma$ if the reasonable assumption is made that 
$ 2 \delta ^{\mathrm{T}}$ is not much smaller than $1\, \mathrm{nm}$
(typical values of $ 2 \delta ^{\mathrm{T}}$ obtained theoretically are of the order of
 half a molecular diameter, see, e.g., Refs. \cite{ThomGu,T2,T7}).      
\par 
In order to analyze experimental data with the help of Eqs. (\ref{lens28}) and (\ref{drop14})
a certain protocol has to be followed as discussed below: 
\begin{enumerate}
\item Giving experimental values of contact angles $\beta$ or $\theta$ requires to 
      define the dividing interfaces relative to which the contact angles are measured.
      (So far this information is missing for basically all published experimental
       efforts to measure the line tension.) 
      After the dividing interfaces have been chosen (iso-density surfaces), the 
      spherical parts of the interface profiles of the drop or lens have to be determined
      by fitting spheres to their central parts (e.g., to the top of the drop). 
      In case of the drop, one such sphere intersects with the planar solid--fluid 
      interface (chosen by convention) at the three-phase-contact line. The contact angle 
      at this line is obtained by a tangential-plane construction to the sphere, i.e., to  
      the spherical extrapolation of the liquid--gas interface down to the solid--fluid interface. 
      In the case of the lens,
      two spheres are defined by the central part of the lens. These spheres intersect at the
      three-phase-contact line, at which tangential 
      planes to the spheres define the contact angles.    
\item The dependence of these angles on the lens or drop sizes, as characterized by $r$, 
      has to be studied in the {\em leading} asymptotic behavior for large $r$. This dependence
      is not only determined by the line tension $\tau$ calculated for a straight 
      contact line but it depends on additional material parameters. Therefore, the 
      common practice of infering $\tau$ from $\beta(r)$ or $\theta(r)$ according to
      Eqs. (\ref{mye10}) or (\ref{mye2}) is not valid. 
\item If one would compare the rhs of Eqs. (\ref{lens28}) and (\ref{drop14}) with theoretical
      quantities, these would have to be evaluated for that choice of the dividing interfaces 
      for which the experimental data on the lhs are given. 
\item The values of certain stiffness constants appearing on the rhs of 
      Eqs. (\ref{lens28}) and (\ref{drop14}) depend sensitively on
      the choice of dividing interfaces. However, the transformation behavior of 
      these material parameters between different such choices
      is known and given by Eqs. (\ref{lens16}) and (\ref{lens17}) for the lens and
      Eq. (\ref{drop11}) for the drop.
\item All quantities on the rhs of Eqs. (\ref{lens28}) and (\ref{drop14}) are accessible to
      theoretical computations, e.g., based on density functional theory. Methods that can 
      be used to calculate the Tolman lengths are described in the literature (see, e.g., Refs. 
      \cite{ThomGu,T1,T1b,T101,T2,T3,T4,T5,T6,T7,T8,T11,T14}). 
\\  
      In calculations of the line tension much care has to be taken not to pick up 
      additional artificial line or edge contributions. In the likely case that 
      within a theoretical approach one has calculated  
      $\tau_{\mathrm{w}}$, its relation to $\tau$ is given by  
      Eqs. (\ref{drop-again-4})
      and (\ref{drop-again-10}), provided both quantities apply to the same thermo\-dynamic
      conditions.   
\\  
      Once the density profiles for a lens or drop 
      have been computed,  
      the stiffness constants $\frac{\mathrm{d} \tau}{\mathrm{d} \theta} \vert$,  
      $\frac{\mathrm{d} \tau}{\mathrm{d} r} \vert$ etc. may be determined from their relations to
      the notional derivatives of $\tau$ (see, e.g., Eq.  (\ref{drop13})) or in the case of 
      $\frac{\mathrm{d} \tau}{\mathrm{d} \theta} \vert$ also to the stiffness constant 
      of the interface (see, e.g., Eq. (\ref{drop8})) which again is given by a notional derivative
      (Eq. (\ref{lensdrop8})). 
      Similar equations which apply to the case of the lens are Eqs. (\ref{lens18}), (\ref{lens19}),
      (\ref{lens25}), (\ref{lens26}), and (\ref{lens27}). $\tau$ itself as well as the  
      notional derivatives of $\tau$ follow from decomposing the grand canonical potential for a series of 
      different choices for the dividing interfaces. For instance one might decompose the grand canonical  
      potential $\Omega$ of a given lens for a series of different values of $[R_1]$ and $[R_2]$.  
      Carrying out this decomposition up to the order $1/r$ and making use of previous knowledge about  
      the notional derivatives of the surface tensions (see Eq. (\ref{lensdrop7})), the notional derivatives 
      $ \left[ \frac{\mathrm{d} \tau}{\mathrm{d} R_1} \right] $ and 
      $ \left[ \frac{\mathrm{d} \tau}{\mathrm{d} R_2} \right] $ 
      can be determined from the dependence of $\tau$ on $[R_1]$ and $[R_2]$. 
      These notional derivatives may be then converted
      into the notional derivatives $ \left[ \frac{\mathrm{d} \tau}{\mathrm{d} \alpha} \right]$     
      and $ \left[ \frac{\mathrm{d} \tau}{\mathrm{d} \gamma} \right]$ via the relations 
      $ \left[ \frac{\mathrm{d} \tau}{\mathrm{d} R_1} \right] = 
                       \frac{ \sin \alpha \cos (\alpha + \gamma ) }{r \sin (\alpha + \gamma )}
                                     \left[ \frac{\mathrm{d} \tau}{\mathrm{d} \alpha} \right] 
                      + \frac{ \sin \gamma }{r \sin (\alpha + \gamma )}
                                     \left[ \frac{\mathrm{d} \tau}{\mathrm{d} \gamma } \right] $  
      and 
      $ \left[ \frac{\mathrm{d} \tau}{\mathrm{d} R_2} \right]  = 
                       \frac{ \sin \alpha }{r \sin (\alpha + \gamma )}
                                     \left[ \frac{\mathrm{d} \tau}{\mathrm{d} \alpha} \right] 
                      + \frac{ \sin \gamma \cos (\alpha + \gamma ) }{r \sin (\alpha + \gamma )}
                                     \left[ \frac{\mathrm{d} \tau}{\mathrm{d} \gamma } \right] $. 
      Similarly the notional derivatives $ \left[ \frac{\mathrm{d} \tau}{\mathrm{d} R} \right] $ and
      $ \left[ \frac{\mathrm{d} \tau}{\mathrm{d} h} \right] $ can be determined in case of the drop
      and converted into $ \left[ \frac{\mathrm{d} \tau}{\mathrm{d} \theta} \right] $ and
      $ \left[ \frac{\mathrm{d} \tau}{\mathrm{d} r} \right] $ via the relations 
      $ \left[ \frac{\mathrm{d} \tau}{\mathrm{d} R} \right] = \frac{ \cos \theta }{r} 
                         \left[ \frac{\mathrm{d} \tau}{\mathrm{d} \theta} \right] + 
          \frac{ 1 }{ \sin \theta } \left[ \frac{\mathrm{d} \tau}{\mathrm{d} h} \right] $  
      and 
      $ \left[ \frac{\mathrm{d} \tau}{\mathrm{d} h} \right] = - \frac{ 1 }{r} 
                         \left[ \frac{\mathrm{d} \tau}{\mathrm{d} \theta} \right] - 
          \frac{ \cos \theta }{ \sin \theta } \left[ \frac{\mathrm{d} \tau}{\mathrm{d} h} \right] $ . 
\end{enumerate}
Although each individual quantity entering the rhs of Eqs. (\ref{lens28}) and (\ref{drop14}) can be
calculated separately in the way indicated above, measurements of the size dependent contact angles
of drops or lenses can provide only certain combinations of material parameters.   
Their values depend sensitively but in a known way on 
the choice of dividing interfaces. 
\par 
There still remain a number of open questions. 
\begin{enumerate}
\item How can the various stiffness constants attributed to the contact line be measured?
      The answer to this question requires the extension of the principles developed
      here for simple geometries to more general geometries. 
      How do the stiffness constants of the contact line influence the equilibrium 
      shapes of more complex structures, e.g., of liquid bridges between solid substrates
      or of lenses and drops distorted from their ideal shape due to external forces? 
\item How can the stiffness constants of the contact line introduced here  
      be related to the 'derivatives' of $\tau$ with respect to contact angles  
      which were introduced by Djikaev and Widom \cite{Widom-n2} in order to restore 
      invariance of their linear adsorption equation against notional shifts of dividing interfaces?  
\end{enumerate} 
\vspace{0.40cm}  
{\large \bf Acknowledgements:}
The authors thank R. Roth and B. Widom for helpful comments. 
One of the authors M.N. expresses his thanks for the hospitality at the Max-Planck-Institut 
f\"ur Metallforschung where most of this work has been done; he also acknowledges the support
via the Polish Ministry of Science and Higher Education grant N 202 076 31/0108. 

\end{document}